\documentclass[9pt]{article} 
\usepackage{graphicx,calc,epsfig,pstricks,bbm} 
\usepackage{a4wide}
\usepackage{amsmath,amssymb,amsfonts} 
\usepackage[bf]{caption2}
\usepackage{rotating}   
\usepackage{psfrag} 
\newcommand{\scr}{\scriptscriptstyle}
\newcommand{\eqa}{\begin{eqnarray}}
\newcommand{\neqa}{\end{eqnarray}}
\newcommand{\equ}{\begin{equation}}
\newcommand{\nequ}{\end{equation}}

\newcommand{\be}{\begin{equation}}
\newcommand{\ee}{\end{equation}}
\newcommand{\ba}{\begin{eqnarray}}
\newcommand{\ea}{\end{eqnarray}}

\def\ket#1{| #1 \rangle}

\def\bb1{\textup{\small{1}} \kern-3.6pt \textup{1}\ }

\DeclareFontFamily{U}{rsfs}{}         % Formal Script            %
\DeclareFontShape{U}{rsfs}{m}{n}{<5> rsfs5 <6><7> rsfs7          %
  <8><9><10><10.95><12><14.4><17.28><20.74><24.88> rsfs10}{}     %
\DeclareMathAlphabet{\mathfs}{U}{rsfs}{m}{n}                     %
\newcommand{\tetraedro}{{\begin{array} {c}
\ifx\JPicScale\undefined\def\JPicScale{1}\fi
\psset{unit=\JPicScale mm}
\psset{linewidth=0.3,dotsep=1,hatchwidth=0.3,hatchsep=1.5,shadowsize=1,dimen=middle}
\psset{dotsize=0.7 2.5,dotscale=1 1,fillcolor=black}
\psset{arrowsize=1 2,arrowlength=1,arrowinset=0.25,tbarsize=0.7 5,bracketlength=0.15,rbracketlength=0.15}
\begin{pspicture}(2,0)(3.33,3.33)
\psline[linewidth=0.15](0.67,0.67)(3.33,0.67)
\psline[linewidth=0.15](0.67,0.67)(2,1.33)
\psline[linewidth=0.15](2,1.33)(3.33,0.67)
\psline[linewidth=0.15](0.67,0.67)(2,2.67)
\psline[linewidth=0.15](2,2.67)(3.33,0.67)
\psline[linewidth=0.15](2,1.33)(2,2.67)
\psline[linewidth=0.15](2,2.67)(2,3.33)
\end{pspicture}
\end{array}\!\!\!}}

\newcommand{\triangolo}{{\begin{array}{c}
\ifx\JPicScale\undefined\def\JPicScale{1}\fi
\psset{unit=\JPicScale mm}
\psset{linewidth=0.3,dotsep=1,hatchwidth=0.3,hatchsep=1.5,shadowsize=1,dimen=middle}
\psset{dotsize=0.7 2.5,dotscale=1 1,fillcolor=black}
\psset{arrowsize=1 2,arrowlength=1,arrowinset=0.25,tbarsize=0.7 5,bracketlength=0.15,rbracketlength=0.15}
\begin{pspicture}(2,0)(3.33,3.33)
\psline[linewidth=0.15](0.67,0.67)(3.33,0.67)
\psline[linewidth=0.15](0.67,0.67)(2,2.67)
\psline[linewidth=0.15](2,2.67)(3.33,0.67)
\psline[linewidth=0.15](2,2.67)(2,3.33)
\end{pspicture}	
\end{array}\!\!\!}}

\author{Emanuele Alesci${}^{1,2}$ and Carlo Rovelli${}^2$\\[2mm]
{\em \normalsize ${}^1$ 
Universit\"at Erlangen, Institut f\"ur Theoretische Physik III,} \\
{\em \normalsize Lehrstuhl f\"ur Quantengravitation Staudtstrasse 7, D-91058 Erlangen, EU
}\\[1mm]
{\em  \normalsize  ${}^2$ Centre de Physique Th\'eorique de Luminy\footnote{Unit\'e mixte de recherche (UMR 6207) du CNRS et des Universit\'es de Provence (Aix-Marseille I), de la M\'editerran\'ee (Aix-Marseille II) et du Sud (Toulon-Var); laboratoire affili\'e \`a la FRUMAM (FR 2291).}, Case 907, F-13288 Marseille, EU}}

\title{\bf A regularization of the hamiltonian constraint compatible with the spinfoam dynamics}
 
 \date{\normalsize \today} 

\begin{document} 
\maketitle  \vspace{-1em}
    \begin{abstract}
\noindent We introduce a new regularization for Thiemann's Hamiltonian constraint. The resulting constraint can generate the 1-4 Pachner moves and is therefore more compatible with the dynamics defined by the spinfoam formalism.  We calculate its matrix elements and observe the appearence of the $15j$ Wigner symbol in these.
     \end{abstract}

%%%%%%%%%%%%%%%%%%%%%%%%%%%%%%%%%%%%%%%%%%%%%%%%%%%%%%%%%%%%%%%%%%%%
%%%%%%%%%                 sec: Introduction              %%%%%%%%%%%
%%%%%%%%%%%%%%%%%%%%%%%%%%%%%%%%%%%%%%%%%%%%%%%%%%%%%%%%%%%%%%%%%%%%

\section{Introduction}

The recent years have seen a steady convergence between the canonical (spin networks) \cite{lqgcan} and the covariant (spinfoams) \cite{lqgcov, BarrettCrane98} versions of loop quantum gravity  (LQG) \cite{lqg}.   In 3-d, the relation between the two formalisms has been clarified in \cite{Noui Perez}. In 4-d, the effort to compute the two-point function of the theory \cite{scattering} has lead to a modification of the covariant theory whose 
background independent \emph{kinematics} matches the canonical one \cite{Engle:2007uq,Livine:2007vk,Kaminski:2009fm}.  Relating the \emph{dynamics} appears to be less obvious \cite{Io karim e francesco}, in spite of the fact that the spinfoam formalism was originally conceived as an exponentiation of the canonical evolution \cite{ReisenbergerRovelli97}.  Here we introduce a small modification of the canonical hamiltonian constraint operator, which brings it closer to the spinfoam dynamics, thus realizing a possible step towards the full convergence of the two formalisms.

The modification we consider is a different regularization of the curvature that appears in the hamiltonian constraint. Instead of point-splitting the curvature over a loop, we point-split it over a tetrahedral graph.  More precisely, instead of writing the curvature as a limit of the holonomy of the connection around a closed loop, we write it as a limit of the 
spin-network function of the connection, associated to a tetrahedral graph.  

The consequences of this alternative regularization are multifold. First, the regularized operator appears to be more natural and more symmetric, especially when acting on four-valent nodes, where it admits a natural simplicial interpretation. In particular, the curvature is evaluated on a plane which appears to be natural from a geometric point of view. Second, and more importantly, when acting on a node the resulting quantum operator generates \emph{three} new nodes, rather than two, as the old Hamiltonian operator. Therefore the constraint implements the 1-4 Pachner move \cite{lqgcov}, which is characteristic of the spinfoam dynamics. Third, the resulting operator creates 4-valent nodes, rather than 3-valent ones, as the old Hamiltonian operator. Since 3-valents nodes have zero volume, the new operator can create nodes with volume.  Finally, when we compute matrix elements of this operator, we find $15j$ Wigner symbols, as well as fusion coefficients \cite{Engle:2007uq,fusion}, namely the basic building blocks of the spinfoam dynamics. 

This paper is organized as follows. In Section II we introduce the tetrahedral regularization of the curvature. In Section III we review the Thiemann's construction of the Hamiltonian constraint and define the new constraint. In Section IV we evaluate some of its matrix elements.  We summarize our results in Section V.  Several Appendices give basic computational tools. 

\section{Tetrahedral regularization of the curvature}

We begin by introducing our main new ingredient for the definition of the hamiltonian constraint: a spin-network regularization of the curvature.  The  hamiltonian constraint is a function of the curvature $F$ of the Ashtekar-Barbero connection $A$ \cite{Barbero95} (see details in the next section).  Since neither $A$ nor $F$ are well defined operators in LQG, the hamiltonian operator is defined by point spitting the curvature in terms of the holonomy $h[\alpha]\in SU(2)$ of $A$ along a closed loop $\alpha$.  Consider a triangular loop $\alpha_{12}$ with vertices $n,n_1, n_2$ and sides $s_{01},s_{02},s_{12}$, and consider the quantity
\begin{equation}
{h}^i_{\triangolo}:= {\rm Tr}\!\left[h_{\alpha_{12}}\tau^i\right] 
\label{tri}
\end{equation}
where $\tau^i$ are the (anti-hermitian) generators of $su(2)$ (which are $i/2$ times the Pauli matrices).  To first order in the area $a$ of the triangle $\alpha_{12}$, this gives curvature at $n$, in the plane defined by the triangle. In locally flat coordinates $x^a, a=1,2,3$, we have, to first order in the area of the triange,
\begin{equation}
{h}^i_{\triangolo} = - \frac14\  F^i_{ab}(n)\, s_{01}^a s_{02}^b .
\end{equation}
More in general \cite{Gaul:2000ba}, we can take the holonomy $h$ and the generators $\tau^i$ in (\ref{tri}) in an arbitrary spin-$m$ representation, and the last equation becomes
\begin{equation}
{h}^i_{\triangolo} = \frac{N^2_m}6\  F^i_{ab}(n)\, s_{01}^a s_{02}^b .
\end{equation}
where $N^2_m={\rm Tr}\!\left[\tau^i\tau^i\right]=-(2m+1)m(m+1)$. Explicitly, calling $h[s]$ the holonomy of $A$ along the segment $s$, writing $s_{ba}$ to indicate the segment $s_{ab}$ with reversed orientation, and using the normalized 3-valent intertwiners $\tau^i{}_{ab}= N_m  i^i{}_{ab}$ (that satisfies $i^i{}_{ab}i^i{}_{ab}=1)$,  we can write \begin{equation}
{h}^{i}_{\triangolo}= N_m\,  i^i{}_{\alpha\beta}\; \;h[{s_{01}}]_{\beta\gamma}\; h[s_{12}]_{\gamma\delta}\; h[s_{20}]_{\delta\alpha}. 
\end{equation}

The idea that we develop in this paper is to replace $h^i_\triangolo\!\!$ with a different object, which we denote 
${h}^i_\tetraedro$, and which is defined by the {\em spin-network function} of the connection $A$, associated to the tetrahedron generated by {\em three} segments $s_{01},s_{02},s_{03}$, emerging from a point $n$, with one open link. Let's define 
\begin{equation}
	 {h}^{i}_\tetraedro
= c_m N_m\ 
i^{i\alpha\beta\gamma}\; i^{\delta\epsilon\zeta}\; i^{\theta\iota\kappa}\; i^{\lambda\mu\nu}\; \;h[{s_{01}}]_{\alpha\delta}\; h[s_{02}]_{\beta\kappa}\; h[s_{03}]_{\gamma\mu}\; h[s_{12}]_{\epsilon\theta}\; h[s_{23}]_{\iota\lambda}\; h[s_{31}]_{\nu\zeta}\; .
\end{equation}
Here $h$ are holomies in a representation of (integer) spin $m$. $i^{\alpha\beta\gamma}$ is the (unique) normalized 3-valent intertwiner between three representations $m$, and $i^{i\alpha\beta\gamma}$ is a normalized 4-valent intertwiner, with the first index in the adjoint representation and the other three in the representation $m$, satisfying $i^{i\alpha\beta\gamma}\;=i^{i\gamma\alpha\beta}\;=i^{i\beta\gamma\alpha}\;$ and $c^{-1}_m = i^{i\alpha\beta\gamma}\; i^{\alpha\epsilon\zeta}\; i^{\epsilon\beta\iota}\; i^{\lambda\gamma\zeta}\;i^{i\lambda\iota}$. The pattern of contractions is given by the following tetrahedral diagram
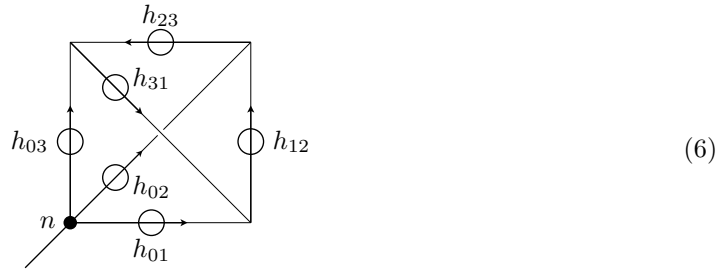
\be
\begin{array}{c}
	\ifx\JPicScale\undefined\def\JPicScale{.6}\fi
\psset{unit=\JPicScale mm}
\psset{linewidth=0.3,dotsep=1,hatchwidth=0.3,hatchsep=1.5,shadowsize=1,dimen=middle}
\psset{dotsize=0.7 2.5,dotscale=1 1,fillcolor=black}
\psset{arrowsize=1 2,arrowlength=1,arrowinset=0.25,tbarsize=0.7 5,bracketlength=0.15,rbracketlength=0.15}
\begin{pspicture}(0,0)(62,58)
\pscircle*(12,12){1.5}
\psline[linewidth=0.3]{*-}(12,12)(2,2)
\psline[linewidth=0.2](12,12)(52,52)
\psline[linewidth=0.2,border=0.8](12,52)(52,12)
\psline[linewidth=0.2](12,12)(12,52)
\psline[linewidth=0.2](12,12)(52,12)
\psline[linewidth=0.2](52,52)(52,12)
\psline[linewidth=0.2](12,52)(52,52)
\psline{<-}(38,12)(12,12)
\psline{<-}(12,38)(12,12)
\psline[linewidth=0.2]{<-}(28,28)(12,12)
\psline{<-}(28,36)(12,52)
\psline{<-}(52,38)(52,12)
\psline{<-}(24,52)(52,52)
\rput(7,12){$n$}
\rput(30,6){$h_{01}$}
\rput(30,20){$h_{02}$}
\rput(3,30){$h_{03}$}
\rput(61,30){$h_{12}$}
\rput(32,58){$h_{23}$}
\rput(30,44){$h_{31}$}
\rput{0}(30,12){\psellipse[](0,0)(2.83,2.83)}
\rput{0}(22,22){\psellipse[](0,0)(2.83,2.83)}
\rput{0}(12,30){\psellipse[](0,0)(2.83,2.83)}
\rput{0}(52,30){\psellipse[](0,0)(2.83,2.83)}
\rput{0}(32,52){\psellipse[](0,0)(2.83,2.83)}
\rput{0}(22,42){\psellipse[](0,0)(2.83,2.83)}
\end{pspicture}
\label{hnew}
\end{array}
\ee
where the circle indicate the presence of the holonomies and the arrows their directions.

In other words, the step we take here is the modification of the regularization of the curvature from a triangle to a tetrahedron:
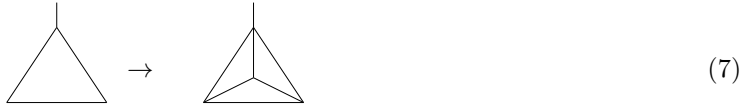
\begin{equation}
{
\begin{array}{c}
\ifx\JPicScale\undefined\def\JPicScale{5}\fi
\psset{unit=\JPicScale mm}
\psset{linewidth=0.03,dotsep=1,hatchwidth=0.3,hatchsep=1.5,shadowsize=1,dimen=middle}
\psset{dotsize=0.7 2.5,dotscale=1 1,fillcolor=black}
\psset{arrowsize=1 2,arrowlength=1,arrowinset=0.25,tbarsize=0.7 5,bracketlength=0.15,rbracketlength=0.15}
\begin{pspicture}(2,0)(3.33,3.33)
\psline[linewidth=0.015](0.67,0.67)(3.33,0.67)
\psline[linewidth=0.015](0.67,0.67)(2,2.67)
\psline[linewidth=0.015](2,2.67)(3.33,0.67)
\psline[linewidth=0.015](2,2.67)(2,3.33)
\end{pspicture}	
\end{array}
}
\to  \hspace{3em}
{\begin{array} {c}
\ifx\JPicScale\undefined\def\JPicScale{5}\fi
\psset{unit=\JPicScale mm}
\psset{linewidth=0.03,dotsep=1,hatchwidth=0.3,hatchsep=1.5,shadowsize=1,dimen=middle}
\psset{dotsize=0.7 2.5,dotscale=1 1,fillcolor=black}
\psset{arrowsize=1 2,arrowlength=1,arrowinset=0.25,tbarsize=0.7 5,bracketlength=0.15,rbracketlength=0.15}
\begin{pspicture}(2,0)(3.33,3.33)
\psline[linewidth=0.015](0.67,0.67)(3.33,0.67)
\psline[linewidth=0.015](0.67,0.67)(2,1.33)
\psline[linewidth=0.015](2,1.33)(3.33,0.67)
\psline[linewidth=0.015](0.67,0.67)(2,2.67)
\psline[linewidth=0.015](2,2.67)(3.33,0.67)
\psline[linewidth=0.015](2,1.33)(2,2.67)
\psline[linewidth=0.015](2,2.67)(2,3.33)
\end{pspicture}
\end{array}}
\end{equation}

The key property of $h^i_\tetraedro$ is that in the limit in which the size of the tetrahedron is small, we have 
\be
h^i_\tetraedro =  h^i_\triangolo
\label{tettri}
\ee
where the triangle is the face of the tetrahedron opposite to the 4-valent node $n$. 

To prove this claim, let's expand the connection $A$ in a Taylor series around $n$. The terms of order zero do not contribute to $h^i_\tetraedro$ because the evaluation of a spinnetwork with a single open end vanishes. The term of order 1 vanishes as well, because $h^i_\tetraedro$  is gauge covariant, and we can always chose a gauge where $A$ vanishes with its first derivatives around a point.  The term of order two does not vanish in general. The only function of $A$ of order two defined at a point is the curvature $F^i_{ab}$. Therefore $h^i_\tetraedro$ must be proportional to the curvature $F^i_{ab}(n)u^a u^b$, for some plane $(u^a\wedge u^b)$.  Since  $h^i_\tetraedro$ does not depend on the coordinates, we can choose coordinates in which the tetrahedron is regular, and then by symmetry the only plane which is selected by the the tetrahedron is the plane of the face opposite to $n$. This proves that $h^i_\tetraedro$ is proportional to $h^i_\triangolo$. To compute the proportionality constant, observe that since $h^i_\tetraedro$ is already of order two, we can deform its sides without affecting it at this order. In particular, we can deform the side 02 to have it run along the sides 01 and 12. Similarly, we deform the side 03 to have it run along the sides 01 and 13. Retracing the holonomies running parallel along 12, and along 13, we obtain (to second order) 
\begin{eqnarray}
	{h}^{i}_\tetraedro
&=&  c_m N_m\ 
i^{i\alpha\beta\gamma}\; i^{\delta\epsilon\zeta}\; i^{\theta\kappa\iota}\; i^{\lambda\mu\nu}\;\;h[{s_{01}}]_{\alpha\delta}\; h[s_{01}]_{\beta\kappa}\; h[s_{01}]_{\gamma\mu}\; \delta_{\nu\zeta}\; \delta_{\epsilon\theta}  \; h[s_{12}]_{\iota x}\; h[s_{23}]_{xy}\; h[s_{31}]_{y\lambda}\; \nonumber \\
&=&  c_m N_m\ 
h[{s_{01}}]^{iw}\; i^{w\alpha\beta\gamma}\; \delta_{\alpha\delta}\; \delta_{\beta\kappa} \delta_{\gamma\mu} \; i^{\delta\epsilon\zeta}\; i^{\theta\kappa\iota}\; i^{\lambda\mu\nu}\; h[s_{12}]_{\iota x}\; h[s_{23}]_{xy}\; h[s_{31}]_{y\lambda}\; \nonumber \\
&=& c_m N_m\ 
i^{i\alpha\beta\gamma}\; i^{\alpha\epsilon\zeta}\; i^{\epsilon\beta\iota}\; i^{\lambda\gamma\zeta}\; h[s_{123}]_{\iota\lambda}\; 
=  c_m\ (i^{i\alpha\beta\gamma}\; i^{\alpha\epsilon\zeta}\; i^{\epsilon\beta\iota}\; i^{\lambda\gamma\zeta}\; i^{i\lambda\iota}) \ {h}^{i}_\triangolo \nonumber \\
&=&   {h}^{i}_\triangolo,  
\end{eqnarray}
where in the first two lines the Kronecher deltas arise from the invariance of the 3-valent and 4-valent nodes respectively, and in the last two lines, we have used the fact that $h[{s_{01}}]$ can be replaced with the identity because the rest is already of second order, $s_{123}=s_{12}\cup s_{23}\cup s_{31}$ and the property \eqref{taglio3} to close the open legs between the recoupling objects and the holonomies. 
Therefore the spin network ${h}^{i}_\tetraedro$ provides a regularization of the curvature alternative to  $ {h}^{i}_\triangle$.\footnote{We have in fact checked this result with an explicit long and tedious calculation, which we do not report here.} In the next section, we show how this can be used in the regularization of the hamiltonian constraint, and the possible advantages it yields.

\section{Definition of the hamiltonian constraint}

The LQG hamiltonian constraint operator was introduced in \cite{loops}, where its main property was found: to act only on the nodes (called ``intersections" at the time) of a LQG state. The operator considered in  \cite{loops}, however, was divergent on general states.  A better behaved operator was defined in \cite{HamiltonianRS}, in the context of a deparametrization of the general relativistic evolution with respect to a ``clock" matter field. Two observations were made in \cite{HamiltonianRS}. First, in order to be well defined in the background independent context, the operator needs to be a density (more precisely, a three-form).  Second, and most importantly, diffeomorphism invariance trivializes the limit where the regulator is removed from the operator. This is the essential property that renders the LQG dynamics finite.  

After many efforts \cite{Early-hamiltonians, Late-hamiltonians,HamiltonianR}, a fully well-defined hamiltonian operator was constructed in the non-deparametrized theory in  \cite{Thiemann96a,Thiemann96b}, using the idea of expressing the inverse triad $e$ as the Poisson bracket between the volume $V$ and the holonomy  $h$ of the Ashtekar connection $A$ (``Thiemann's trick"): $ e\sim h^{-1}[h,V]$ and using a regularization based on a given triangulation $T$ of the space manifold. (But see also \cite{dubbi}.) Let us reviewing Thieman's construction. 

\subsection{Thiemann hamiltonian constraint}

The hamiltonian constraint, which codes the dynamics of general relativity, is 
\be
{\cal C}=-2{\rm Tr}[(F-4K\wedge K)\wedge e]
\label{costraint}\ee
where $e=e^i_a dx^a\tau_i$ is the inverse triad (which we assume having positive determinant) and $K $ is the extrinsic curvature. Here we consider only the ``euclidean"  hamiltonian constraint ${\cal H}=-2{\rm Tr}[F \wedge e]$.   Following Thiemann \cite{Thiemann96a}, and choosing units where $8\pi Gc^{-3}\gamma=1$, we can write 
\begin{equation}
  \label{keyID1}
  e_a^i(x) = \{A_a^i(x),V\} 
\end{equation}
where $V$ is the volume of an arbitrary region $\Sigma$ containing the point $x$. Using this, and smearing the constraint with a lapse function $N(x)$, we have 
\begin{eqnarray}
  \mathcal{H}[N] &=&   
       \int_\Sigma d^3x \, N(x)\, \mathcal{H}(x)\nonumber \\
     &=&  - 2 \int_\Sigma  
      \, N  \ {\rm Tr}(F \wedge 
       \{ A, V \})~\nonumber
             \label{H_E_N}   .
\end{eqnarray} 
In order to regularize this expression, Thiemann introduces a triangulation $T$ of the manifold $\Sigma$ into elementary tetrahedra with analytic edges.   We can proceed as follows.

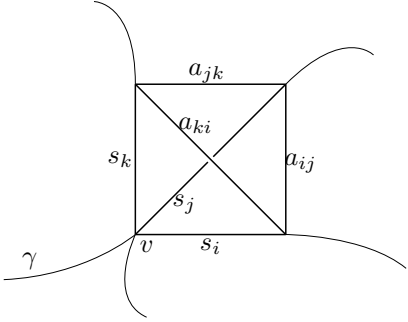
\begin{figure}
  \begin{center}$
\begin{array}{c}
	\ifx\JPicScale\undefined\def\JPicScale{0.5}\fi
\psset{unit=\JPicScale mm}
\psset{linewidth=0.3,dotsep=1,hatchwidth=0.3,hatchsep=1.5,shadowsize=1,dimen=middle}
\psset{dotsize=0.7 2.5,dotscale=1 1,fillcolor=black}
\psset{arrowsize=1 2,arrowlength=1,arrowinset=0.25,tbarsize=0.7 5,bracketlength=0.15,rbracketlength=0.15}
\begin{pspicture}(0,0)(109,84)
\psline[linewidth=0.4](37,22)(77,62)
\psline[linewidth=0.4,border=0.8](37,62)(77,22)
\psline[linewidth=0.4](37,22)(37,62)
\psline[linewidth=0.4](37,22)(77,22)
\psline[linewidth=0.4](77,62)(77,22)
\psline[linewidth=0.4](37,62)(77,62)
\pscustom[linewidth=0.2]{\psbezier(37,62)(37,84)(26,84)(26,84)
\psbezier(26,84)(26,84)(26,84)
}
\pscustom[linewidth=0.2]{\psbezier(77,62)(92.56,77.56)(100.33,69.78)(100.33,69.78)
\psbezier(100.33,69.78)(100.33,69.78)(100.33,69.78)
}
\pscustom[linewidth=0.2]{\psbezier(109,13)(98,22)(77,22)(77,22)
\psbezier(77,22)(77,22)(77,22)
}
\pscustom[linewidth=0.2]{\psbezier(2,10)(24,11)(37,22)(37,22)
\psbezier(37,22)(37,22)(37,22)
}
\pscustom[linewidth=0.2]{\psbezier(40,0)(29,5)(37,22)(37,22)
\psbezier(37,22)(37,22)(37,22)
}
\rput(40,19){$v$}
\rput(57,19){$s_i$}
\rput(50,30){$s_j$}
\rput(33,42){$s_k$}
\rput(81,41){$a_{ij}$}
\rput(56,65){$a_{jk}$}
\rput(53,51){$a_{ki}$}
\rput(9,15){$\gamma$}
\end{pspicture}
\end{array}
$
    \parbox{9cm}{\caption[]{\label{tetrahedron} \small
       An elementary tetrahedron $\Delta \in T$ constructed
      by adapting it to a graph $\gamma$ which
       underlies a cylindrical function.}}
  \end{center}
\end{figure}

Take a tetrahedron $\Delta$, and a vertex $v$ of this tetrahedron. Call the three edges that meet at $v$ as $s_i$, $i=1,2,3$ and denote $a_{ij}$ the edge connecting the two end-points of $s_{i}$ and $s_{j}$ opposite to $v$; in this way $s_{i}$, $s_{j}$ and $a_{ij}$ form a triangle.  Denote this triangle as $\alpha_{ij} := s_{i} \circ a_{ij} \circ
s_j^{-1}$. Figure \ref{tetrahedron} illustrates the construction.
Decompose the smeared Euclidean constraint (\ref{H_E_N}) into a sum of one term per each tetrahedron of the triangulation
\begin{eqnarray}
  \label{Ham_T1}
  \mathcal{H}[N] = \sum_{\Delta \in T}
        {-2}\, \int_\Delta d^3x \, N 
            \ \epsilon^{abc}\ {\rm Tr}(F_{ab}
            \{ A_c, V \}) ~.
          \label{Ham_T2}
\end{eqnarray}
Define the 
classical {\em regularized\/} hamiltonian constraint as 
\begin{equation}
   \mathcal{H}_{T}[N]
     :=  \sum_{\Delta \in T} \mathcal{H}_{\Delta} [N] ~,
        \label{H_delta:class_equiv}
 \end{equation} 
where 
\be
\begin{split}
      \mathcal{H}_{\Delta}[N]
     :&= - \frac{1}{3} \, 
        N(v) \, \epsilon^{ijk} \,
        \mbox{Tr}\Big[h_{\alpha_{ij}} 
        h_{s_{k}} \big\{ 
        h^{-1}_{s_{k}},V\big\}\Big]\\
         &=- \frac{1}{3} \, 
        N(v) \ \epsilon^{ijk} \ h^l_{\triangolo ij}\ 
        {\rm Tr}\Big[\tau^l h_{s_{k}} \big\{ 
        h^{-1}_{s_{k}},V\big\}\Big]
        \label{H_delta}
\end{split}
\ee
where $\triangolo{\scriptscriptstyle ab}=\alpha_{ab}$ and we use the notation $h_{s}:=h[s]$. This expression converges to the Hamiltonian constraint (\ref{Ham_T2}) if the triangulation is sufficiently fine. The expression (\ref{H_delta:class_equiv}) can finally  be promoted to a quantum operator,  since volume and holonomy have corresponding well-defined operators in LQG.  The lattice spacing of the triangulation $T$ that acts as a regularization parameter.  For a sufficiently fine triangulation, $  \mathcal{H}_{T}[N] $, converges to $\mathcal{H}[N]$.

This concludes the review of the definition of the Thiemann's operator.  Before continuing, recall also that  in  \cite{Gaul:2000ba} it was pointed out that the operator can be immediately generalized by replacing the trace in the first line of  (\ref{Ham_T2}) with a trace in an arbitrary representation $m$. ($\mbox{Tr}_{m}[U]=\mbox{Tr}[R^{(m)}(U)]$ where where $R^{(m)}$ is the matrix representing $U$ in the representation $m$.)   Equation \eqref{H_delta} can thus be replaced by
 \begin{equation}
   \label{Hm_delta:classical2}
   \mathcal{H}^m_{\Delta}[N]
     := \frac{N(v)}{2 N^2_m} \, 
         \, \epsilon^{ijk} \,
        \mbox{Tr}\Big[h^{(m)}_{\alpha_{ij}} 
        h^{(m)}_{s_{k}} \big\{ 
        h^{(m)-1}_{s_{k}},V\big\}\Big] ~,
         \end{equation}
where here we have indicated explicitly the representation in which the holonomy is taken: $h^{(m)}=R^{(m)}(h)$.  As shown in \cite{Gaul:2000ba}, this converges to $\mathcal{H}[N]$.

\subsection{The new hamiltonian constraint}

We are finally ready to define the new constraint. In this paper we only consider the sector of the theory states formed states with only 4-valent nodes. Generalizations will be considered elsewhere.  

Fix a fiducial flat metric in the space manifold $\Sigma$. Consider a triangulation $T$ of $\Sigma$.   Consider the dual of $T$, and in particular its one-skeleton $\Gamma$. $\Gamma$ is a graph with nodes $v$ in the center of the tetrahedra $v$ of $T$, and straight links that cut the triangles of $T$. Fix a tetrahedron $v$ and one of its vertices, say $s$. Let $s^a$ be the segment that joins the center of $v$ to $s$, and $u^a$ and $v^a$ two of the sides of the triangle $s$ opposite to the tetrahedron's vertex $s$. The volume of the tetrahedron can be written as $V=\sum_s \frac 1{18}\epsilon_{abc}s^a u^b v^c$ where the sum is over the four vertices of the tetrahedron. Consider now the quantity 
\be
H_{\Delta}= \sum_s \ h^i_{\triangolo s}\ {\rm Tr}[\tau^i\, h_s^{-1}\{h_s, V\}]
 \label{alternativa1}
\ee
where $\triangolo\,s$ is the triangle opposite to the vertex $s$. 
If $A$ and $e$ are constant on the tetrahedron, it is easy to see (for instance using coordinates in which the tetrahedron is regular) that this gives 
\be
H_{\Delta}= \sum_s \  F^i_{ab}u^a v^b   s^c e_c^i=18 \;{\rm Tr}(F_{ab}e_c)\epsilon^{abc}V=18 \; \int_v {\rm Tr}(F\wedge e). 
 \ee
Therefore we can replace (\ref{Hm_delta:classical2}) with 
\be
\begin{split}
      \mathcal{H}_{\Delta}[N]
     :&= \frac{N(v)}{36 N_m^2} \, 
         \sum_s \,\ h^i_{\triangolo s}\ 
        {\rm Tr}\Big[\tau^i\, h_{s} \big\{ 
        h^{-1}_{s},V\big\}\Big]
        \label{H_delta2}
\end{split}
\ee
 where all holomonies are taken in the representation $m$.  Notice that the sum is over the four links emerging from $v$ in $T$ and $\triangolo\, s$ is a triangle that joins the three points sitting on the three other links emerging from $v$. Notice also that this triangle and the center $v$ define a tetrahedron, which we shall denote $\tetraedro\,s$. It is then natural to replace the triangle regularization of the curvature with the tetrahedral one that we have defined in the previous section. 
Therefore we can replace (\ref{H_delta2}) with 
\be
      \mathcal{H}_{\Delta}[N]
     :=\frac{N(v)}{36N_m^2} \,  \, \sum_s \,\ h^l_{\tetraedro s}\ 
        {\rm Tr}\Big[\tau^l h_{s} \big\{ 
        h^{-1}_{s},V\big\}\Big]
        \label{H_delta3}
\ee
Since spin networks are well defined operator in the quantum theory, this expression provides the basis for an alternative definition of the hamiltonian constraint. The quantum operator corresponding to \eqref{H_delta3} can be obtained simply by replacing the volume $V$ with the volume operator, and replace the Poisson brackets with $-i/\hbar$ the quantum commutator. It is easy to see that the first term of the commutator vanishes. 
Thus the quantum operator is
\be
\begin{split}
      \mathcal{\hat{H}_{\Delta}}[N]
      :&= \frac{-i}{36\cdot 8\pi \gamma l^2_p} \, 
    \sum_v
        \frac{N(v)}{N_m^2} \, \sum_s \,\ \hat{h}^i_{\tetraedro s}\ 
        {\rm Tr}\Big[\tau^i \hat{h}_{s} 
        \hat{V} \hat{h}^{-1}_{s} \Big]
        \label{H_delta4}
\end{split}
\ee
The action of the operator on a spin network state with support on a graph $\gamma$ can then be defined, following  \cite{Thiemann96a,Borissovetal97}, by choosing a regularizing triangulation $T$ adapted to $\gamma$. Here we must chose $T$ such that $\gamma$ is a subgraph of $T^*$.  This completes the definition of the operator we were looking for. 

The main differences between the operator (\ref{H_delta4}) and the old one are the following.
\begin{enumerate}
\item The curvature is computed on a surface that is properly \emph{dual} to the direction indicuated by $e$. Notice in fact that it is computed on a triangle that \emph{surrounds} the direction of the link $s$.

\item Notice that this is different from the old case; there, $\gamma$ had to be a subgraph of the triangulation $T$ itself, not its dual. 

\item The new operator creates {\em three} new links instead than one. Therefore generically it transforms a 4-valent node into \emph{four} nodes.  This is precisely the action of the dynamics in the simplicial spinfoam models. 

\item The nodes created are themselves 4-valent. Thus are ``of the same kind" as the original node.  This is not the case as the old operator. 

\end{enumerate}

In the following section we  compute the action of the operator on a quantum state. 

\section{Action of the quantum constraint}

\subsection{Old constraint}

Let us begin by reviewing some elements of the action of the old constraint, before discussing the new one. 

It is immediate to see that when acting on a spin network state, the operator reduces to a sum over terms each acting on individual nodes. Acting on nodes of valence $n$ the operator gives
\begin{equation}
   {\mathcal{\hat{H}}}^m_{\gamma}[N] \, \psi_\gamma = 
  \sum_{v \in \mathcal{V}(\gamma)} 8 N_v
  \sum_{v(\Delta) = v}  {\mathcal{\hat{H}}}^m_{\Delta} 
  \,\frac{p_{\Delta}}{E(v)} \, \psi_\gamma ~,
\end{equation}
where ${\mathcal{H}}^m_{\Delta}$ is the quantum version of \eqref{Hm_delta:classical2}, $\mathcal{V}(\gamma)$ is the set of nodes of $\gamma$ and $N_v$ is the value of the lapse at the vertex.  $E(v) =\binom{n}{3}$ is the number of unordered triples of edges adjacent to $v$. Moreover, $p_{\Delta}=1$, whenever 
$\Delta$ is a tetrahedron having three edges coinciding with three
edges of the spin network state, that meet at the node $v$.
In the other cases $p_{\Delta}=0$.

As first realized in \cite{HamiltonianRS}, the continuum limit of the
operator turns out to be trivial in the quantum theory.  On 
diffeo\-mor\-phism-invariant (bra) states the regulator dependence drops out trivially. 
Indeed, take two operators $\mathcal{H}$ and $\mathcal{H}'$ that are related by a refinement of the triangulation (adapted to a given state); they differ only in the size of the loops $\alpha_{ij}$; therefore the resulting states are in the same equivalence class under diffeomorphisms. If $\phi$ is a diffeomorphism invariant state, we have $\langle \phi \mathcal{H}\psi \rangle = 
\langle \phi \mathcal{H}' \psi \rangle$, and then the (dual) 
action of  the two constraints $\mathcal{H}$ and
$\mathcal{H}'$ on $\phi$ is the same.  The restriction of
the hamiltonian constraint on the (dual) diffeomorphism invariant states is independent from the refinement of the triangulation.

Let us compute in particular the result in the action of the operator ${\mathcal{H}}^m_{\Delta}$ on trivalent nodes,  following \cite{Gaul:2000ba}.  We denote a trivalent node as  
$\ket{v(j_i,j_j,j_k)} \equiv \ket{v_3}$, whereas $j_i,j_j,j_k$ are the
spins of the adjacent edges $e_i,\, e_j,\, e_k$:
\be
\left|v_3\right\rangle=\begin{array}{c}
	\ifx\JPicScale\undefined\def\JPicScale{0.6}\fi
\psset{unit=\JPicScale mm}
\psset{linewidth=0.3,dotsep=1,hatchwidth=0.3,hatchsep=1.5,shadowsize=1,dimen=middle}
\psset{dotsize=0.7 2.5,dotscale=1 1,fillcolor=black}
\psset{arrowsize=1 2,arrowlength=1,arrowinset=0.25,tbarsize=0.7 5,bracketlength=0.15,rbracketlength=0.15}
\begin{pspicture}(0,0)(88,79)
\psline[linewidth=0.2,linestyle=dashed,dash=0.5 0.5](12,17)(38,43)
\psline[linewidth=0.2,linestyle=dashed,dash=0.5 0.5](62,43)(79,26)
\psline[linewidth=0.2,linestyle=dashed,dash=0.5 0.5](55,74)(55,56)
\psline[linewidth=0.2,linestyle=dashed,dash=0.5 0.5](45,74)(45,56)
\psline[linewidth=0.2,linestyle=dashed,dash=0.5 0.5](46,35)(28,17)
\psline[linewidth=0.2,linestyle=dashed,dash=0.5 0.5](54,35)(72,17)
\psline[dotsize=1.3 2.5]{-*}(20,17)(50,47)
\psline(50,47)(80,17)
\psline(50,47)(50,74)
\rput{0}(50,42.5){\psellipticarc[linewidth=0.25,linestyle=dashed,dash=0.5 0.5](0,0)(8.5,8.5){-118.07}{-61.93}}
\rput{0}(49.83,44.83){\psellipticarc[linewidth=0.2,linestyle=dashed,dash=0.5 0.5](0,0)(12.3,12.3){-8.57}{65.17}}
\rput{0}(49.58,45.15){\psellipticarc[linewidth=0.2,linestyle=dashed,dash=0.5 0.5](0,0)(11.78,11.78){112.89}{190.51}}
\psline[linewidth=0.2,linestyle=dashed,dash=0.5 0.5,border=0.3](79,26)(88,17)
\rput(17,14){$j_i$}
\rput(83,14){$j_j$}
\rput(50,79){$j_k$}
\rput(50,41){$v$}
\rput(30,31){$\scr{e_i}$}
\rput(70,31){$\scr{e_j}$}
\rput(47,69){$\scr{e_k}$}
\end{pspicture}
\end{array}
\ee 
With this notation we represent only the node and its adjacencies. In particular everything contained in the dashed circle belongs to the node. 
Let us compute:
\begin{equation}
  \label{H_m_Delta} 
    \hat{\mathcal{H}}^m_{\Delta} \, \ket{v_3} = \frac{-2 i}{3 l_0^2 N_m^2} \, 
  \epsilon^{ijk} \, \mbox{Tr} \left(
  \frac{\hat{h}^{(m)}[\alpha_{ij}] - \hat{h}^{(m)}[\alpha_{ji}]}{2} \,
  \hat{h}^{(m)}[s_{k}]\, \hat{V} \,  
  \hat{h}^{(m)}[s^{-1}_{k}] \right) \ket{v_3} ~. 
\end{equation}
where we have indicated explicitly that the holonomies are in the $m$ representation.
The operator $ {h}^{(m)}[s^{-1}_{k}]$, corresponding to the holonomy along a segment 
$s_k$ with reversed orientation, attaches an open spin-$m$ segment to the 
edge $e_k$. Since the segment $s_k$, with one end at the node, is entirely contained in $e_k$, the two holonomies are given by the same group element in two different representations; the original $j_k$ and the new $m$ and can then be tensorized, using the recoupling identity \eqref{sro}.
There is then a free index in the color-$m$ representation located at the node (inside the dashed circle), making it non-gauge-invariant. A new node has also been created on the edge $e_k$:

\be
  \label{holonomy_action1} 
  \hat{h}^{(m)}[s^{-1}_{k}]~\left|v_3\right\rangle
      ~= \;\;\sum_c \,\dim c
         \begin{array}{c}
	\ifx\JPicScale\undefined\def\JPicScale{0.5}\fi
\psset{unit=\JPicScale mm}
\psset{linewidth=0.3,dotsep=1,hatchwidth=0.3,hatchsep=1.5,shadowsize=1,dimen=middle}
\psset{dotsize=0.7 2.5,dotscale=1 1,fillcolor=black}
\psset{arrowsize=1 2,arrowlength=1,arrowinset=0.25,tbarsize=0.7 5,bracketlength=0.15,rbracketlength=0.15}
\begin{pspicture}(20,0)(90,74)
\psline[linewidth=0.2,linestyle=dashed,dash=0.5 0.5](10,17)(36,43)
\psline[linewidth=0.2,linestyle=dashed,dash=0.5 0.5](64,43)(90,17)
\psline[linewidth=0.2,linestyle=dashed,dash=0.5 0.5](57,74)(57,56)
\psline[linewidth=0.2,linestyle=dashed,dash=0.5 0.5](43,74)(43,56)
\psline[linewidth=0.2,linestyle=dashed,dash=0.5 0.5](46,33)(30,17)
\psline[linewidth=0.2,linestyle=dashed,dash=0.5 0.5](54,33)(70,17)
\psline[dotsize=1.3 2.5]{-*}(20,17)(50,47)
\psline(50,47)(80,17)
\psline(50,47)(50,74)
\rput{0}(50,40.5){\psellipticarc[linewidth=0.25,linestyle=dashed,dash=0.5 0.5](0,0)(8.5,8.5){-118.07}{-61.93}}
\rput{0}(52,44.7){\psellipticarc[linewidth=0.2,linestyle=dashed,dash=0.5 0.5](0,0)(12.3,12.3){-8.57}{65.17}}
\rput{0}(47.58,45.15){\psellipticarc[linewidth=0.2,linestyle=dashed,dash=0.5 0.5](0,0)(11.78,11.78){112.89}{190.51}}
\rput(30,31){$\scr{j_i}$}
\rput(70,31){$\scr{j_j}$}
\rput(46,69){$\scr{j_k}$}
\rput(55,65){$\scr{m}$}
\psline(50,53)(55,53)
\psline(50,63)(55,63)
\psline{->}(55,53)(52,53)
\psline{<-}(53,63)(50,63)
\rput(55,55){$\scr{m}$}
\rput(46,51){$\scr{j_k}$}
\rput(47,59){$\scr{c}$}
\end{pspicture}
\end{array}
\ee 
The range of the sum over the spin $c$ is determined by the Clebsh-Gordan conditions. 
The next operator acts then on a 4-valent non-invariant node with virtual link in $j_k$ representation.

Next step is the calculation of $ {V}\left( {h}^{(m)}[s^{-1}_{k}]\, \ket{v}\right)$. To this aim, we need the matrix elements of the volume operator \cite{RovelliSmolin95,AshtekarLewand98,Lewandowski96}, computed in  \cite{DePietriRovelli96} and applied to Thiemann's Hamiltonian constraint operator in \cite{Borissovetal97,Gaul:2000ba}, see Appendix \ref{Volume section}). Here we need the action of the volume on a 4-valent not gauge invariant node; reviewed in Appendix \ref{4-valent not gauge} (In the general case of the Hamiltonian constraint acting on a $n$ valent vertex, the volume operator will act on a non gauge invariant node of valence $n+1$.)

The last step is the evaluation of the holonomies on the left of the volume.
The operators $ {h}^{(m)}[\alpha_{ij}]\,  {h}^{(m)}[s_{k}]$ and $ {h}^{(m)}[\alpha_{ij}]\,  {h}^{(m)}[s_{k}]$ add open loops with opposite orientations $\alpha_{ij}$ and $\alpha_{ji}$ to the not gauge invariant 4-node. The action is concluded by the contraction of the free ends of the open loops to the two open links in $ {h}^{(m)}[s^{-1}_{k}]\, \ket{v}$ taking into account the orientations. (Note that is the $m$-trace that ensure connection and summation on the last index).

The final result of the action of the trace part of the operator is then:
\begin{eqnarray}
 \label{trace_part}
 \lefteqn{
  \mbox{Tr}\left(
  \frac{ {h}^{(m)}[\alpha_{ij}] -  {h}^{(m)}[\alpha_{ji}]}{2} \:
   {h}^{(m)}[s_{k}]\,  {V} \,  {h}^{(m)}[s^{-1}_{k}] 
  \right) \, |v (j_i,j_j,j_k) \rangle} \hspace{2.5cm} \nonumber \\
  & &=~ \: \frac{l_0^3}{8} \, \sum_{a,b} 
        \, A^{(m)}(j_i,a|j_j,b|j_k) 
      \begin{array}{c}
\ifx\JPicScale\undefined\def\JPicScale{0.4}\fi
\psset{unit=\JPicScale mm}
\psset{linewidth=0.3,dotsep=1,hatchwidth=0.3,hatchsep=1.5,shadowsize=1,dimen=middle}
\psset{dotsize=0.7 2.5,dotscale=1 1,fillcolor=black}
\psset{arrowsize=1 2,arrowlength=1,arrowinset=0.25,tbarsize=0.7 5,bracketlength=0.15,rbracketlength=0.15}
\begin{pspicture}(0,0)(90,74)
\psline[linewidth=0.2,linestyle=dashed,dash=0.5 0.5](10,17)(36,43)
\psline[linewidth=0.2,linestyle=dashed,dash=0.5 0.5](64,43)(90,17)
\psline[linewidth=0.2,linestyle=dashed,dash=0.5 0.5](57,74)(57,56)
\psline[linewidth=0.2,linestyle=dashed,dash=0.5 0.5](43,74)(43,56)
\psline[linewidth=0.2,linestyle=dashed,dash=0.5 0.5](33,20)(30,17)
\psline[linewidth=0.2,linestyle=dashed,dash=0.5 0.5](54,33)(60,27)
\psline(50,47)(50,74)
\rput{0}(50,40.5){\psellipticarc[linewidth=0.25,linestyle=dashed,dash=0.5 0.5](0,0)(8.5,8.5){-118.07}{-61.93}}
\rput{0}(52,44.69){\psellipticarc[linewidth=0.2,linestyle=dashed,dash=0.5 0.5](0,0)(12.3,12.3){-8.57}{65.17}}
\rput{0}(47.58,45.15){\psellipticarc[linewidth=0.2,linestyle=dashed,dash=0.5 0.5](0,0)(11.78,11.78){112.89}{190.51}}
\rput(16,17){$\scr{j_i}$}
\rput(84,17){$\scr{j_j}$}
\rput(46,69){$\scr{j_k}$}
\rput(32,33){$\scr{a}$}
\psline[linestyle=dotted](40,27)(60,27)
\psline[linestyle=dotted](33,20)(67,20)
\psline[linewidth=0.2,linestyle=dashed,dash=0.5 0.5](46,33)(40,27)
\psline[linewidth=0.2,linestyle=dashed,dash=0.5 0.5](67,20)(70,17)
\psline(73,24)(27,24)
\rput(50,22){$\scr{m}$}
\psline(50,47)(80,17)
\psline[dotsize=1.3 2.5]{-*}(20,17)(50,47)
\rput(68,33){$\scr{b}$}
\end{pspicture}
\end{array}
\vspace{-1em}
\end{eqnarray}
where the range of the sums over $a,b$ is determined by the  Clebsh Gordan conditions, and
\begin{eqnarray}
  \lefteqn{
  A^{(m)}(j_i,a|j_j,b|j_k) := \sum_c\; \lambda^{mj_i}_a 
        \lambda^{mj_j}_b d_a d_b d_c  \:
               ~~\times} \nonumber \\
   &&\hspace{-0.5cm} \times \sum_{\beta(j_i,j_j,m,c)} 
     \!\!\!\! V{}_{j_k}{}^\beta (j_i,j_j,m,c) 
 \left[
   \lambda_c^{m\beta} \;
   \{
\begin{smallmatrix}
                      a      & j_j & c \\
                      \beta  & m   & j_i                  
\end{smallmatrix}
\}
\{\begin{smallmatrix}
                      a      & b & j_k \\
                      m      & c   & j_j                  
\end{smallmatrix}
\}
  -
   \lambda^{mj_k}_c
\{\begin{smallmatrix}
                      j_i      & b & c \\
                      m  & \beta  & j_j                  
\end{smallmatrix}
\}
   \{\begin{smallmatrix}
                      a      & b & j_k \\
                      c      & m   & j_i                  
\end{smallmatrix}
\} 
    \right],
   \label{amplitude}
\end{eqnarray}
The summation index $\beta = \beta(j_i,j_j,m,c)$ which appears due to 
the non-diagonal action of the volume operator, ranges on the values which are determined by the simultaneous
admissibility of the 3 valent nodes $\{j_i,j_j,\beta \}$ and 
$\{m,c,\beta\}$ namely on the space of intertwiners on which 
the $ {W}$ operator acts. See the Appendix \ref{Volume section}. 
The complete action of the operator on a three valent state  $|v (j_i,j_j,j_k) \rangle$ is then given by contracting the trace part (\ref{trace_part}) with 
$\epsilon^{ijk}$. We get the sum of three terms:

\begin{eqnarray}
  \label{final_action_of_H}
  \lefteqn{
   {\mathcal{H}}^m_{\Delta} \, \big| v (j_i,j_j,j_k) \big\rangle 
    = \: \frac{i l_0}{12 C(m)} \, 
      \left[\rule{0cm}{1.1cm}\right. \sum_{a,b} 
A^{(m)}(j_i,a|j_j,b|j_k)  )   
     \begin{array}{c} 
\ifx\JPicScale\undefined\def\JPicScale{0.4}\fi
\psset{unit=\JPicScale mm}
\psset{linewidth=0.3,dotsep=1,hatchwidth=0.3,hatchsep=1.5,shadowsize=1,dimen=middle}
\psset{dotsize=0.7 2.5,dotscale=1 1,fillcolor=black}
\psset{arrowsize=1 2,arrowlength=1,arrowinset=0.25,tbarsize=0.7 5,bracketlength=0.15,rbracketlength=0.15}
\begin{pspicture}(0,0)(90,80)
\psline[linewidth=0.2,linestyle=dashed,dash=0.5 0.5](30,37)(43,50)
\psline[linewidth=0.2,linestyle=dashed,dash=0.5 0.5](57,50)(70,37)
\psline[linewidth=0.2,linestyle=dashed,dash=0.5 0.5](57,77)(57,68)
\psline[linewidth=0.2,linestyle=dashed,dash=0.5 0.5](43,77)(43,58)
\psline[linewidth=0.2,linestyle=dashed,dash=0.5 0.5](50,38)(60,27)
\psline(50,47)(50,77)
\rput(17,17){$\scr{j_i}$}
\rput(83,17){$\scr{j_j}$}
\rput(50,80){$\scr{j_k}$}
\rput(39,32){$\scr{a}$}
\psline[linewidth=0.2,linestyle=dashed,dash=0.5 0.5](33,20)(30,17)
\psline(50,47)(80,17)
\psline[dotsize=1.3 2.5]{-*}(20,17)(50,47)
\rput(33,52){$\scr{m}$}
\psline[linewidth=0.2,linestyle=dashed,dash=0.5 0.5](57,68)(57,50)
\psline[linewidth=0.2,linestyle=dashed,dash=0.5 0.5](70,37)(90,17)
\psline[linewidth=0.2,linestyle=dashed,dash=0.5 0.5](43,58)(43,50)
\psline[linewidth=0.2,linestyle=dashed,dash=0.5 0.5](10,17)(30,37)
\psline[linestyle=dotted](40,27)(60,27)
\psline[linestyle=dotted](33,20)(67,20)
\psline(73,24)(27,24)
\rput(50,22){$\scr{m}$}
\psline[linewidth=0.2,linestyle=dashed,dash=0.5 0.5](50,38)(39,27)
\psline[linewidth=0.2,linestyle=dashed,dash=0.5 0.5](66,20)(70,17)
\rput(62,32){$\scr{b}$}
\end{pspicture}
       \end{array}  } 
        \hspace{3cm} \nonumber \\
   && +~ \sum_{b,c} A^{(m)}(j_j,b|j_k,c|j_i) 
         \begin{array}{c}
         \ifx\JPicScale\undefined\def\JPicScale{0.4}\fi
\psset{unit=\JPicScale mm}
\psset{linewidth=0.3,dotsep=1,hatchwidth=0.3,hatchsep=1.5,shadowsize=1,dimen=middle}
\psset{dotsize=0.7 2.5,dotscale=1 1,fillcolor=black}
\psset{arrowsize=1 2,arrowlength=1,arrowinset=0.25,tbarsize=0.7 5,bracketlength=0.15,rbracketlength=0.15}
\begin{pspicture}(0,0)(90,81)
\psline[linewidth=0.2,linestyle=dashed,dash=0.5 0.5](10,17)(43,50)
\psline[linewidth=0.2,linestyle=dashed,dash=0.5 0.5](57,50)(63,44)
\psline[linewidth=0.2,linestyle=dashed,dash=0.5 0.5](57,81)(57,68)
\psline[linewidth=0.2,linestyle=dashed,dash=0.5 0.5](43,81)(43,50)
\psline[linewidth=0.2,linestyle=dashed,dash=0.5 0.5](50,38)(70,17)
\psline(50,47)(50,81)
\rput(17,17){$\scr{j_i}$}
\rput(83,17){$\scr{j_j}$}
\rput(47,81){$\scr{j_k}$}
\rput(60,33){$\scr{b}$}
\rput(47,62){$\scr{c}$}
\psline[linewidth=0.2,linestyle=dashed,dash=0.5 0.5](50,38)(30,17)
\psline(50,47)(80,17)
\psline[dotsize=1.3 2.5]{-*}(20,17)(50,47)
\rput(66,52){$\scr{m}$}
\psline[linewidth=0.2,linestyle=dashed,dash=0.5 0.5](57,58)(63,44)
\psline[linewidth=0.2,linestyle=dashed,dash=0.5 0.5](57,68)(70,37)
\psline(50,78)(73,24)
\psline[linewidth=0.2,linestyle=dashed,dash=0.5 0.5](57,58)(57,50)
\psline[linewidth=0.2,linestyle=dashed,dash=0.5 0.5](70,37)(90,17)
\end{pspicture}
\end{array}
  \nonumber \\
   && +~ \sum_{a,c} A^{(m)}(j_k,c|j_i,a|j_j)  
   \begin{array}{c} \ifx\JPicScale\undefined\def\JPicScale{0.4}\fi
\psset{unit=\JPicScale mm}
\psset{linewidth=0.3,dotsep=1,hatchwidth=0.3,hatchsep=1.5,shadowsize=1,dimen=middle}
\psset{dotsize=0.7 2.5,dotscale=1 1,fillcolor=black}
\psset{arrowsize=1 2,arrowlength=1,arrowinset=0.25,tbarsize=0.7 5,bracketlength=0.15,rbracketlength=0.15}
\begin{pspicture}(0,0)(90,81)
\psline[linewidth=0.2,linestyle=dashed,dash=0.5 0.5](37,44)(43,50)
\psline[linewidth=0.2,linestyle=dashed,dash=0.5 0.5](57,50)(70,37)
\psline[linewidth=0.2,linestyle=dashed,dash=0.5 0.5](57,81)(57,68)
\psline[linewidth=0.2,linestyle=dashed,dash=0.5 0.5](43,81)(43,68)
\psline[linewidth=0.2,linestyle=dashed,dash=0.5 0.5](50,38)(70,17)
\psline(50,47)(50,81)
\rput(17,17){$\scr{j_i}$}
\rput(83,17){$\scr{j_j}$}
\rput(47,81){$\scr{j_k}$}
\rput(39,32){$\scr{a}$}
\rput(47,62){$\scr{c}$}
\psline[linewidth=0.2,linestyle=dashed,dash=0.5 0.5](50,38)(30,17)
\psline(50,47)(80,17)
\psline[dotsize=1.3 2.5]{-*}(20,17)(50,47)
\rput(33,52){$\scr{m}$}
\psline(50,78)(27,24)
\psline[linewidth=0.2,linestyle=dashed,dash=0.5 0.5](57,68)(57,50)
\psline[linewidth=0.2,linestyle=dashed,dash=0.5 0.5](70,37)(90,17)
\psline[linewidth=0.2,linestyle=dashed,dash=0.5 0.5](43,58)(37,44)
\psline[linewidth=0.2,linestyle=dashed,dash=0.5 0.5](43,68)(30,37)
\psline[linewidth=0.2,linestyle=dashed,dash=0.5 0.5](43,58)(43,50)
\psline[linewidth=0.2,linestyle=dashed,dash=0.5 0.5](10,17)(30,37)
\end{pspicture} 
     \end{array}  
        \left.\rule{0cm}{1.1cm}\right] ~.
\label{fine thomas}
\end{eqnarray}
These yield the original state with the new link $m$ between all the possible pairs of edges adjacent to the  node, with corresponding amplitudes (\ref{amplitude}) given by cyclic permutations of argument pairs. 

\section{New constraint}
To analyze the new constraint we restrict our attention to 4-valent nodes.
On a single 4-valent node, it reads
\be
H_{\Delta} \left|v_4\right\rangle= \frac{-i}{36\cdot 8\pi \gamma l^2_p N_m^2} \, 
     \sum_s \;\; {h}^{i}_{\tetraedro s}
\;\, 
{\rm Tr}\big[\tau^i\;  {h}^{(m)}[s]\,  {V} \,
         {h}^{(m)}[s^{-1}] \big] \left|v_4\right\rangle  \ Ê :=\ 
-i \frac{C}{N_m^2} \ \sum_s\  H^s_{\Delta} \left|v_4\right\rangle.
\label{Hmio}
\ee
where $C=\frac{1}{36\cdot 8\pi \gamma\; l^2_p}$ is a constant and the sum runs over the four edges that emerge from the node. The action of this constraint begins, as with old one, by adding a new holonomy $ {h}^{(m)}[s^{-1}]$ in the $s$ direction. This creates two free legs and increases the valence of the node by one.  Then the volume operator acts on the resulting  non-gauge-invariant 5-valent node (see Appendix \ref{5-valent not gauge}).
That is
\be
\begin{split}
{\rm Tr}\big[\tau^i\;         {h}^{(m)}[s]\,  {V} \,
         {h}^{(m)}[s^{-1}] \big] \left|v_4\right\rangle=
&\; N_m \;i^{i\delta\epsilon} 
\, 
\big[         {h}^{(m)}[s]\,  {V} \,
         {h}^{(m)}[s^{-1}] \big]_{\delta\epsilon}
          \hspace{-1em}
\begin{array}{c}
	\ifx\JPicScale\undefined\def\JPicScale{0.5}\fi
\psset{unit=\JPicScale mm}
\psset{linewidth=0.3,dotsep=1,hatchwidth=0.3,hatchsep=1.5,shadowsize=1,dimen=middle}
\psset{dotsize=0.7 2.5,dotscale=1 1,fillcolor=black}
\psset{arrowsize=1 2,arrowlength=1,arrowinset=0.25,tbarsize=0.7 5,bracketlength=0.15,rbracketlength=0.15}
\begin{pspicture}(20,0)(102,78)
\psline[linewidth=0.2,linestyle=dashed,dash=0.5 0.5](26,5)(37,40)
\psline[linewidth=0.2,linestyle=dashed,dash=0.5 0.5](62,39)(80,4)
\psline[linewidth=0.2,linestyle=dashed,dash=0.5 0.5](36,75)(44,58)
\psline[linewidth=0.2,linestyle=dashed,dash=0.5 0.5](26,75)(37,52)
\psline[linewidth=0.2,linestyle=dashed,dash=0.5 0.5](45,33)(36,5)
\psline[linewidth=0.2,linestyle=dashed,dash=0.5 0.5](55,33)(70,4)
\psline(31,5)(44,46)
\psline(53,46)(75,4)
\psline(44,46)(31,75)
\psline(53,46)(100,32)
\psline[linewidth=0.6,linestyle=dashed,dash=1 1](44,46)(53,46)
\psline[linewidth=0.2,linestyle=dashed,dash=0.5 0.5](62,39)(98,28)
\psline[linewidth=0.2,linestyle=dashed,dash=0.5 0.5](64,48)(101,37)
\rput{90}(50.2,45.54){\psellipticarc[linewidth=0.2,linestyle=dashed,dash=0.5 0.5](0,0)(14.5,-14){-27.38}{80.22}}
\rput{0}(50,46){\psellipticarc[linewidth=0.2,linestyle=dashed,dash=0.5 0.5](0,0)(14,-14){67.38}{111.04}}
\rput{49.96}(50.1,46.02){\psellipticarc[linewidth=0.2,linestyle=dashed,dash=0.5 0.5](0,0)(14.06,-13.96){-152.39}{-105.39}}
\rput(30,58){}
\rput(30,3){$\scr{j_i}$}
\rput(77,2){$\scr{j_j}$}
\rput(102,32){$\scr{j_k}$}
\rput(29,78){$\scr{j_l}$}
\rput(49,49){$\scr{i_n}$}
\end{pspicture}
\end{array}
\\
&=N_m \, \sum_{c,d,e} \,d_c\, \frac{l_0^3}{4} \,  
                \, V{}_{j_l,i_n}{}^{d,e} (j_i,j_j,j_k,m,c)
                \hspace{-4em}
\begin{array}{c}
	\ifx\JPicScale\undefined\def\JPicScale{0.5}\fi
\psset{unit=\JPicScale mm}
\psset{linewidth=0.3,dotsep=1,hatchwidth=0.3,hatchsep=1.5,shadowsize=1,dimen=middle}
\psset{dotsize=0.7 2.5,dotscale=1 1,fillcolor=black}
\psset{arrowsize=1 2,arrowlength=1,arrowinset=0.25,tbarsize=0.7 5,bracketlength=0.15,rbracketlength=0.15}
\begin{pspicture}(0,0)(102,78)
\psline[linewidth=0.2,linestyle=dashed,dash=0.5 0.5](26,5)(37,40)
\psline[linewidth=0.2,linestyle=dashed,dash=0.5 0.5](62,39)(80,4)
\psline[linewidth=0.2,linestyle=dashed,dash=0.5 0.5](36,75)(44,58)
\psline[linewidth=0.2,linestyle=dashed,dash=0.5 0.5](26,75)(37,52)
\psline[linewidth=0.2,linestyle=dashed,dash=0.5 0.5](45,33)(36,5)
\psline[linewidth=0.2,linestyle=dashed,dash=0.5 0.5](55,33)(70,4)
\psline(31,5)(44,46)
\psline(53,46)(75,4)
\psline(44,46)(31,75)
\psline(53,46)(100,32)
\psline[linewidth=0.6,linestyle=dashed,dash=1 1](44,46)(53,46)
\psline[linewidth=0.2,linestyle=dashed,dash=0.5 0.5](62,39)(98,28)
\psline[linewidth=0.2,linestyle=dashed,dash=0.5 0.5](64,48)(101,37)
\rput{90}(50.2,45.54){\psellipticarc[linewidth=0.2,linestyle=dashed,dash=0.5 0.5](0,0)(14.5,-14){-27.38}{80.22}}
\rput{0}(50,46){\psellipticarc[linewidth=0.2,linestyle=dashed,dash=0.5 0.5](0,0)(14,-14){67.38}{111.04}}
\rput{50.55}(50.1,46.02){\psellipticarc[linewidth=0.2,linestyle=dashed,dash=0.5 0.5](0,0)(14.06,-13.96){-151.8}{-104.8}}
\rput(30,58){}
\rput(30,3){$\scr{j_i}$}
\rput(77,2){$\scr{j_j}$}
\rput(102,32){$\scr{j_k}$}
\rput(29,78){$\scr{j_l}$}
\rput(49,44){$\scr{e}$}
\psline(35,66)(38,68)
\psline(41,53)(42,54)
\psline(38,68)(45,54)
\psline(42,54)(43,52)
\psline(45,54)(45,52)
\psline(43,52)(45,52)
\psline(46,50)(45,52)
\rput(48,51){$\scr{1}$}
\rput(36,69){$\scr{m}$}
\rput(36,59){$\scr{c}$}
\rput(40,48){$\scr{d}$}
\rput(42,55){$\scr{m}$}
\end{pspicture}
\end{array}
\end{split}
\ee
where we have converted the generator in $m$ representation  $\tau^i$, in the corresponding normalized intertwiner $i^{i\delta\epsilon}$. Note that the trace part leaves a free leg in representation $1$. We are now ready to give the complete action of the operator:
\be
\begin{split}
&	 {h}^{i}_{\tetraedro s }
	\; \;i^{i\delta\epsilon} 
	\, 
	\big[         {h}^{(m)}[s]\,  {V} \,
	         {h}^{(m)}[s^{-1}] \big]_{\delta\epsilon} \left|v_4\right\rangle=
	        \\
	        &
	        =N_m^2 c_m
\sum_{c,d,e} \,d_c\, \frac{l_0^3}{4} \,  
                \, V{}_{j_l,i_n}{}^{d,e} (j_i,j_j,j_k,m,c)
\begin{array}{c}
\ifx\JPicScale\undefined\def\JPicScale{0.5}\fi
\psset{unit=\JPicScale mm}
\psset{linewidth=0.3,dotsep=1,hatchwidth=0.3,hatchsep=1.5,shadowsize=1,dimen=middle}
\psset{dotsize=0.7 2.5,dotscale=1 1,fillcolor=black}
\psset{arrowsize=1 2,arrowlength=1,arrowinset=0.25,tbarsize=0.7 5,bracketlength=0.15,rbracketlength=0.15}
\begin{pspicture}(0,0)(102,78)
\psline[linewidth=0.2,linestyle=dashed,dash=0.5 0.5](26,5)(37,40)
\psline[linewidth=0.2,linestyle=dashed,dash=0.5 0.5](77,11)(80,4)
\psline[linewidth=0.2,linestyle=dashed,dash=0.5 0.5](37,8)(36,5)
\psline[linewidth=0.2,linestyle=dashed,dash=0.5 0.5](55,33)(64,16)
\psline(31,5)(44,46)
\psline[linewidth=0.2,linestyle=dashed,dash=0.5 0.5](62,39)(70,36)
\psline[linewidth=0.2,linestyle=dashed,dash=0.5 0.5](84,32)(73,19)
\psline[linewidth=0.2,linestyle=dashed,dash=0.5 0.5](93,30)(77,11)
\psline[linewidth=0.2,linestyle=dashed,dash=0.5 0.5](56,30)(42,25)
\psline[linewidth=0.2,linestyle=dashed,dash=0.5 0.5](59,23)(40,16)
\psline[linewidth=0.2,linestyle=dashed,dash=0.5 0.5](70,36)(65,34)
\psline[linewidth=0.2,linestyle=dashed,dash=0.5 0.5](82,33)(68,28)
\psline[linewidth=0.2,linestyle=dashed,dash=0.5 0.5](64,16)(40,16)
\psline[linewidth=0.2,linestyle=dashed,dash=0.5 0.5](68,8)(37,8)
\psline[linewidth=0.2,linestyle=dashed,dash=0.5 0.5](36,75)(44,58)
\psline[linewidth=0.2,linestyle=dashed,dash=0.5 0.5](26,75)(37,52)
\psline(44,46)(31,75)
\psline[linewidth=0.2,linestyle=dashed,dash=0.5 0.5](64,48)(101,37)
\rput{90}(50.2,45.54){\psellipticarc[linewidth=0.2,linestyle=dashed,dash=0.5 0.5](0,0)(14.5,-14){-27.38}{80.22}}
\rput{0}(50,46){\psellipticarc[linewidth=0.2,linestyle=dashed,dash=0.5 0.5](0,0)(14,-14){67.38}{111.04}}
\rput{50.55}(50.1,46.02){\psellipticarc[linewidth=0.2,linestyle=dashed,dash=0.5 0.5](0,0)(14.06,-13.96){-151.8}{-104.8}}
\rput(30,58){}
\rput(29,78){$\scr{j_l}$}
\rput(49,44){$\scr{e}$}
\psline(35,66)(38,68)
\psline(41,53)(42,54)
\psline(38,68)(45,54)
\psline(42,54)(43,52)
\psline(45,55)(45,52)
\psline(43,52)(45,52)
\psline(46,50)(45,52)
\rput(36,69){$\scr{m}$}
\rput(36,59){$\scr{c}$}
\rput(40,48){$\scr{d}$}
\psline[linewidth=0.2,linestyle=dashed,dash=0.5 0.5](45,33)(42,25)
\psline[linewidth=0.2,linestyle=dashed,dash=0.5 0.5](68,8)(70,4)
\psline[linewidth=0.2,linestyle=dashed,dash=0.5 0.5](62,39)(73,18)
\psline[linewidth=0.2,linestyle=dashed,dash=0.5 0.5](92,30)(98,28)
\psline[linewidth=0.2,linestyle=dashed,dash=0.5 0.5](82,33)(84,32)
\psline(37,16)(49,49)
\psline(68,12)(49,50)
\psline(68,12)(37,16)
\psline(59,25)(37,16)
\psline(90,37)(68,29)
\psline(90,37)(68,12)
\psline(90,37)(51,50)
\psline[dotsize=1.8 2.5]{*-}(50,50)(46,50)
\rput(29,3){$\scr{j_i}$}
\rput(77,2){$\scr{j_j}$}
\rput(102,31){$\scr{j_k}$}
\psline[border=0.5](53,46)(100,32)
\psline[border=0.5](53,46)(75,4)
\psline[linewidth=0.6,linestyle=dashed,dash=1 1,border=0.4](44,46)(53,46)
\rput(46,36){$\scr{m}$}
\rput(49,11){$\scr{m}$}
\rput(82,23){$\scr{m}$}
\rput(53,36){$\scr{m}$}
\rput(61,49){$\scr{m}$}
\rput(52,24){$\scr{m}$}
\rput(47,52){$\scr{1}$}
\rput(42,55){$\scr{m}$}
\end{pspicture}
\end{array}
\\
&=N_m^2 c_m
\sum_{c,d,e,f,g,h} \,d_c \,d_f\,d_g\,d_h\, \frac{l_0^3}{4} \,  
                \, V{}_{j_l,i_n}{}^{d,e} (j_i,j_j,j_k,m,c) \begin{array}{c}
	\ifx\JPicScale\undefined\def\JPicScale{0.5}\fi
\psset{unit=\JPicScale mm}
\psset{linewidth=0.3,dotsep=1,hatchwidth=0.3,hatchsep=1.5,shadowsize=1,dimen=middle}
\psset{dotsize=0.7 2.5,dotscale=1 1,fillcolor=black}
\psset{arrowsize=1 2,arrowlength=1,arrowinset=0.25,tbarsize=0.7 5,bracketlength=0.15,rbracketlength=0.15}
\begin{pspicture}(0,0)(102,78)
\psline[linewidth=0.2,linestyle=dashed,dash=0.5 0.5](26,5)(37,40)
\psline[linewidth=0.2,linestyle=dashed,dash=0.5 0.5](77,11)(80,4)
\psline[linewidth=0.2,linestyle=dashed,dash=0.5 0.5](37,8)(36,5)
\psline[linewidth=0.2,linestyle=dashed,dash=0.5 0.5](55,33)(64,16)
\psline(42,40)(44,46)
\psline[linewidth=0.2,linestyle=dashed,dash=0.5 0.5](62,39)(70,36)
\psline[linewidth=0.2,linestyle=dashed,dash=0.5 0.5](84,32)(73,19)
\psline[linewidth=0.2,linestyle=dashed,dash=0.5 0.5](93,30)(77,11)
\psline[linewidth=0.2,linestyle=dashed,dash=0.5 0.5](56,30)(42,25)
\psline[linewidth=0.2,linestyle=dashed,dash=0.5 0.5](59,23)(40,16)
\psline[linewidth=0.2,linestyle=dashed,dash=0.5 0.5](70,36)(65,34)
\psline[linewidth=0.2,linestyle=dashed,dash=0.5 0.5](82,33)(68,28)
\psline[linewidth=0.2,linestyle=dashed,dash=0.5 0.5](64,16)(40,16)
\psline[linewidth=0.2,linestyle=dashed,dash=0.5 0.5](68,8)(37,8)
\psline[linewidth=0.2,linestyle=dashed,dash=0.5 0.5](36,75)(44,58)
\psline[linewidth=0.2,linestyle=dashed,dash=0.5 0.5](26,75)(37,52)
\psline(44,46)(31,75)
\psline[linewidth=0.2,linestyle=dashed,dash=0.5 0.5](64,48)(101,37)
\rput{90}(50.2,45.54){\psellipticarc[linewidth=0.2,linestyle=dashed,dash=0.5 0.5](0,0)(14.5,-14){-27.38}{80.22}}
\rput{0}(50,46){\psellipticarc[linewidth=0.2,linestyle=dashed,dash=0.5 0.5](0,0)(14,-14){67.38}{111.04}}
\rput{50.55}(50.1,46.02){\psellipticarc[linewidth=0.2,linestyle=dashed,dash=0.5 0.5](0,0)(14.06,-13.96){-151.8}{-104.8}}
\rput(30,58){}
\rput(29,78){$\scr{j_l}$}
\rput(49,44){$\scr{e}$}
\psline(35,66)(38,68)
\psline(41,53)(42,54)
\psline(38,68)(45,54)
\psline(42,54)(43,52)
\psline(45,55)(45,52)
\psline(43,52)(45,52)
\psline(46,50)(45,52)
\rput(36,69){$\scr{m}$}
\rput(36,59){$\scr{c}$}
\rput(40,48){$\scr{d}$}
\psline[linewidth=0.2,linestyle=dashed,dash=0.5 0.5](45,33)(42,25)
\psline[linewidth=0.2,linestyle=dashed,dash=0.5 0.5](68,8)(70,4)
\psline[linewidth=0.2,linestyle=dashed,dash=0.5 0.5](62,39)(73,18)
\psline[linewidth=0.2,linestyle=dashed,dash=0.5 0.5](92,30)(98,28)
\psline[linewidth=0.2,linestyle=dashed,dash=0.5 0.5](82,33)(84,32)
\psline(37,16)(39,22)
\psline(55,38)(49,50)
\psline(68,12)(37,16)
\psline(59,25)(37,16)
\psline(90,37)(68,29)
\psline(90,37)(68,12)
\psline(90,37)(78,41)
\psline[dotsize=1.8 2.5]{*-}(50,50)(46,50)
\rput(29,3){$\scr{j_i}$}
\rput(77,2){$\scr{j_j}$}
\rput(102,31){$\scr{j_k}$}
\psline[border=0.5](77,39)(100,32)
\psline[border=0.5](53,46)(57,38)
\psline[linewidth=0.6,linestyle=dashed,dash=1 1,border=0.4](44,46)(53,46)
\rput(46,38){$\scr{m}$}
\rput(49,11){$\scr{m}$}
\rput(82,23){$\scr{m}$}
\rput(53,38){$\scr{m}$}
\rput(61,49){$\scr{m}$}
\rput(52,24){$\scr{m}$}
\psline(46,40)(49,49)
\psline(43,37)(46,40)
\psline(43,37)(42,40)
\psline(39,25)(43,37)
\psline(39,22)(39,25)
\psline(31,5)(37,24)
\psline(39,25)(37,24)
\psline(55,38)(57,36)
\rput(37,32){$\scr{f}$}
\psline(68,12)(64,20)
\psline[border=0.5](66,21)(75,4)
\psline(57,38)(57,36)
\psline(57,36)(64,23)
\psline(64,23)(66,21)
\psline(64,23)(64,20)
\psline(61,46)(50,49)
\psline(63,44)(61,46)
\psline[border=0.5](53,46)(60,44)
\psline(60,44)(63,44)
\psline(75,41)(78,41)
\psline(75,41)(77,39)
\psline(63,44)(75,41)
\rput(63,30){$\scr{g}$}
\rput(71,44){$\scr{h}$}
\rput(39,41){$\scr{j_i}$}
\rput(59,39){$\scr{j_j}$}
\rput(60,42){$\scr{j_k}$}
\rput(47,52){$\scr{1}$}
\rput(42,55){$\scr{m}$}
\end{pspicture}
\end{array}
\end{split}
\ee
\vskip1cm
\noindent  We have inserted the tetrahedral spinnetwork of $h_{\tetraedro}$, contracted with the free leg of the trace part of the operator and in the last line we have recoupled the holonomies on the same edges. The construction of the matrix elements of the volume acting on 5-valent nodes $V{}_{j_l,i_n}{}^{d,e} (j_i,j_j,j_k,m,c)$, is reviewed in the Appendix \ref{5-valent not gauge} (but we stress that a complete analythic formula is lacking). We can then simplify the last expression using the recoupling identity \eqref{taglio2} in the upper part of the graph. We obtain
\be
\begin{split}
&\hat{H}^{s}_{\Delta}|v_4\rangle=N_m^2 c_m\!\!\! \sum_{c,d,e,f,g,h} 
 \hspace{-1em}d_c \,d_f\,d_g\,d_h\, \frac{l_0^3}{4} \,  
                \, V_{j_l,i_n}\!\!\!{}^{d,e} (j_i,j_j,j_k,m,c) 
                \;
                \left\{\begin{matrix}
                      j_l      & d & 1 \\
                      m      & m   & c                  
\end{matrix}
\right\} \hspace{-4em}
\begin{array}{c}
	\ifx\JPicScale\undefined\def\JPicScale{0.5}\fi
\psset{unit=\JPicScale mm}
\psset{linewidth=0.3,dotsep=1,hatchwidth=0.3,hatchsep=1.5,shadowsize=1,dimen=middle}
\psset{dotsize=0.7 2.5,dotscale=1 1,fillcolor=black}
\psset{arrowsize=1 2,arrowlength=1,arrowinset=0.25,tbarsize=0.7 5,bracketlength=0.15,rbracketlength=0.15}
\begin{pspicture}(0,0)(102,78)
\psline[linewidth=0.2,linestyle=dashed,dash=0.5 0.5](26,5)(37,40)
\psline[linewidth=0.2,linestyle=dashed,dash=0.5 0.5](77,11)(80,4)
\psline[linewidth=0.2,linestyle=dashed,dash=0.5 0.5](37,8)(36,5)
\psline[linewidth=0.2,linestyle=dashed,dash=0.5 0.5](55,33)(64,16)
\psline(42,40)(44,46)
\psline[linewidth=0.2,linestyle=dashed,dash=0.5 0.5](62,39)(70,36)
\psline[linewidth=0.2,linestyle=dashed,dash=0.5 0.5](84,32)(73,19)
\psline[linewidth=0.2,linestyle=dashed,dash=0.5 0.5](93,30)(77,11)
\psline[linewidth=0.2,linestyle=dashed,dash=0.5 0.5](56,30)(42,25)
\psline[linewidth=0.2,linestyle=dashed,dash=0.5 0.5](59,23)(40,16)
\psline[linewidth=0.2,linestyle=dashed,dash=0.5 0.5](70,36)(65,34)
\psline[linewidth=0.2,linestyle=dashed,dash=0.5 0.5](82,33)(68,28)
\psline[linewidth=0.2,linestyle=dashed,dash=0.5 0.5](64,16)(40,16)
\psline[linewidth=0.2,linestyle=dashed,dash=0.5 0.5](68,8)(37,8)
\psline[linewidth=0.2,linestyle=dashed,dash=0.5 0.5](36,75)(44,58)
\psline[linewidth=0.2,linestyle=dashed,dash=0.5 0.5](26,75)(37,52)
\psline(44,46)(31,75)
\psline[linewidth=0.2,linestyle=dashed,dash=0.5 0.5](64,48)(101,37)
\rput{90}(50.2,45.54){\psellipticarc[linewidth=0.2,linestyle=dashed,dash=0.5 0.5](0,0)(14.5,-14){-27.38}{80.22}}
\rput{0}(50,46){\psellipticarc[linewidth=0.2,linestyle=dashed,dash=0.5 0.5](0,0)(14,-14){67.38}{111.04}}
\rput{50.55}(50.1,46.02){\psellipticarc[linewidth=0.2,linestyle=dashed,dash=0.5 0.5](0,0)(14.06,-13.96){-151.8}{-104.8}}
\rput(30,58){}
\rput(29,78){$\scr{j_l}$}
\rput(49,44){$\scr{e}$}
\psline(36,64)(38,65)
\psline(46,50)(38,65)
\rput(39,53){$\scr{d}$}
\psline[linewidth=0.2,linestyle=dashed,dash=0.5 0.5](45,33)(42,25)
\psline[linewidth=0.2,linestyle=dashed,dash=0.5 0.5](68,8)(70,4)
\psline[linewidth=0.2,linestyle=dashed,dash=0.5 0.5](62,39)(73,18)
\psline[linewidth=0.2,linestyle=dashed,dash=0.5 0.5](92,30)(98,28)
\psline[linewidth=0.2,linestyle=dashed,dash=0.5 0.5](82,33)(84,32)
\psline(37,16)(39,22)
\psline(55,38)(49,50)
\psline(68,12)(37,16)
\psline(59,25)(37,16)
\psline(90,37)(68,29)
\psline(90,37)(68,12)
\psline(90,37)(78,41)
\psline[dotsize=1.8 2.5]{*-}(50,50)(46,50)
\rput(29,3){$\scr{j_i}$}
\rput(77,2){$\scr{j_j}$}
\rput(102,31){$\scr{j_k}$}
\psline[border=0.5](77,39)(100,32)
\psline[border=0.5](53,46)(57,38)
\psline[linewidth=0.6,linestyle=dashed,dash=1 1,border=0.4](44,46)(53,46)
\rput(46,38){$\scr{m}$}
\rput(49,11){$\scr{m}$}
\rput(82,23){$\scr{m}$}
\rput(53,38){$\scr{m}$}
\rput(61,49){$\scr{m}$}
\rput(52,24){$\scr{m}$}
\psline(46,40)(49,49)
\psline(43,37)(46,40)
\psline(43,37)(42,40)
\psline(39,25)(43,37)
\psline(39,22)(39,25)
\psline(31,5)(37,24)
\psline(39,25)(37,24)
\psline(55,38)(57,36)
\rput(37,32){$\scr{f}$}
\psline(68,12)(64,20)
\psline[border=0.5](66,21)(75,4)
\psline(57,38)(57,36)
\psline(57,36)(64,23)
\psline(64,23)(66,21)
\psline(64,23)(64,20)
\psline(61,46)(50,49)
\psline(63,44)(61,46)
\psline[border=0.5](53,46)(60,44)
\psline(60,44)(63,44)
\psline(75,41)(78,41)
\psline(75,41)(77,39)
\psline(63,44)(75,41)
\rput(63,30){$\scr{g}$}
\rput(71,44){$\scr{h}$}
\rput(39,41){$\scr{j_i}$}
\rput(59,39){$\scr{j_j}$}
\rput(60,42){$\scr{j_k}$}
\rput(45,55){$\scr{1}$}
\end{pspicture}
\end{array}
\end{split}
\ee

The final result is then 
\be
\begin{split}
&	 \hat{H}^{s}_{\Delta}|v_4\rangle
=
%	        \\&=
N_m^2 c_m \!\!\!\!
\sum_{c,d,e,f,g,h,k} \,d_c \,d_f\,d_g\,d_h\,d_k\, \frac{l_0^3}{4} \,  
                \, V{}_{j_l,i_n}{}^{d,e} (j_i,j_j,j_k,m,c) \nonumber \\& \hspace{5em}\times \left\{
\begin{matrix}
                      j_l      & d & 1 \\
                      m      & m   & c                  
\end{matrix}
\right\} \ \ 
F_{\times}(m,d,e,f,g,h,k,j_i,j_j,j_k,j_l) \;  
           \hspace{-3em}     \begin{array}{c}
\ifx\JPicScale\undefined\def\JPicScale{0.5}\fi
\psset{unit=\JPicScale mm}
\psset{linewidth=0.3,dotsep=1,hatchwidth=0.3,hatchsep=1.5,shadowsize=1,dimen=middle}
\psset{dotsize=0.7 2.5,dotscale=1 1,fillcolor=black}
\psset{arrowsize=1 2,arrowlength=1,arrowinset=0.25,tbarsize=0.7 5,bracketlength=0.15,rbracketlength=0.15}
\begin{pspicture}(0,0)(102,78)
\psline[linewidth=0.2,linestyle=dashed,dash=0.5 0.5](26,5)(37,40)
\psline[linewidth=0.2,linestyle=dashed,dash=0.5 0.5](77,11)(80,4)
\psline[linewidth=0.2,linestyle=dashed,dash=0.5 0.5](37,8)(36,5)
\psline[linewidth=0.2,linestyle=dashed,dash=0.5 0.5](55,33)(64,16)
\psline(34,18)(44,46)
\psline[linewidth=0.2,linestyle=dashed,dash=0.5 0.5](62,39)(70,36)
\psline[linewidth=0.2,linestyle=dashed,dash=0.5 0.5](84,32)(73,19)
\psline[linewidth=0.2,linestyle=dashed,dash=0.5 0.5](93,30)(77,11)
\psline[linewidth=0.2,linestyle=dashed,dash=0.5 0.5](56,30)(42,25)
\psline[linewidth=0.2,linestyle=dashed,dash=0.5 0.5](59,23)(45,18)
\psline[linewidth=0.2,linestyle=dashed,dash=0.5 0.5](70,36)(65,34)
\psline[linewidth=0.2,linestyle=dashed,dash=0.5 0.5](82,33)(68,28)
\psline[linewidth=0.2,linestyle=dashed,dash=0.5 0.5](64,16)(44,16)
\psline[linewidth=0.2,linestyle=dashed,dash=0.5 0.5](68,8)(37,8)
\psline[linewidth=0.2,linestyle=dashed,dash=0.5 0.5](36,75)(44,58)
\psline[linewidth=0.2,linestyle=dashed,dash=0.5 0.5](26,75)(37,52)
\psline(44,46)(31,75)
\psline[linewidth=0.2,linestyle=dashed,dash=0.5 0.5](64,48)(101,37)
\rput{90}(50.2,45.54){\psellipticarc[linewidth=0.2,linestyle=dashed,dash=0.5 0.5](0,0)(14.5,-14){-27.38}{80.22}}
\rput{0}(50,46){\psellipticarc[linewidth=0.2,linestyle=dashed,dash=0.5 0.5](0,0)(14,-14){67.38}{111.04}}
\rput{50.55}(50.1,46.02){\psellipticarc[linewidth=0.2,linestyle=dashed,dash=0.5 0.5](0,0)(14.06,-13.96){-151.8}{-104.8}}
\rput(30,58){}
\rput(29,78){$\scr{j_l}$}
\rput(48,49){$\scr{k}$}
\psline[linewidth=0.2,linestyle=dashed,dash=0.5 0.5](45,33)(42,25)
\psline[linewidth=0.2,linestyle=dashed,dash=0.5 0.5](68,8)(70,4)
\psline[linewidth=0.2,linestyle=dashed,dash=0.5 0.5](62,39)(73,18)
\psline[linewidth=0.2,linestyle=dashed,dash=0.5 0.5](92,30)(98,28)
\psline[linewidth=0.2,linestyle=dashed,dash=0.5 0.5](82,33)(84,32)
\psline(69,12)(36,15)
\psline(59,25)(36,15)
\psline(90,37)(68,29)
\psline(90,37)(69,12)
\rput(29,3){$\scr{j_i}$}
\rput(77,2){$\scr{j_j}$}
\rput(102,31){$\scr{j_k}$}
\psline[border=0.5](94,38)(100,32)
\psline[border=0.5](53,46)(74,10)
\psline[linewidth=0.6,linestyle=dashed,dash=1 1,border=0.4](44,46)(53,46)
\rput(49,11){$\scr{m}$}
\rput(82,23){$\scr{m}$}
\rput(52,24){$\scr{m}$}
\psline(31,5)(34,18)
\rput(37,32){$\scr{f}$}
\psline[border=0.5](74,10)(75,4)
\psline[border=0.5](53,46)(94,38)
\rput(61,28){$\scr{g}$}
\rput(71,40){$\scr{h}$}
\psline[linewidth=0.6,linestyle=dashed,dash=1 1](36,15)(34,18)
\psline[linewidth=0.6,linestyle=dashed,dash=1 1](68,12)(74,10)
\psline[linewidth=0.6,linestyle=dashed,dash=1 1](94,38)(90,37)
\psline[linewidth=0.2,linestyle=dashed,dash=0.5 0.5](45,18)(44,16)
\rput(71,9){$\scr{m}$}
\rput(37,19){$\scr{m}$}
\rput(92,35){$\scr{m}$}
\end{pspicture}
\end{array}
\end{split}
\label{finale}
\ee
where we have used the recoupling identity \eqref{taglio3} in the dashed circle representing the node and $F_{\times}$ is the evaluation of the following recoupling object:
\be
F_{\times}(m,d,e,f,g,h,k,j_i,j_j,j_k,j_l)=
\begin{array}{c}
\ifx\JPicScale\undefined\def\JPicScale{0.8}\fi
\psset{unit=\JPicScale mm}
\psset{linewidth=0.3,dotsep=1,hatchwidth=0.3,hatchsep=1.5,shadowsize=1,dimen=middle}
\psset{dotsize=0.7 2.5,dotscale=1 1,fillcolor=black}
\psset{arrowsize=1 2,arrowlength=1,arrowinset=0.25,tbarsize=0.7 5,bracketlength=0.15,rbracketlength=0.15}
\begin{pspicture}(0,0)(54,49)
\psline(16,28)(18,34)
\psline(18,34)(11,49)
\rput(8,33){$\scr{j_l}$}
\rput(28,32){$\scr{e}$}
\psline(12,47)(15,48)
\psline(25,39)(15,48)
\rput(14,36){$\scr{d}$}
\rput(21,26){$\scr{m}$}
\psline(20,28)(25,37)
\psline(17,25)(20,28)
\psline(17,25)(16,28)
\psline(16,21)(17,25)
\rput(19,18){$\scr{f}$}
\rput(14,29){$\scr{j_i}$}
\psline(36,26)(27,37)
\psline[border=0.5](34,34)(38,26)
\psline[linewidth=0.6,linestyle=dashed,dash=1 1,border=0.4](18,34)(34,34)
\rput(33,26){$\scr{m}$}
\rput(42,37){$\scr{m}$}
\psline(36,26)(38,24)
\psline(38,26)(38,24)
\psline(38,24)(39,22)
\psline(42,34)(28,38)
\psline(44,32)(42,34)
\psline[border=0.5](34,34)(41,32)
\psline(41,32)(44,32)
\psline(44,32)(51,30)
\rput(36,18){$\scr{g}$}
\rput(54,18){$\scr{h}$}
\rput(40,27){$\scr{j_j}$}
\rput(41,30){$\scr{j_k}$}
\psline(18,10)(11,25)
\psline[border=0.5](34,10)(39,1)
\psline[linewidth=0.6,linestyle=dashed,dash=1 1,border=0.4](18,10)(34,10)
\psline(34,10)(51,4)
\psline(11,49)(11,25)
\psline[border=0.3](16,4)(16,21)
\psline(16,4)(18,10)
\psline[border=0.3](39,22)(39,1)
\psline(51,30)(51,4)
\rput(26,7){$\scr{k}$}
\rput(21,48){$\scr{1}$}
\rput{0}(26,38){\psellipse[fillstyle=solid](0,0)(2,2)}
\end{pspicture}
\end{array}
\label{F}
\ee 
The form of the previous coefficient and of the normalizating factor $c_m$ depend on the specific intertwiner $i^{i\alpha\beta\gamma}=
\begin{array}{c}
	\ifx\JPicScale\undefined\def\JPicScale{0.2}\fi
\psset{unit=\JPicScale mm}
\psset{linewidth=0.3,dotsep=1,hatchwidth=0.3,hatchsep=1.5,shadowsize=1,dimen=middle}
\psset{dotsize=0.7 2.5,dotscale=1 1,fillcolor=black}
\psset{arrowsize=1 2,arrowlength=1,arrowinset=0.25,tbarsize=0.7 5,bracketlength=0.15,rbracketlength=0.15}
\begin{pspicture}(0,0)(40,44)
\psline[linewidth=0.2,arrowsize=2.8 2,dotsize=3 3.5]{*-}(12,18)(0,6)
\psline(38,18)(12,18)
\psline(12,44)(12,18)
\psline[linewidth=0.2](28,34)(12,18)
\end{pspicture}
\end{array}
$ , symmetric for cyclic permutations on the last three index,
chosen for the regularization of the curvature.
We complete the analysis of the constraint making the following choice for $i^{i\alpha\beta\gamma}$:\\
\be
\begin{array}{c}
\ifx\JPicScale\undefined\def\JPicScale{0.4}\fi
\psset{unit=\JPicScale mm}
\psset{linewidth=0.3,dotsep=1,hatchwidth=0.3,hatchsep=1.5,shadowsize=1,dimen=middle}
\psset{dotsize=0.7 2.5,dotscale=1 1,fillcolor=black}
\psset{arrowsize=1 2,arrowlength=1,arrowinset=0.25,tbarsize=0.7 5,bracketlength=0.15,rbracketlength=0.15}
\begin{pspicture}(0,0)(166,39)
\psline[linewidth=0.2,arrowsize=2.8 2,dotsize=1.5 2.5]{*-}(12,12)(0,0)
\psline(38,12)(12,12)
\psline(12,38)(12,12)
\psline[linewidth=0.2](28,28)(12,12)
\rput(45,11){=}
\psline[linewidth=0.2](60,13)(48,1)
\psline(86,13)(60,13)
\psline(60,39)(60,22)
\psline[linewidth=0.2](76,29)(69,22)
\psline(69,22)(60,22)
\psline[linestyle=dotted](60,22)(60,13)
\psline[linewidth=0.2](100,13)(88,1)
\psline(126,13)(109,13)
\psline(100,39)(100,13)
\psline[linewidth=0.2](116,29)(109,22)
\psline(109,22)(109,13)
\psline[linestyle=dotted](109,13)(100,13)
\psline(166,12)(140,12)
\psline(140,38)(140,12)
\psline[linewidth=0.2](156,28)(149,21)
\psline[linewidth=0.2,border=1.05](149,21)(128,0)
\psline[linestyle=dotted](149,21)(140,21)
\rput(91,13){+}
\rput(132,13){+}
\rput(14,8){Sym}
\end{pspicture}
\end{array}
\label{nodo simmetrizzato}
\ee
(a better choice may probably be obtained using the Livine-Speziale coherent intertwiners \cite{Livine:2007vk}; this will be considered elsewhere.) This choice consists in a symmetrization over pairings (that we denote $i_x,i_y,i_z$) of a 4-valent intertwiner between a representation $1$ and three representations $m$ (we stress that the virtual link can only take the values $m\pm1$).
 
In this case $F_{Sym}$ is the sum of three terms, one for each pairing of the 4-valent intertwiner:
\be
\begin{split}
&F_{Sym}(m,d,e,f,g,h,k, j_i,j_j,j_k,j_l)\\
&=
\begin{array}{c}
	\ifx\JPicScale\undefined\def\JPicScale{0.8}\fi
\psset{unit=\JPicScale mm}
\psset{linewidth=0.3,dotsep=1,hatchwidth=0.3,hatchsep=1.5,shadowsize=1,dimen=middle}
\psset{dotsize=0.7 2.5,dotscale=1 1,fillcolor=black}
\psset{arrowsize=1 2,arrowlength=1,arrowinset=0.25,tbarsize=0.7 5,bracketlength=0.15,rbracketlength=0.15}
\begin{pspicture}(0,0)(54,49)
\psline(16,28)(18,34)
\psline(18,34)(11,49)
\rput(8,33){$\scr{j_l}$}
\rput(28,32){$\scr{e}$}
\psline(12,47)(15,48)
\psline(23,38)(15,48)
\rput(14,36){$\scr{d}$}
\rput(21,26){$\scr{m}$}
\psline(20,28)(23,38)
\psline(17,25)(20,28)
\psline(17,25)(16,28)
\psline(16,21)(17,25)
\rput(19,18){$\scr{f}$}
\rput(14,29){$\scr{j_i}$}
\psline(36,26)(30,38)
\psline[linewidth=0.55,linestyle=dashed,dash=1 1](30,38)(23,38)
\psline[border=0.5](34,34)(38,26)
\psline[linewidth=0.6,linestyle=dashed,dash=1 1,border=0.4](18,34)(34,34)
\rput(33,26){$\scr{m}$}
\rput(42,37){$\scr{m}$}
\psline(36,26)(38,24)
\psline(38,26)(38,24)
\psline(38,24)(39,22)
\psline(42,34)(30,38)
\psline(44,32)(42,34)
\psline[border=0.5](34,34)(41,32)
\psline(41,32)(44,32)
\psline(44,32)(51,30)
\rput(36,18){$\scr{g}$}
\rput(54,18){$\scr{h}$}
\rput(40,27){$\scr{j_j}$}
\rput(41,30){$\scr{j_k}$}
\psline(18,10)(11,25)
\psline[border=0.5](34,10)(39,1)
\psline[linewidth=0.6,linestyle=dashed,dash=1 1,border=0.4](18,10)(34,10)
\psline(34,10)(51,4)
\psline(11,49)(11,25)
\psline[border=0.3](16,4)(16,21)
\psline(16,4)(18,10)
\psline[border=0.3](39,22)(39,1)
\psline(51,30)(51,4)
\rput(27,41){$\scr{i_x}$}
\rput(26,7){$\scr{k}$}
\rput(21,48){$\scr{1}$}
\end{pspicture}
\end{array}
+\hspace{-1em}
\begin{array}{c}
	\ifx\JPicScale\undefined\def\JPicScale{0.8}\fi
\psset{unit=\JPicScale mm}
\psset{linewidth=0.3,dotsep=1,hatchwidth=0.3,hatchsep=1.5,shadowsize=1,dimen=middle}
\psset{dotsize=0.7 2.5,dotscale=1 1,fillcolor=black}
\psset{arrowsize=1 2,arrowlength=1,arrowinset=0.25,tbarsize=0.7 5,bracketlength=0.15,rbracketlength=0.15}
\begin{pspicture}(0,0)(54,49)
\psline(16,28)(18,34)
\psline(18,34)(11,49)
\rput(8,33){$\scr{j_l}$}
\rput(28,32){$\scr{e}$}
\psline(12,47)(15,48)
\psline(26,43)(15,48)
\rput(14,36){$\scr{d}$}
\rput(21,26){$\scr{m}$}
\psline(20,28)(26,38)
\psline(17,25)(20,28)
\psline(17,25)(16,28)
\psline(16,21)(17,25)
\rput(19,18){$\scr{f}$}
\rput(14,29){$\scr{j_i}$}
\psline(36,26)(26,43)
\psline[linewidth=0.55,linestyle=dashed,dash=1 1](26,43)(26,38)
\psline[border=0.5](34,34)(38,26)
\psline[linewidth=0.6,linestyle=dashed,dash=1 1,border=0.4](18,34)(34,34)
\rput(33,26){$\scr{m}$}
\rput(42,37){$\scr{m}$}
\psline(36,26)(38,24)
\psline(38,26)(38,24)
\psline(38,24)(39,22)
\psline[border=0.3](42,34)(26,38)
\psline(44,32)(42,34)
\psline[border=0.5](34,34)(41,32)
\psline(41,32)(44,32)
\psline(44,32)(51,30)
\rput(36,18){$\scr{g}$}
\rput(54,18){$\scr{h}$}
\rput(40,27){$\scr{j_j}$}
\rput(41,30){$\scr{j_k}$}
\psline(18,10)(11,25)
\psline[border=0.5](34,10)(39,1)
\psline[linewidth=0.6,linestyle=dashed,dash=1 1,border=0.4](18,10)(34,10)
\psline(34,10)(51,4)
\psline(11,49)(11,25)
\psline[border=0.3](16,4)(16,21)
\psline(16,4)(18,10)
\psline[border=0.3](39,22)(39,1)
\psline(51,30)(51,4)
\rput(23,39){$\scr{i_z}$}
\rput(26,7){$\scr{k}$}
\rput(21,48){$\scr{1}$}
\end{pspicture}
\end{array}
+\hspace{-1em}
\begin{array}{c}
	\ifx\JPicScale\undefined\def\JPicScale{0.8}\fi
\psset{unit=\JPicScale mm}
\psset{linewidth=0.3,dotsep=1,hatchwidth=0.3,hatchsep=1.5,shadowsize=1,dimen=middle}
\psset{dotsize=0.7 2.5,dotscale=1 1,fillcolor=black}
\psset{arrowsize=1 2,arrowlength=1,arrowinset=0.25,tbarsize=0.7 5,bracketlength=0.15,rbracketlength=0.15}
\begin{pspicture}(0,0)(54,49)
\psline(16,28)(18,34)
\psline(18,34)(11,49)
\rput(8,33){$\scr{j_l}$}
\rput(28,32){$\scr{e}$}
\psline(12,47)(15,48)
\psline(26,43)(15,48)
\rput(14,36){$\scr{d}$}
\rput(21,26){$\scr{m}$}
\psline(20,28)(26,38)
\psline(17,25)(20,28)
\psline(17,25)(16,28)
\psline(16,21)(17,25)
\rput(19,17){$\scr{f}$}
\rput(14,29){$\scr{j_i}$}
\psline(36,26)(26,38)
\psline[linewidth=0.55,linestyle=dashed,dash=1 1](26,43)(26,38)
\psline[border=0.5](34,34)(38,26)
\psline[linewidth=0.6,linestyle=dashed,dash=1 1,border=0.4](18,34)(34,34)
\rput(33,26){$\scr{m}$}
\rput(42,37){$\scr{m}$}
\psline(36,26)(38,24)
\psline(38,26)(38,24)
\psline(38,24)(39,22)
\psline(42,34)(26,43)
\psline(44,32)(42,34)
\psline[border=0.5](34,34)(41,32)
\psline(41,32)(44,32)
\psline(44,32)(51,30)
\rput(36,17){$\scr{g}$}
\rput(54,17){$\scr{h}$}
\rput(40,27){$\scr{j_j}$}
\rput(41,30){$\scr{j_k}$}
\psline(18,10)(11,25)
\psline[border=0.5](34,10)(39,1)
\psline[linewidth=0.6,linestyle=dashed,dash=1 1,border=0.4](18,10)(34,10)
\psline(34,10)(51,4)
\psline(11,49)(11,25)
\psline[border=0.3](16,4)(16,21)
\psline(16,4)(18,10)
\psline[border=0.3](39,22)(39,1)
\psline(51,30)(51,4)
\rput(23,40){$\scr{i_y}$}
\rput(26,7){$\scr{k}$}
\rput(21,48){$\scr{1}$}
\end{pspicture}
\end{array}
\end{split}
\label{F_Sym}
\ee
Each of the three terms depends on fifteen spins but the first one is topologically different from the other two and in fact it can be simplified using \eqref{taglio2} on the three dashed lines
\be
\begin{array}{c}
	\ifx\JPicScale\undefined\def\JPicScale{0.8}\fi
\psset{unit=\JPicScale mm}
\psset{linewidth=0.3,dotsep=1,hatchwidth=0.3,hatchsep=1.5,shadowsize=1,dimen=middle}
\psset{dotsize=0.7 2.5,dotscale=1 1,fillcolor=black}
\psset{arrowsize=1 2,arrowlength=1,arrowinset=0.25,tbarsize=0.7 5,bracketlength=0.15,rbracketlength=0.15}
\begin{pspicture}(0,0)(54,49)
\psline(16,28)(18,34)
\psline(18,34)(11,49)
\rput(8,33){$\scr{j_l}$}
\rput(28,32){$\scr{e}$}
\psline(12,47)(15,48)
\psline(23,38)(15,48)
\rput(14,36){$\scr{d}$}
\rput(21,26){$\scr{m}$}
\psline(20,28)(23,38)
\psline(17,25)(20,28)
\psline(17,25)(16,28)
\psline(16,21)(17,25)
\rput(19,18){$\scr{f}$}
\rput(14,29){$\scr{j_i}$}
\psline(36,26)(30,38)
\psline[linewidth=0.55,linestyle=dashed,dash=1 1](30,38)(23,38)
\psline[border=0.5](34,34)(38,26)
\psline[linewidth=0.6,linestyle=dashed,dash=1 1,border=0.4](18,34)(34,34)
\rput(33,26){$\scr{m}$}
\rput(42,37){$\scr{m}$}
\psline(36,26)(38,24)
\psline(38,26)(38,24)
\psline(38,24)(39,22)
\psline(42,34)(30,38)
\psline(44,32)(42,34)
\psline[border=0.5](34,34)(41,32)
\psline(41,32)(44,32)
\psline(44,32)(51,30)
\rput(36,18){$\scr{g}$}
\rput(54,18){$\scr{h}$}
\rput(40,27){$\scr{j_j}$}
\rput(41,30){$\scr{j_k}$}
\psline(18,10)(11,25)
\psline[border=0.5](34,10)(39,1)
\psline[linewidth=0.6,linestyle=dashed,dash=1 1,border=0.4](18,10)(34,10)
\psline(34,10)(51,4)
\psline(11,49)(11,25)
\psline[border=0.3](16,4)(16,21)
\psline(16,4)(18,10)
\psline[border=0.3](39,22)(39,1)
\psline(51,30)(51,4)
\rput(27,41){$\scr{i_x}$}
\rput(26,7){$\scr{k}$}
\rput(21,48){$\scr{1}$}
\end{pspicture}
\end{array}
\hspace{-.5em}
=\hspace{-2.5em}
\begin{array}{c}
	\ifx\JPicScale\undefined\def\JPicScale{0.8}\fi
\psset{unit=\JPicScale mm}
\psset{linewidth=0.3,dotsep=1,hatchwidth=0.3,hatchsep=1.5,shadowsize=1,dimen=middle}
\psset{dotsize=0.7 2.5,dotscale=1 1,fillcolor=black}
\psset{arrowsize=1 2,arrowlength=1,arrowinset=0.25,tbarsize=0.7 5,bracketlength=0.15,rbracketlength=0.15}
\begin{pspicture}(0,0)(74,49)
\psline(16,28)(18,34)
\psline(18,34)(11,49)
\rput(8,33){$\scr{j_l}$}
\rput(26,28){$\scr{e}$}
\psline(12,47)(15,48)
\psline(23,38)(15,48)
\rput(14,36){$\scr{d}$}
\rput(21,26){$\scr{m}$}
\psline(20,28)(23,38)
\psline(17,25)(20,28)
\psline(17,25)(16,28)
\psline(16,21)(17,25)
\rput(19,18){$\scr{f}$}
\rput(14,29){$\scr{j_i}$}
\psline[linewidth=0.55,linestyle=dashed,dash=1 1](33,27)(23,38)
\psline(18,10)(11,25)
\psline[linewidth=0.6,linestyle=dashed,dash=1 1,border=0.4](18,10)(33,27)
\psline(11,49)(11,25)
\psline[border=0.3](16,4)(16,21)
\psline(16,4)(18,10)
\rput(30,34){$\scr{i_x}$}
\rput(29,17){$\scr{k}$}
\rput(21,48){$\scr{1}$}
\psline[linewidth=0.6,linestyle=dashed,dash=1 1,border=0.4](18,34)(33,27)
\psline(56,26)(50,38)
\psline[linewidth=0.55,linestyle=dashed,dash=1 1](50,38)(38,27)
\psline[border=0.5](54,34)(58,26)
\psline[linewidth=0.6,linestyle=dashed,dash=1 1,border=0.4](38,27)(54,34)
\rput(53,26){$\scr{m}$}
\rput(62,37){$\scr{m}$}
\psline(56,26)(58,24)
\psline(58,26)(58,24)
\psline(58,24)(59,22)
\psline(62,34)(50,38)
\psline(64,32)(62,34)
\psline[border=0.5](54,34)(61,32)
\psline(61,32)(64,32)
\psline(64,32)(71,30)
\rput(56,18){$\scr{g}$}
\rput(74,18){$\scr{h}$}
\rput(60,27){$\scr{j_j}$}
\rput(61,30){$\scr{j_k}$}
\psline[border=0.5](54,10)(59,1)
\psline[linewidth=0.6,linestyle=dashed,dash=1 1,border=0.4](38,27)(54,10)
\psline(54,10)(71,4)
\psline[border=0.3](59,22)(59,1)
\psline(71,30)(71,4)
\rput(43,35){$\scr{i_x}$}
\rput(42,17){$\scr{k}$}
\rput(46,28){$\scr{e}$}
\end{pspicture}
\end{array}
\hspace{-.5em}
=
\left\{\begin{matrix}
j_i&e&d\\
m&i_x&1\\
f&k&j_l
\end{matrix}\right\}
\left\{\begin{matrix}
k&e&i_x\\
g&j_j&m\\
h&j_k&m
\end{matrix}\right\}
\ee

The first term in the sum \eqref{F_Sym} is then the product of two 9j symbols (which are strictly related to the fusion coefficients used in the spinfoam vertex amplitude \cite{Engle:2007uq,fusion}). The second and third term are, instead, two different $15j$ symbols as can be seen rearranging the graphs to the more familiar shape.
The final expression is then 
\be
\begin{split}
&F_{Sym}(m,d,e,f,g,h,k, j_i,j_j,j_k,j_l)\\
&=\ \left\{\begin{matrix}
j_i&e&d\\
m&i_x&1\\
f&k&j_l
\end{matrix}\right\}
\left\{\begin{matrix}
k&e&i_x\\
g&j_j&m\\
h&j_k&m
\end{matrix}\right\}\; +\!\!
\begin{array}{c}
\ifx\JPicScale\undefined\def\JPicScale{0.6}\fi
\psset{unit=\JPicScale mm}
\psset{linewidth=0.3,dotsep=1,hatchwidth=0.3,hatchsep=1.5,shadowsize=1,dimen=middle}
\psset{dotsize=0.7 2.5,dotscale=1 1,fillcolor=black}
\psset{arrowsize=1 2,arrowlength=1,arrowinset=0.25,tbarsize=0.7 5,bracketlength=0.15,rbracketlength=0.15}
\begin{pspicture}(0,0)(61,47)
\psline(40,45)(31,45)
\psline(54.9,5)(59.9,37)
\psline(27.9,15)(16.9,5)
\psline(31,45)(12,37)
\psline(40,45)(59.9,37)
\psline(11.9,37)(23.9,29)
\psline(59.9,37)(47.9,29)
\psline(11.9,37)(16.9,5)
\psline(54.9,5)(43.9,15)
\psline(16.9,5)(54.9,5)
\psline(40,45)(46,13)
\psline(47.9,29)(27.9,15)
\psline(43.9,15)(23.9,29)
\psline(20,32)(52,32)
\psline(26,13)(31,45)
\rput(19,43){$\scr{j_l}$}
\rput(35,47){$\scr{k}$}
\rput(51,43){$\scr{h}$}
\rput(32,39){$\scr{f}$}
\rput(39,39){$\scr{g}$}
\rput(35,30){$\scr{i_z}$}
\rput(17,35){$\scr{1}$}
\rput(54,35){$\scr{m}$}
\rput(32,25){$\scr{m}$}
\rput(39,25){$\scr{m}$}
\rput(36,3){$\scr{e}$}
\rput(50,12){$\scr{j_j}$}
\rput(21,12){$\scr{j_i}$}
\rput(11,23){$\scr{d}$}
\rput(61,23){$\scr{j_k}$}
\end{pspicture}
\end{array}
\quad
+\!\!
\begin{array}{c}
\ifx\JPicScale\undefined\def\JPicScale{0,6}\fi
\psset{unit=\JPicScale mm}
\psset{linewidth=0.3,dotsep=1,hatchwidth=0.3,hatchsep=1.5,shadowsize=1,dimen=middle}
\psset{dotsize=0.7 2.5,dotscale=1 1,fillcolor=black}
\psset{arrowsize=1 2,arrowlength=1,arrowinset=0.25,tbarsize=0.7 5,bracketlength=0.15,rbracketlength=0.15}
\begin{pspicture}(0,0)(61,47)
\psline(40,45)(31,45)
\psline(54.9,5)(59.9,37)
\psline(27.9,15)(16.9,5)
\psline(31,45)(12,37)
\psline(40,45)(59.9,37)
\psline(11.9,37)(23.9,29)
\psline(59.9,37)(47.9,29)
\psline(11.9,37)(16.9,5)
\psline(54.9,5)(43.9,15)
\psline(16.9,5)(54.9,5)
\psline(40,45)(46,13)
\psline(47.9,29)(27.9,15)
\psline(43.9,15)(23.9,29)
\psline(20,32)(52,32)
\psline(26,13)(31,45)
\rput(19,43){$\scr{j_l}$}
\rput(35,47){$\scr{k}$}
\rput(51,43){$\scr{g}$}
\rput(32,39){$\scr{f}$}
\rput(39,39){$\scr{h}$}
\rput(35,30){$\scr{i_y}$}
\rput(17,35){$\scr{1}$}
\rput(54,35){$\scr{m}$}
\rput(32,25){$\scr{m}$}
\rput(39,25){$\scr{m}$}
\rput(36,3){$\scr{e}$}
\rput(50,12){$\scr{j_k}$}
\rput(21,12){$\scr{j_i}$}
\rput(11,23){$\scr{d}$}
\rput(61,23){$\scr{j_j}$}
\end{pspicture}
\end{array}
\end{split}
\ee
The inverse of the normalization coefficient $c_m$, in the case of the symmetrized node \eqref{nodo simmetrizzato} is
\be
\begin{array} {c}
\ifx\JPicScale\undefined\def\JPicScale{0.5}\fi
\psset{unit=\JPicScale mm}
\psset{linewidth=0.3,dotsep=1,hatchwidth=0.3,hatchsep=1.5,shadowsize=1,dimen=middle}
\psset{dotsize=0.7 2.5,dotscale=1 1,fillcolor=black}
\psset{arrowsize=1 2,arrowlength=1,arrowinset=0.25,tbarsize=0.7 5,bracketlength=0.15,rbracketlength=0.15}
\begin{pspicture}(0,0)(54,57)
\psline[linewidth=0.2,dotsize=2 2.5]{*-}(12,12)(7,7)
\psline[linewidth=0.2](12,12)(52,52)
\psline[linewidth=0.2,border=0.3](12,52)(52,12)
\psline[linewidth=0.2](12,12)(12,52)
\psline[linewidth=0.2](12,12)(52,12)
\psline[linewidth=0.2](52,52)(52,12)
\psline[linewidth=0.2](12,52)(52,52)
\psline(52,38)(52,12)
\psline(32,52)(32,57)
\rput(4,32){$\scriptstyle{1}$}
\psline(32,57)(7,57)
\psline(7,57)(7,7)
\rput(32,10){$\scriptstyle{m}$}
\rput(25,22){$\scriptstyle{m}$}
\rput(14,27){$\scriptstyle{m}$}
\rput(23,50){$\scriptstyle{m}$}
\rput(41,50){$\scriptstyle{m}$}
\rput(25,42){$\scriptstyle{m}$}
\rput(13,8){$\scriptstyle{Sym}$}
\rput(54,32){$\scriptstyle{m}$}
\end{pspicture}
\end{array}
=
\begin{array}{c}
\ifx\JPicScale\undefined\def\JPicScale{0.5}\fi
\psset{unit=\JPicScale mm}
\psset{linewidth=0.3,dotsep=1,hatchwidth=0.3,hatchsep=1.5,shadowsize=1,dimen=middle}
\psset{dotsize=0.7 2.5,dotscale=1 1,fillcolor=black}
\psset{arrowsize=1 2,arrowlength=1,arrowinset=0.25,tbarsize=0.7 5,bracketlength=0.15,rbracketlength=0.15}
\begin{pspicture}(0,0)(54,57)
\psline(21,21)(52,52)
\psline[linewidth=0.2,border=0.3](12,52)(52,12)
\psline[linewidth=0.2](52,52)(52,12)
\psline[linewidth=0.2](12,52)(52,52)
\psline(52,38)(52,12)
\psline(32,52)(32,57)
\rput(4,32){$\scriptstyle{1}$}
\psline(32,57)(7,57)
\psline(7,57)(7,7)
\rput(32,10){$\scriptstyle{m}$}
\rput(25,22){$\scriptstyle{m}$}
\rput(14,27){$\scriptstyle{m}$}
\rput(23,50){$\scriptstyle{m}$}
\rput(41,50){$\scriptstyle{m}$}
\rput(25,42){$\scriptstyle{m}$}
\rput(54,32){$\scriptstyle{m}$}
\psline(12,12)(7,7)
\psline(52,12)(12,12)
\psline(12,52)(12,21)
\psline(21,21)(12,21)
\psline[linestyle=dotted](12,21)(12,12)
\rput(14,16){$\scriptstyle{i_x}$}
\end{pspicture}	
\end{array}
+
\begin{array}{c}
\ifx\JPicScale\undefined\def\JPicScale{0.5}\fi
\psset{unit=\JPicScale mm}
\psset{linewidth=0.3,dotsep=1,hatchwidth=0.3,hatchsep=1.5,shadowsize=1,dimen=middle}
\psset{dotsize=0.7 2.5,dotscale=1 1,fillcolor=black}
\psset{arrowsize=1 2,arrowlength=1,arrowinset=0.25,tbarsize=0.7 5,bracketlength=0.15,rbracketlength=0.15}
\begin{pspicture}(0,0)(54,57)
\psline(21,21)(52,52)
\psline[linewidth=0.2,border=0.3](12,52)(52,12)
\psline[linewidth=0.2](52,52)(52,12)
\psline[linewidth=0.2](12,52)(52,52)
\psline(52,38)(52,12)
\psline(32,52)(32,57)
\rput(4,32){$\scriptstyle{1}$}
\psline(32,57)(7,57)
\psline(7,57)(7,7)
\rput(32,10){$\scriptstyle{m}$}
\rput(25,22){$\scriptstyle{m}$}
\rput(14,27){$\scriptstyle{m}$}
\rput(23,50){$\scriptstyle{m}$}
\rput(41,50){$\scriptstyle{m}$}
\rput(25,42){$\scriptstyle{m}$}
\rput(54,32){$\scriptstyle{m}$}
\psline(12,12)(7,7)
\psline(12,52)(12,12)
\rput(16,15){$\scriptstyle{i_y}$}
\psline(52,12)(21,12)
\psline(21,21)(21,12)
\psline[linestyle=dotted](21,12)(12,12)
\end{pspicture}
\end{array}
+
\begin{array}{c}
	\ifx\JPicScale\undefined\def\JPicScale{0.5}\fi
\psset{unit=\JPicScale mm}
\psset{linewidth=0.3,dotsep=1,hatchwidth=0.3,hatchsep=1.5,shadowsize=1,dimen=middle}
\psset{dotsize=0.7 2.5,dotscale=1 1,fillcolor=black}
\psset{arrowsize=1 2,arrowlength=1,arrowinset=0.25,tbarsize=0.7 5,bracketlength=0.15,rbracketlength=0.15}
\begin{pspicture}(0,0)(54,57)
\psline(21,21)(52,52)
\psline[linewidth=0.2,border=0.3](12,52)(52,12)
\psline[linewidth=0.2](52,52)(52,12)
\psline[linewidth=0.2](12,52)(52,52)
\psline(52,38)(52,12)
\psline(32,52)(32,57)
\rput(4,32){$\scriptstyle{1}$}
\psline(32,57)(7,57)
\psline(7,57)(7,7)
\rput(32,10){$\scriptstyle{m}$}
\rput(25,22){$\scriptstyle{m}$}
\rput(14,29){$\scriptstyle{m}$}
\rput(23,50){$\scriptstyle{m}$}
\rput(41,50){$\scriptstyle{m}$}
\rput(25,42){$\scriptstyle{m}$}
\rput(54,32){$\scriptstyle{m}$}
\psline(52,12)(12,12)
\rput(17,24){$\scriptstyle{i_z}$}
\psline(12,52)(12,12)
\psline[linewidth=0.2](28,28)(21,21)
\psline[border=1.05](21,21)(7,7)
\psline[linestyle=dotted](21,21)(12,21)
\end{pspicture}
\end{array}
\ee
The evaluation of the previous coefficients gives
\be
(c_m^{Sym})^{-1}=
\left\{
\begin{array}{ccc}
                      1      & m & i_x \\
                      m      & m   & m \\
                      m      & m   & m           
\end{array}
\right\} 
+
\left\{
\begin{matrix}
                      m      & m & i_y \\
                      m      & m   & m                  
\end{matrix}
\right\} 
\left\{
\begin{matrix}
                      m      & 1 & i_y \\
                      m      & m   & m                  
\end{matrix}
\right\} 
+
\left\{
\begin{matrix}
                      m      & m & i_z \\
                      m      & m   & m                  
\end{matrix}
\right\} 
\left\{
\begin{matrix}
                      m      & 1 & i_z \\
                      m      & m   & m                  
\end{matrix}
\right\}
\ee
The complete constraint is then obtained summing over the four legs $s$. 
%, this gives a linear combination of term of the kind \eqref{finale} with a permutation in the labels of the coefficients and of the direction of the new added triangle to the 4-valent node, giving exacly the analog of $\eqref{fine thomas}$.
 
\section{Conclusions}

We have studied a new regularization of the hamiltonian constraint operator. The main idea is to replace the holonomy of the connection around a \emph{triangle}, used for regularizing the curvature, by a spin-network function of the connection, defined over a \emph{tetrahedron}.  

We have pointed out two possible advantages with this alternative. First, it seems to us that the regularization is geometrically better motivated, at least from a simplicial point of view: the two factors of the Hamiltonian constraint ${\cal H}={\rm Tr}\,F\wedge e$ can  naturally be related to dual geometrical objects. The tetrad $e$ is determined by a link $s$ at a node. It seems reasonable to expect that the curvature $F$ must be taken around a loop that \emph{circles} the direction determined by $s$. This is what the new regularization does in general. 

Second, the new operators transforms a 4-valent node into \emph{four} 4-valent nodes. (The old operator added two 3-valent nodes to any node.) Therefore the new operator implements the 1-4 Pachner moves that characterizes the simplicial spinfoam evolution. We view this as a step that could simplify the long sought bridge between the canonical and the covariant definition of the dynamics.  We also notice the appearances of 15-j Wigner symbols, which are characteristic of the spinfoam theory amplitudes.

Thiemann's proof that the Hamiltonian operator is anomaly free  \cite{Thiemann96a} does not go through with the new operator, at least at first sight.  The proof was indeed based on the fact that the nodes generated were rather ``special" and had no volume, and therefore the Hamiltonian could not act on them. If one judges this ``special" form of the nodes generated to be a part of the theory that we need to keep, then this could be a problem for the operator considered here. On the other hand, the fact that the old operator generated such special nodes, on which it could not act again, is sometime viewed as an unconvincing aspect of the old construction, and this is partially corrected with the constraint considered here. (On different ways to address the issue, see in particular \cite{master,Giesel:2006uj}.)

Finally, in this paper we have restricted our attention to 4-valent nodes and the 1-4 Pachner move, but we think that the construction given here could be extended to include nodes of arbitrary valence.

\section*{Acknowledgments}

Thanks to Thomas Thiemann for helpful discussions and for an accurate reading of the draft.

\newpage

\appendix

\section{Recoupling theory}\label{recoupling}

We give here the definitions at the basis of recoupling theory and the graphical notation
that is used in the text. Our main reference source is \cite{BrinkSatchler68}. 

\begin{itemize}\setlength{\itemsep}{.2in}

\item{\em Wigner 3j-symbols.} These are represented by a 3-valent node, the three lines stand for the angular momenta wich are coupled by the  3j-symbol.
We denote the anti-clockwise orientation with a + sign and the clockwise orientation with a sign -. in index notation $v^{\alpha\beta\gamma}$: 
\begin{equation}
\begin{array}{cccc}
\left(\begin{array}{ccc}     a & b & c \\          \alpha & \beta & \gamma \end{array}\right)&=
	\begin{array}{c}
	\ifx\JPicScale\undefined\def\JPicScale{1}\fi
\psset{unit=\JPicScale mm}
\psset{linewidth=0.2,dotsep=1,hatchwidth=0.3,hatchsep=1.5,shadowsize=1,dimen=middle}
\psset{dotsize=0.7 2.5,dotscale=1 1,fillcolor=black}
\psset{arrowsize=1 2,arrowlength=1,arrowinset=0.25,tbarsize=0.7 5,bracketlength=0.15,rbracketlength=0.15}
\begin{pspicture}(0,0)(25,17.5)
\psline(10,10)(17.5,10)
\psline(17.5,10)(22.5,15)
\psline(17.5,10)(22.5,5)
\rput(10,12.5){$a\, \alpha$}
\rput(25,2.5){$b\,\beta$}
\rput(25,17.5){$c\,\gamma$}
\rput(24,10){$+$}
\end{pspicture}
\end{array}&=
\begin{array}{c}
	\ifx\JPicScale\undefined\def\JPicScale{1}\fi
\psset{unit=\JPicScale mm}
\psset{linewidth=0.2,dotsep=1,hatchwidth=0.3,hatchsep=1.5,shadowsize=1,dimen=middle}
\psset{dotsize=0.7 2.5,dotscale=1 1,fillcolor=black}
\psset{arrowsize=1 2,arrowlength=1,arrowinset=0.25,tbarsize=0.7 5,bracketlength=0.15,rbracketlength=0.15}
\begin{pspicture}(0,0)(25,17.5)
\psline(10,10)(17.5,10)
\psline(17.5,10)(22.5,15)
\psline(17.5,10)(22.5,5)
\rput(10,12.5){$a\, \alpha$}
\rput(25,2.5){$c\,\gamma$}
\rput(25,17.5){$b\,\beta$}
\rput(24,10){-}
\end{pspicture}
\end{array}
\end{array}\label{3j}
\end{equation}
The symmetry relation $v^{\alpha \beta\gamma}=(-1)^{a+b+c}\;v^{\alpha\gamma\beta}=\lambda_a^{bc}\;v^{\alpha\gamma\beta}$
\begin{equation}
	\left(\begin{array}{ccc}     a & b & c \\          \alpha & \beta & \gamma \end{array}\right)=
	(-1)^{a+b+c}\left(\begin{array}{ccc}     a & c & b \\          \alpha & \gamma & \beta \end{array}\right)
\end{equation}
implies

\begin{equation}
			\begin{array}{c}
\ifx\JPicScale\undefined\def\JPicScale{1}\fi
\psset{unit=\JPicScale mm}
\psset{linewidth=0.2,dotsep=1,hatchwidth=0.3,hatchsep=1.5,shadowsize=1,dimen=middle}
\psset{dotsize=0.7 2.5,dotscale=1 1,fillcolor=black}
\psset{arrowsize=1 2,arrowlength=1,arrowinset=0.25,tbarsize=0.7 5,bracketlength=0.15,rbracketlength=0.15}
\begin{pspicture}(0,0)(18,19)
\psline(0,10)(10,10)
\psline(10,10)(17,3)
\psline(10,10)(17,17)
\rput(13,10){+}
\rput(-1,12){$a$}
\rput(18,19){$b$}
\rput(18,1){$c$}
\rput(18,1){}
\end{pspicture}
\end{array}\;\;=(-1)^{a+b+c}\;\;
		\begin{array}{c}
\ifx\JPicScale\undefined\def\JPicScale{1}\fi
\psset{unit=\JPicScale mm}
\psset{linewidth=0.2,dotsep=1,hatchwidth=0.3,hatchsep=1.5,shadowsize=1,dimen=middle}
\psset{dotsize=0.7 2.5,dotscale=1 1,fillcolor=black}
\psset{arrowsize=1 2,arrowlength=1,arrowinset=0.25,tbarsize=0.7 5,bracketlength=0.15,rbracketlength=0.15}
\begin{pspicture}(0,0)(18,19)
\psline(0,10)(10,10)
\psline(10,10)(17,3)
\psline(10,10)(17,17)
\rput(13,10){-}
\rput(-1,12){$a$}
\rput(18,19){$b$}
\rput(18,1){$c$}
\rput(18,1){}
\end{pspicture}
\end{array}
\label{cambio orientazione}
\end{equation}

\item{\em The Kroneker delta.}

\begin{equation}
\delta_{ab}\;\delta^{\alpha}_{\beta}=\;\;	\begin{array}{c}	\ifx\JPicScale\undefined\def\JPicScale{1}\fi
\psset{unit=\JPicScale mm}
\psset{linewidth=0.2,dotsep=1,hatchwidth=0.3,hatchsep=1.5,shadowsize=1,dimen=middle}
\psset{dotsize=0.7 2.5,dotscale=1 1,fillcolor=black}
\psset{arrowsize=1 2,arrowlength=1,arrowinset=0.25,tbarsize=0.7 5,bracketlength=0.15,rbracketlength=0.15}
\begin{pspicture}(0,10)(20,12.5)
\psline(20,10)(0,10)
\rput(0,12.5){$a\alpha$}
\rput(20,12.5){$b\beta$}
\rput(15,10){}
\end{pspicture}
	\end{array}. 
\end{equation}

\item{\em First orthogonality relation for 3j-symbols.} 
\begin{equation}
	\sum_{\alpha,\beta}\left(\begin{array}{ccc}     a & b & c \\          \alpha & \beta & \gamma \end{array}\right)
	\left(\begin{array}{ccc}     a & b & c' \\          \alpha & \beta  & \gamma' \end{array}\right)=\frac{1}{2c+1}\;\delta_{cc'}\; \delta^{\gamma}_{\;\gamma'}
\end{equation}
\begin{equation}
\begin{array}{c}
		\ifx\JPicScale\undefined\def\JPicScale{0.8}\fi
	\psset{unit=\JPicScale mm}
	\psset{linewidth=0.2,dotsep=1,hatchwidth=0.3,hatchsep=1.5,shadowsize=1,dimen=middle}
	\psset{dotsize=0.7 2.5,dotscale=1 1,fillcolor=black}
	\psset{arrowsize=1 2,arrowlength=1,arrowinset=0.25,tbarsize=0.7 5,bracketlength=0.15,rbracketlength=0.15}
	\begin{pspicture}(0,0)(50,16)
	\rput{0}(24.5,7){\psellipse[](0,0)(10.5,-7)}
	\rput(25,2){$a$}
	\rput(25,16){$b	$}
	\rput(37,4){-}
	\rput(12,4){+}
	\psline(14,7)(0,7)
	\psline(35,7)(49,7)
	\rput(3,9){$c$}
	\rput(46,10){$c'$}
		\end{pspicture}
\end{array}
=\frac{1}{2c+1}
\begin{array}{c}	\ifx\JPicScale\undefined\def\JPicScale{1}\fi
\psset{unit=\JPicScale mm}
\psset{linewidth=0.2,dotsep=1,hatchwidth=0.3,hatchsep=1.5,shadowsize=1,dimen=middle}
\psset{dotsize=0.7 2.5,dotscale=1 1,fillcolor=black}
\psset{arrowsize=1 2,arrowlength=1,arrowinset=0.25,tbarsize=0.7 5,bracketlength=0.15,rbracketlength=0.15}
\begin{pspicture}(0,10)(20,12.5)
\psline(20,10)(0,10)
\rput(4,12.5){c$\gamma$}
\rput(20,12.5){c$'$$\gamma'$}
\rput(15,10){}
\end{pspicture}
	\end{array}
	\label{second ortogonality}
\end{equation}
This implies
\begin{equation}
\begin{array}{c}
		\ifx\JPicScale\undefined\def\JPicScale{0.8}\fi
	\psset{unit=\JPicScale mm}
	\psset{linewidth=0.3,dotsep=1,hatchwidth=0.3,hatchsep=1.5,shadowsize=1,dimen=middle}
	\psset{dotsize=0.7 2.5,dotscale=1 1,fillcolor=black}
	\psset{arrowsize=1 2,arrowlength=1,arrowinset=0.25,tbarsize=0.7 5,bracketlength=0.15,rbracketlength=0.15}
	\begin{pspicture}(0,0)(35,16)
	\rput{0}(16.5,7){\psellipse[](0,0)(10.5,-7)}
	\psline(27,7)(6,7)
	\rput(3,7){-}
	\rput(30,7){+}
	\rput(16,2){$a$}
	\rput(16,9){$c$}
	\rput(16,16){$b$}
	\rput(35,7){}
	\rput(40,7){}
	\rput[l](35,7){}
	\end{pspicture}
\end{array}
=1 \label{theta=1}
\end{equation}

\item{\em Second orthogonality relation.}
\begin{equation}
		\sum_{c\gamma}(2c+1)\left(\begin{array}{ccc}     a & b & c \\          \alpha & \beta & \gamma \end{array}\right)
	\left(\begin{array}{ccc}     a & b & c \\          \alpha' & \beta'  & \gamma \end{array}\right)=\;\delta^{\alpha}_{\;\alpha'}\; \delta^{\beta}_{\;\beta'}
\end{equation}
Graphically
\begin{equation}
\sum_{c}(2c+1)
\begin{array}{c}
	\ifx\JPicScale\undefined\def\JPicScale{0.9}\fi
\psset{unit=\JPicScale mm}
\psset{linewidth=0.2,dotsep=1,hatchwidth=0.3,hatchsep=1.5,shadowsize=1,dimen=middle}
\psset{dotsize=0.7 2.5,dotscale=1 1,fillcolor=black}
\psset{arrowsize=1 2,arrowlength=1,arrowinset=0.25,tbarsize=0.7 5,bracketlength=0.15,rbracketlength=0.15}
\begin{pspicture}(0,0)(27,18)
\psline(3,3)(10,10)
\psline(10,10)(3,17)
\psline(10,10)(20,10)
\psline(20,10)(27,17)
\psline(20,10)(27,3)
\rput(11,8){+}
\rput(19,8){-}
\rput(15,12){c}
\rput(5,18){$a\alpha$}
\rput(5,2){$b\beta$}
\rput(25,18){$a\alpha'$}
\rput(25,2){$b\beta'$}
\end{pspicture}\end{array}
=
\begin{array}{c}
	\ifx\JPicScale\undefined\def\JPicScale{0.9}\fi
\psset{unit=\JPicScale mm}
\psset{linewidth=0.2,dotsep=1,hatchwidth=0.3,hatchsep=1.5,shadowsize=1,dimen=middle}
\psset{dotsize=0.7 2.5,dotscale=1 1,fillcolor=black}
\psset{arrowsize=1 2,arrowlength=1,arrowinset=0.25,tbarsize=0.7 5,bracketlength=0.15,rbracketlength=0.15}
\begin{pspicture}(0,0)(27,18)
\psline(3,5)(27,5)
\psline(3,15)(27,15)
\rput(5,18){$a\alpha$}
\rput(5,2){$b\beta$}
\rput(25,18){$a\alpha'$}
\rput(25,2){$b\beta'$}
\end{pspicture}\end{array}
\label{sro}
\end{equation}

\item{\em The ``basic rule".}

\begin{equation}
\begin{split}
	&\sum_{\delta\epsilon\phi}(-1)^{d+e+f-\delta-\epsilon-\phi}
	 \left(\begin{array}{ccc}     d  & e & c \\          -\delta & \epsilon & \gamma \end{array}\right)
	\left(\begin{array}{ccc}     e  & f & a \\          -\epsilon & \phi & \alpha \end{array}\right) 
	\left(\begin{array}{ccc}     f  & d & b \\          -\phi & \delta & \beta \end{array}\right) \\
	\\ &\hspace{5em} =
\left\{\begin{array}{ccc}                      a  & b & c \\                      d & e & f                  \end{array}\right\} 	
	\left(\begin{array}{ccc}     a  & b & c \\          \alpha & \beta & \gamma \end{array}\right) 
\end{split}
\end{equation}

\begin{equation}\label{basic rule}
\begin{array}{c}
	\ifx\JPicScale\undefined\def\JPicScale{1}\fi
\psset{unit=\JPicScale mm}
\psset{linewidth=0.3,dotsep=1,hatchwidth=0.3,hatchsep=1.5,shadowsize=1,dimen=middle}
\psset{dotsize=0.7 2.5,dotscale=1 1,fillcolor=black}
\psset{arrowsize=1 2,arrowlength=1,arrowinset=0.25,tbarsize=0.7 5,bracketlength=0.15,rbracketlength=0.15}
\begin{pspicture}(0,0)(32,29)
\psline(3,3)(10,10)
\psline(10,22)(3,29)
\psline(10,22)(16,16)
\psline(10,10)(16,16)
\psline(9,9)(9,23)
\psline(16,16)(31,16)
\rput(10,8){-}
\rput(10,24){}
\rput(10,24){-}
\rput(17,14){-}
\rput(2,1){$a$}
\rput(2,31){$c$}
\rput(32,14){$b$}
\rput(6,16){$e$}
\rput(15,11){$f$}
\rput(15,21){$d$}
\rput(9,16){}
\psline{<-}(12,20)(16,16)
\psline{<-}(13,13)(9,9)
\psline{<-}(9,15)(9,23)
\end{pspicture}
\end{array}=\quad
\begin{array}{c}
	\ifx\JPicScale\undefined\def\JPicScale{1}\fi
\psset{unit=\JPicScale mm}
\psset{linewidth=0.3,dotsep=1,hatchwidth=0.3,hatchsep=1.5,shadowsize=1,dimen=middle}
\psset{dotsize=0.7 2.5,dotscale=1 1,fillcolor=black}
\psset{arrowsize=1 2,arrowlength=1,arrowinset=0.25,tbarsize=0.7 5,bracketlength=0.15,rbracketlength=0.15}
\begin{pspicture}(0,0)(29,25)
\psline(3,3)(3,23)
\psline(11,13)(3,23)
\psline(11,13)(3,3)
\psline(11,13)(27,13)
\psline(3,23)(27,13)
\psline{<-}(15,18)(27,13)
\psline(27,13)(3,3)
\psline{<-}(15,8)(3,3)
\psline{<-}(3,12)(3,23)
\rput(8,13){+}
\rput(29,13){+}
\rput(13,21){$d$}
\rput(13,5){$f$}
\rput(1,11){$e$}
\rput(14,15){$b$}
\rput(6,16){$c$}
\rput(6,10){$a$}
\rput(2,25){+}
\rput(2,1){+}
\end{pspicture}
\end{array}\begin{array}{c}
	\ifx\JPicScale\undefined\def\JPicScale{1}\fi
\psset{unit=\JPicScale mm}
\psset{linewidth=0.3,dotsep=1,hatchwidth=0.3,hatchsep=1.5,shadowsize=1,dimen=middle}
\psset{dotsize=0.7 2.5,dotscale=1 1,fillcolor=black}
\psset{arrowsize=1 2,arrowlength=1,arrowinset=0.25,tbarsize=0.7 5,bracketlength=0.15,rbracketlength=0.15}
\begin{pspicture}(0,0)(23,19)
\psline(3,3)(10,10)
\psline(3,17)(10,10)
\psline(10,10)(21,10)
\rput(7,10){+}
\rput(2,1){$a$}
\rput(2,19){$c$}
\rput(23,10){$b$}
\end{pspicture}
\end{array}
\end{equation}

\end{itemize}

\section{Simplify graphs}

Here are some useful recoupling theory relations used for simplfying spin network evaluations.\\

\be
\begin{array}{c}
\ifx\JPicScale\undefined\def\JPicScale{.8}\fi
\psset{unit=\JPicScale mm}
\psset{linewidth=0.3,dotsep=1,hatchwidth=0.3,hatchsep=1.5,shadowsize=1,dimen=middle}
\psset{dotsize=0.7 2.5,dotscale=1 1,fillcolor=black}
\psset{arrowsize=1 2,arrowlength=1,arrowinset=0.25,tbarsize=0.7 5,bracketlength=0.15,rbracketlength=0.15}
\begin{pspicture}(0,0)(70,20)
\pspolygon[fillstyle=vlines](0,20)(15,20)(15,0)(0,0)
\psline(15,6)(23,6)
\psline(15,14)(23,14)
\rput(28,10){$=$}
\pspolygon[fillstyle=vlines](34,20)(49,20)(49,0)(34,0)
\psline(49,14)(52,14)
\psline(49,6)(52,6)
\rput{0}(52,9.99){\psellipticarc[](0,0)(4,4){-92.52}{92.52}}
\rput{0}(63,10){\psellipticarc[](0,0)(4,4){90}{270}}
\rput(70,10){$\frac{\delta_{ j_1j_2}}{\dim j_1}$}
\rput(19,4){$j_1$}
\rput(19,16){$j_2$}
\rput(51,4){$j_1$}
\rput(62,4){$j_1$}
\end{pspicture}
\end{array}
\label{taglio1}
\ee\\

\be
\begin{array}{c}
	\ifx\JPicScale\undefined\def\JPicScale{1}\fi
\psset{unit=\JPicScale mm}
\psset{linewidth=0.3,dotsep=1,hatchwidth=0.3,hatchsep=1.5,shadowsize=1,dimen=middle}
\psset{dotsize=0.7 2.5,dotscale=1 1,fillcolor=black}
\psset{arrowsize=1 2,arrowlength=1,arrowinset=0.25,tbarsize=0.7 5,bracketlength=0.15,rbracketlength=0.15}
\begin{pspicture}(0,0)(67,20)
\pspolygon[fillstyle=vlines](0,20)(15,20)(15,0)(0,0)
\psline(15,5)(23,5)
\psline(15,15)(23,15)
\rput(28,10){$=$}
\pspolygon[fillstyle=vlines](34,20)(49,20)(49,0)(34,0)
\psline(49,15)(53,15)
\psline(49,5)(53,5)
\rput(19,3){$j_1$}
\rput(19,13){$j_3$}
\rput(52,3){$j_1$}
\rput(66,3){$j_1$}
\psline(15,10)(23,10)
\rput(19,8){$j_2$}
\psline(49,10)(57,10)
\psline(53,15)(57,10)
\psline(57,10)(53,5)
\psline(67,5)(64,5)
\psline(67,15)(64,15)
\rput(66,13){$j_3$}
\psline(67,10)(60,10)
\psline(64,5)(60,10)
\psline(60,10)(64,15)
\rput(52,13){$j_3$}
\rput(52,8){$j_2$}
\rput(66,8){$j_2$}
\end{pspicture}
\end{array}
\label{taglio2}
\ee\\

\be
\begin{array}{c}
	\ifx\JPicScale\undefined\def\JPicScale{1}\fi
\psset{unit=\JPicScale mm}
\psset{linewidth=0.3,dotsep=1,hatchwidth=0.3,hatchsep=1.5,shadowsize=1,dimen=middle}
\psset{dotsize=0.7 2.5,dotscale=1 1,fillcolor=black}
\psset{arrowsize=1 2,arrowlength=1,arrowinset=0.25,tbarsize=0.7 5,bracketlength=0.15,rbracketlength=0.15}
\begin{pspicture}(0,0)(78,19)
\pspolygon[fillstyle=vlines](0,19)(15,19)(15,0)(0,0)
\psline(15,2)(23,2)
\psline(15,12)(23,12)
\rput(28,10){$=$}
\pspolygon[fillstyle=vlines](38,19)(53,19)(53,0)(38,0)
\psline(53,12)(61,12)
\psline(53,2)(57,2)
\rput(19,0){$j_1$}
\rput(19,10){$j_3$}
\rput(56,0){$j_1$}
\rput(77,0){$j_1$}
\psline(15,7)(23,7)
\rput(19,5){$j_2$}
\psline(53,7)(61,7)
\psline(61,12)(61,7)
\psline(61,7)(57,2)
\psline(78,2)(75,2)
\psline(78,12)(71,12)
\rput(77,10){$j_3$}
\psline(78,7)(71,7)
\psline(75,2)(71,7)
\psline(71,7)(71,12)
\rput(56,10){$j_3$}
\rput(56,5){$j_2$}
\rput(77,5){$j_2$}
\psline(15,17)(23,17)
\rput(19,15){$j_4$}
\psline(53,17)(57,17)
\rput(56,15){$j_4$}
\psline(57,17)(61,12)
\psline(78,17)(74,17)
\psline(74,17)(71,12)
\rput(77,15){$j_4$}
\rput(63,9){$i$}
\rput(69,9){$i$}
\rput(33,10){$\sum_i d_i$}
\end{pspicture}
\end{array}
\label{taglio3}
\ee\\

\noindent In these relations the dashed block represent a completely contracted graph with no free legs.

\section{Volume}\label{Volume section}

We collect here some basic elements about the definition and the calculation of volume matrix elements, see also \cite{Brunnemann:2004xi}. 
Acting on a spinnetwork state, $ {V}$  doesn't change the graph nor the edge spins. It  only acts on the intertwiners at the node.
Let us restrict the attention here to the Ashtekar-Lewandowski volume operator. 
On a cylindrical function $\psi_\gamma$, it is given by
\begin{equation}
  \label{volume}
   {\hat{V}} \, \psi_\gamma = \sum_{v \in \mathcal{V}(\gamma)}
         {\hat{V}}_v \, \psi_\gamma ~,
\end{equation}
where 
\begin{equation}
  \label{V_v}
   {V}_v = l^3_0\, \sqrt{\left| \frac{i}{16\cdot3!} 
        \sum_{e_{I}\cap e_{J} \cap e_{K}=v} 
        \epsilon(e_{I},e_{J},e_{K})\,  {W}_{[IJK]} 
        \right|} ~.
\end{equation}
The first sum extends over the set $\mathcal{V}(\gamma)$ of nodes
of the underlying graph, while the sum in (\ref{V_v}) extends
over all triples $(e_I, e_J, e_K)$ of edges adjacent to
a node. The orientation factor 
$\epsilon(e_{I},e_{J},e_{K})$ is
$+1$ if the tangents $(\dot{e}_{I},\dot{e}_{J},\dot{e}_{K})$ 
at the node are positively oriented, $-1$ for negative 
orientation, and $0$ in the case of degenerate, i.e. linearly 
dependent or planar edges.
Besides, edges meeting in an $n$-node are assumed to be 
outgoing. 

The core of the operator (\ref{V_v}) is given by
$ {W}_{[IJK]}$ 
that acts on the finite dimensional intertwiner space of 
an $n$-valent node $v_n$. Its action is described in terms 
of the `grasping' \cite{AshtekarLewand98,RovelliSmolin95} of any three distinct edges $e_I,\,e_J$ and 
$e_K$ adjacent to $v_n$: a triple grasping operator represented as follows:
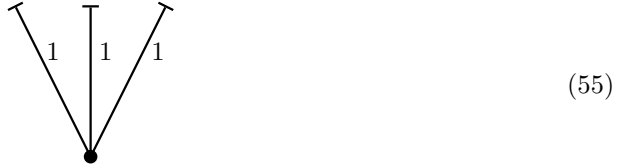
\be
\begin{array}{c}
\ifx\JPicScale\undefined\def\JPicScale{1}\fi
\psset{unit=\JPicScale mm}
\psset{linewidth=0.3,dotsep=1,hatchwidth=0.3,hatchsep=1.5,shadowsize=1,dimen=middle}
\psset{dotsize=0.7 2.5,dotscale=1 1,fillcolor=black}
\psset{arrowsize=1 2,arrowlength=1,arrowinset=0.25,tbarsize=0.7 5,bracketlength=0.15,rbracketlength=0.15}
\begin{pspicture}(0,0)(30,20)
\psline[dotsize=0.7 3.7]{-|*}(20,0)(10,20)
\psline[dotsize=0.7 3.7]{*-|*}(20,0)(20,20)
\psline[dotsize=0.7 3.7]{-|*}(20,0)(30,20)
\rput(15,14){1}
\rput(22,14){1}
\rput(29,14){1}
\end{pspicture}
\end{array}
\ee
It is a three valent node in rep $1$ that has three "free hands" that will be attached to three distinct adjacent edges $(e_I, e_J, e_K)$ of $v_n$, creating a single new three valent node on each of these edges. Note that for every triple of edges, $ {W}_{[IJK]}$ in (\ref{V_v}) affects only the intertwiner associated to $v_n$; in the graphical notation it will add links and nodes only inside the dashed circles that represent the node. 
Restricting the action to \emph{real} edges only, 
the volume operator is equally well-defined on non-gauge-invariant 
nodes.

The volume operator `grasps' triples of real edges 
$(e_I, e_J, e_K)$ adjacent to a node. In the case of a non-gauge invariant n-valent node we can easily calculate its action looking at the action on gauge invariants n+1-valent nodes. In fact for any valence of the node the operator contributes one term for each triple: 
the action on non-gauge invariant nodes is simply given by the action on gauge invariant ones done grasping only the real edges and not the free index edges of the node.

\subsection{4-valent not gauge invariant case}
\label{4-valent not gauge}
We are now ready to compute the action of the volume operator on the 4-valent non gauge invariant node generated by the action of the first holonomy contained in the hamiltonian operator. 
We obtain
\begin{eqnarray}
   {V} \left(  {h}^{(m)}[s^{-1}_{k}] \, \ket{v} \right)
        &=& \sum_c d_c \; {V}_v\; 
        \begin{array}{c}
	\ifx\JPicScale\undefined\def\JPicScale{1}\fi
\psset{unit=\JPicScale mm}
\psset{linewidth=0.3,dotsep=1,hatchwidth=0.3,hatchsep=1.5,shadowsize=1,dimen=middle}
\psset{dotsize=0.7 2.5,dotscale=1 1,fillcolor=black}
\psset{arrowsize=1 2,arrowlength=1,arrowinset=0.25,tbarsize=0.7 5,bracketlength=0.15,rbracketlength=0.15}
\begin{pspicture}(0,0)(30.31,27)
\psline[dotsize=1.3 2.5]{-*}(8,7)(16,15)
\psline(16,15)(24,7)
\psline(16,15)(16,24)
\rput{0}(16,8.5){\psellipticarc[linewidth=0.25,linestyle=dashed,dash=0.5 0.5](0,0)(8.5,8.5){-118.07}{-61.93}}
\rput{0}(18,12.69){\psellipticarc[linewidth=0.2,linestyle=dashed,dash=0.5 0.5](0,0)(12.3,12.3){-8.57}{65.17}}
\rput{0}(13.58,13.15){\psellipticarc[linewidth=0.2,linestyle=dashed,dash=0.5 0.5](0,0)(11.78,11.78){112.89}{190.51}}
\rput(5,6){$j_i$}
\rput(27,6){$j_j$}
\psline(16,21)(21,21)
\psline{->}(21,21)(18,21)
\rput(21,23){$m$}
\rput(13,19){$j_k$}
\rput(14,26){$c$}
\end{pspicture}
\end{array}
=
        \nonumber \\
        &=& \sum_c \,d_c\, \frac{l_0^3}{4} \,  
                \, \sqrt{\left| i  {W}_{[j_ij_jc]} \right|}
           \;
            \begin{array}{c}
	\ifx\JPicScale\undefined\def\JPicScale{1}\fi
\psset{unit=\JPicScale mm}
\psset{linewidth=0.3,dotsep=1,hatchwidth=0.3,hatchsep=1.5,shadowsize=1,dimen=middle}
\psset{dotsize=0.7 2.5,dotscale=1 1,fillcolor=black}
\psset{arrowsize=1 2,arrowlength=1,arrowinset=0.25,tbarsize=0.7 5,bracketlength=0.15,rbracketlength=0.15}
\begin{pspicture}(0,0)(30.31,27)
\psline[dotsize=1.3 2.5]{-*}(8,7)(16,15)
\psline(16,15)(24,7)
\psline(16,15)(16,24)
\rput{0}(16,8.5){\psellipticarc[linewidth=0.25,linestyle=dashed,dash=0.5 0.5](0,0)(8.5,8.5){-118.07}{-61.93}}
\rput{0}(18,12.69){\psellipticarc[linewidth=0.2,linestyle=dashed,dash=0.5 0.5](0,0)(12.3,12.3){-8.57}{65.17}}
\rput{0}(13.58,13.15){\psellipticarc[linewidth=0.2,linestyle=dashed,dash=0.5 0.5](0,0)(11.78,11.78){112.89}{190.51}}
\rput(5,6){$j_i$}
\rput(27,6){$j_j$}
\psline(16,21)(21,21)
\psline{->}(21,21)(18,21)
\rput(21,23){$m$}
\rput(13,19){$j_k$}
\rput(14,26){$c$}
\end{pspicture}
\end{array}
    \label{vol_on_ngi_3vertex}
\end{eqnarray}
In the last equation we have used the fact that the volume operator is linear and on a single 4-valent non-gauge invariant node  
both sums in (\ref{volume}) and (\ref{V_v}) reduce to a
single term. The operator $ {W}_{[j_ij_jc]}$ 
denotes the grasping of the three real edges of the 
non-gauge-invariant 3-node, colored $j_i\,j_j$ and $c$ in 
this order. The $3!$ factor in (\ref{V_v}) is canceled out by 
those terms that appear due to permutations of the three grasped 
edges, since they are all equal up a sign.
The action of $ {W}$ (before taking the absolute value and the square root) on a non-gauge-invariant 3-node can generally 
be expressed as
\begin{equation}
  \label{grapheqn_for_W}
   {\hat{W}}_{[j_ij_jc]} 
    \;\; \begin{array}{c}
	\ifx\JPicScale\undefined\def\JPicScale{1}\fi
\psset{unit=\JPicScale mm}
\psset{linewidth=0.3,dotsep=1,hatchwidth=0.3,hatchsep=1.5,shadowsize=1,dimen=middle}
\psset{dotsize=0.7 2.5,dotscale=1 1,fillcolor=black}
\psset{arrowsize=1 2,arrowlength=1,arrowinset=0.25,tbarsize=0.7 5,bracketlength=0.15,rbracketlength=0.15}
\begin{pspicture}(0,0)(30.31,27)
\psline[dotsize=1.3 2.5]{-*}(8,7)(16,15)
\psline(16,15)(24,7)
\psline(16,15)(16,24)
\rput{0}(16,8.5){\psellipticarc[linewidth=0.25,linestyle=dashed,dash=0.5 0.5](0,0)(8.5,8.5){-118.07}{-61.93}}
\rput{0}(18,12.69){\psellipticarc[linewidth=0.2,linestyle=dashed,dash=0.5 0.5](0,0)(12.3,12.3){-8.57}{65.17}}
\rput{0}(13.58,13.15){\psellipticarc[linewidth=0.2,linestyle=dashed,dash=0.5 0.5](0,0)(11.78,11.78){112.89}{190.51}}
\rput(5,6){$j_i$}
\rput(27,6){$j_j$}
\psline(16,21)(21,21)
\psline{->}(21,21)(18,21)
\rput(21,23){$m$}
\rput(14,18){$\alpha$}
\rput(14,26){$c$}
\end{pspicture}
\end{array}
=
 \sum_\beta W^{(4)}_{[j_ij_jc]}(j_i,j_j,m,c){}_\alpha{}^\beta
    \;\; \begin{array}{c}
	\ifx\JPicScale\undefined\def\JPicScale{1}\fi
\psset{unit=\JPicScale mm}
\psset{linewidth=0.3,dotsep=1,hatchwidth=0.3,hatchsep=1.5,shadowsize=1,dimen=middle}
\psset{dotsize=0.7 2.5,dotscale=1 1,fillcolor=black}
\psset{arrowsize=1 2,arrowlength=1,arrowinset=0.25,tbarsize=0.7 5,bracketlength=0.15,rbracketlength=0.15}
\begin{pspicture}(0,0)(30.31,27)
\psline[dotsize=1.3 2.5]{-*}(8,7)(16,15)
\psline(16,15)(24,7)
\psline(16,15)(16,24)
\rput{0}(16,8.5){\psellipticarc[linewidth=0.25,linestyle=dashed,dash=0.5 0.5](0,0)(8.5,8.5){-118.07}{-61.93}}
\rput{0}(18,12.69){\psellipticarc[linewidth=0.2,linestyle=dashed,dash=0.5 0.5](0,0)(12.3,12.3){-8.57}{65.17}}
\rput{0}(13.58,13.15){\psellipticarc[linewidth=0.2,linestyle=dashed,dash=0.5 0.5](0,0)(11.78,11.78){112.89}{190.51}}
\rput(5,6){$j_i$}
\rput(27,6){$j_j$}
\psline(16,21)(21,21)
\psline{->}(21,21)(18,21)
\rput(21,23){$m$}
\rput(14,18){$\beta$}
\rput(14,26){$c$}
\end{pspicture}
\end{array}
\end{equation}
or using vector notation for the state vectors, as
\begin{equation}
  \label{eqn_for_W}
   {\hat{W}}_{[j_ij_jc]} \; \ket{v_\alpha}
   = \sum_\beta W^{(4)}_{[j_ij_jc]}(j_i,j_j,m,c){}_\alpha{}^\beta \; 
        \ket{v_\beta} ~.
\end{equation}
The sum over $\beta$ range on the dimension of the intertwiner space determined by the Clebsh-Gordan at the two three valent nodes.
$W^{(4)}_{[j_ij_jc]}$ are the matrix elements of the operator $ {W}_{[j_ij_jc]}$ that acts on the triple of edges colored $(j_i,j_j,c)$, in a basis of 4-valent nodes.
It has been calculated for the first time in
\cite{DePietriRovelli96}, where the general case concerning the
volume operator acting on $n$-valent nodes is considered.
In our case it reads
\be
\begin{array}{c}
	\ifx\JPicScale\undefined\def\JPicScale{1}\fi
\psset{unit=\JPicScale mm}
\psset{linewidth=0.3,dotsep=1,hatchwidth=0.3,hatchsep=1.5,shadowsize=1,dimen=middle}
\psset{dotsize=0.7 2.5,dotscale=1 1,fillcolor=black}
\psset{arrowsize=1 2,arrowlength=1,arrowinset=0.25,tbarsize=0.7 5,bracketlength=0.15,rbracketlength=0.15}
\begin{pspicture}(0,0)(27,29)
\psline[dotsize=1.3 2.5]{-*}(8,7)(16,15)
\psline(16,15)(24,7)
\psline(16,15)(16,27)
\rput(5,6){$j_i$}
\rput(27,6){$j_j$}
\psline(16,21)(21,21)
\rput(21,23){$m$}
\rput(14,18){$j_k$}
\rput(16,29){$c$}
\psline{-)}(2,18)(11,10)
\psline{-)}(2,18)(22,9)
\psline{-)}(2,18)(16,25)
\rput(14,22){$c$}
\rput(5,22){$1$}
\rput(7,18){$1$}
\rput(4,14){$1$}
\end{pspicture}
\end{array}
=N^c N^{j_i}N^{j_j}\sum_i\;d_i
\begin{array}{c}
\ifx\JPicScale\undefined\def\JPicScale{1}\fi
\psset{unit=\JPicScale mm}
\psset{linewidth=0.3,dotsep=1,hatchwidth=0.3,hatchsep=1.5,shadowsize=1,dimen=middle}
\psset{dotsize=0.7 2.5,dotscale=1 1,fillcolor=black}
\psset{arrowsize=1 2,arrowlength=1,arrowinset=0.25,tbarsize=0.7 5,bracketlength=0.15,rbracketlength=0.15}
\begin{pspicture}(0,0)(70.31,28)
\psline(16,15)(22,9)
\psline(16,15)(16,25)
\rput(28,16){$\scr{j_i}$}
\rput(30,12){$\scr{j_j}$}
\psline(16,21)(21,21)
\rput(23,20){$\scr{m}$}
\rput(14,18){$\scr{j_k}$}
\rput(24,25){$\scr{c}$}
\psline(2,18)(22,9)
\psline(2,18)(16,25)
\rput(15,23){$\scr{c}$}
\rput(5,22){$\scr{1}$}
\rput(7,18){$\scr{1}$}
\rput(4,14){$\scr{1}$}
\psline[border=0.45](11,10)(16,15)
\psline(2,18)(11,10)
\psline[border=0.3](11,10)(32,15)
\psline(32,15)(22,9)
\psline(32,15)(32,21)
\psline(32,21)(21,21)
\rput(35,18){$i$}
\psline(16,25)(32,21)
\rput(19,15){$\scr{j_j}$}
\rput(13,15){$\scr{j_i}$}
\psline[dotsize=1.3 2.5]{-*}(48,8)(56,16)
\psline(56,16)(64,8)
\psline(56,16)(56,25)
\rput{0}(56,9.5){\psellipticarc[linewidth=0.25,linestyle=dashed,dash=0.5 0.5](0,0)(8.5,8.5){-118.07}{-61.93}}
\rput{0}(58,13.69){\psellipticarc[linewidth=0.2,linestyle=dashed,dash=0.5 0.5](0,0)(12.3,12.3){-8.57}{65.17}}
\rput{0}(53.58,14.15){\psellipticarc[linewidth=0.2,linestyle=dashed,dash=0.5 0.5](0,0)(11.78,11.78){112.89}{190.51}}
\rput(45,7){$j_i$}
\rput(67,7){$j_j$}
\psline(56,22)(61,22)
\psline{->}(61,22)(58,22)
\rput(61,24){$m$}
\rput(53,19){$i$}
\rput(53,28){$c$}
\end{pspicture}	
\end{array}
\label{4valente graspato}
\ee
Where we have used the grasping operators in the left hand side and the relation \eqref{taglio3} in the right hand side of \eqref{4valente graspato}
  
It has also been shown that in an appropriate basis the operators 
$i  {W}$ are represented by antisymmetric, purely 
imaginary, i.e. hermitian matrices, which are diagonalizable 
and have real eigenvalues \cite{DePietriRovelli96}. Hence the absolute value and the square root in (\ref{vol_on_ngi_3vertex}) are well-defined. 

This basis is realized by a 
rescaling, or node normalization respectively.
The virtual internal edge is multiplied by $\sqrt{\dim}$
\be
\left|v_\alpha\right\rangle_N=\sqrt{\dim\alpha}\left|v_\alpha\right\rangle
\label {rescaling}
\ee,
With this normalization, (\ref{eqn_for_W}) is 
rewritten as
\begin{equation}
  \label{norm_eqn_for_W:2}
   {\hat{W}}_{[j_ij_jc]} \; \ket{v_\alpha}_N
   = \sum_\beta \frac{\sqrt{\dim\alpha}}{\sqrt{\dim\beta}}W^{(4)}_{[j_ij_jc]}(j_i,j_j,m,c){}_\alpha{}^\beta \; 
        \ket{v_\beta}_N =\sum_\beta \tilde{W}^{(4)}_{[j_ij_jc]}(j_i,j_j,m,c){}_\alpha{}^\beta \; 
        \ket{v_\beta}_N
\end{equation}
where $\tilde{W}^{(4)}_{[j_ij_jc]}$ are the matrix elements of $ {W}_{[j_ij_jc]}$ between two normalized states.

\be
\begin{split}
&\tilde{W}^{(4)}_{[j_ij_jc]}(j_i,j_j,m,c){}_\alpha{}^\beta
=\frac{\sqrt{\dim\alpha}}{\sqrt{\dim\beta}}N^c N^{j_i}N^{j_j}\;
\begin{array}{c}
\ifx\JPicScale\undefined\def\JPicScale{1}\fi
\psset{unit=\JPicScale mm}
\psset{linewidth=0.3,dotsep=1,hatchwidth=0.3,hatchsep=1.5,shadowsize=1,dimen=middle}
\psset{dotsize=0.7 2.5,dotscale=1 1,fillcolor=black}
\psset{arrowsize=1 2,arrowlength=1,arrowinset=0.25,tbarsize=0.7 5,bracketlength=0.15,rbracketlength=0.15}
\begin{pspicture}(0,0)(40,28)
\psline(16,15)(22,9)
\psline(16,15)(16,25)
\rput(28,16){$\scr{j_i}$}
\rput(30,12){$\scr{j_j}$}
\psline(16,21)(21,21)
\rput(23,20){$\scr{m}$}
\rput(14,18){$\scr{\alpha}$}
\rput(24,25){$\scr{c}$}
\psline(2,18)(22,9)
\psline(2,18)(16,25)
\rput(15,23){$\scr{c}$}
\rput(5,22){$\scr{1}$}
\rput(7,18){$\scr{1}$}
\rput(4,14){$\scr{1}$}
\psline[border=0.45](11,10)(16,15)
\psline(2,18)(11,10)
\psline[border=0.3](11,10)(32,15)
\psline(32,15)(22,9)
\psline(32,15)(32,21)
\psline(32,21)(21,21)
\rput(35,18){$\beta$}
\psline(16,25)(32,21)
\rput(19,15){$\scr{j_j}$}
\rput(13,15){$\scr{j_i}$}
\end{pspicture}	
\end{array}
=
\\
&
=\frac{\sqrt{\dim\alpha}}{\sqrt{\dim\beta}}N^c N^{j_i}N^{j_j} \left\{
\begin{array}{ccc}
\alpha  & \beta & 1 \\
c & c & m
\end{array}
\right\}
\begin{array}{c}
	\ifx\JPicScale\undefined\def\JPicScale{1}\fi
\psset{unit=\JPicScale mm}
\psset{linewidth=0.3,dotsep=1,hatchwidth=0.3,hatchsep=1.5,shadowsize=1,dimen=middle}
\psset{dotsize=0.7 2.5,dotscale=1 1,fillcolor=black}
\psset{arrowsize=1 2,arrowlength=1,arrowinset=0.25,tbarsize=0.7 5,bracketlength=0.15,rbracketlength=0.15}
\begin{pspicture}(0,0)(30,25)
\psline(16,15)(22,9)
\psline(16,15)(16,25)
\rput(28,16){$\scr{j_i}$}
\rput(30,12){$\scr{j_j}$}
\rput(14,18){$\scr{\alpha}$}
\psline(2,18)(22,9)
\psline(2,18)(16,25)
\rput(5,22){$\scr{1}$}
\rput(7,18){$\scr{1}$}
\rput(4,14){$\scr{1}$}
\psline[border=0.45](11,9)(16,15)
\psline(2,18)(11,9)
\psline[border=0.3](11,9)(30,18)
\psline(30,18)(22,9)
\rput(27,22){$\scr{\beta}$}
\psline(16,25)(30,18)
\rput(19,14){$\scr{j_j}$}
\rput(13,14){$\scr{j_i}$}
\end{pspicture}
\end{array}\\
&=
\frac{\sqrt{\dim\alpha}}{\sqrt{\dim\beta}}N^c N^{j_i}N^{j_j} 
\left\{
\begin{array}{ccc}
\alpha  & \beta & 1 \\
c & c & m
\end{array}
\right\}
\left\{
\begin{array}{ccc}                    
  1  & 1 & 1 \\     
   j_i & j_j & \beta \\ 
      j_i & j_j & \alpha
\end{array}
\right\} 	
\label{4valente graspato fine}
\end{split}
\ee
Where in the last equation we have used the basic rule \eqref{basic rule}
on the upper triangle reducing the network to a $9j-$symbol. The symmetry properties of the 9j symbol (or equivalently the Clebsh Gordan condition in the node $(1,\alpha,\beta)$) imply that the antisymmetric matrix 
$\tilde{W}^{(4)}{}_\alpha{}^\beta$ has non-zero elements only in the entries that are subject to \mbox{$|\alpha-\beta| = 1$}. 
Hence $\tilde{W}^{(4)}{}_\alpha{}^\beta$ has only sub- and superdiagonal non-zero entries and is real and antisymmetric in the rescaled basis.
Since the required $i \tilde{W}^{(4)}$ is hermitian, it can be diagonalized and from this form we can extract the required absolute 
value and square root in a well-defined way.
The action of the volume operator is in general 
\emph{not} diagonal.\footnote{One exception 
turns out to be given by Thiemann's original 
$m=1$ operator. In this case, the action of the volume is indeed diagonal,
and $\tilde{W}^{(4)}$ is a $(2 \times 2)$ matrix, allowing 
explicit calculations \cite{Borissovetal97}}
However for arbitrary $m$ and spins of the node there is not an explicit general analytic formula.

The relation between the node operator $ {V}_v$ and 
the square root of the local grasp $i  {W}$ reads
in the trivalent case
\begin{equation}
  \label{V_def}
  \sqrt{|i W|}\: {}_\alpha{}^\beta = (V_v){}_\alpha{}^\beta 
        \equiv V{}_\alpha{}^\beta ~.
\end{equation}
Inserting this in (\ref{vol_on_ngi_3vertex}), we obtain
for the non-diagonal action of the volume operator
\begin{eqnarray}
  \label{final_action_of_V}
   {V} \left(  {h}^{(m)}[s^{-1}_{k}] \, \ket{v} \right)
        &=&\sum_c \,d_c\, \frac{l_0^3}{4} \,  
                \, \sqrt{\left| i  {W}_{[j_ij_jc]} \right|}
           \;
            \begin{array}{c}
	\ifx\JPicScale\undefined\def\JPicScale{1}\fi
\psset{unit=\JPicScale mm}
\psset{linewidth=0.3,dotsep=1,hatchwidth=0.3,hatchsep=1.5,shadowsize=1,dimen=middle}
\psset{dotsize=0.7 2.5,dotscale=1 1,fillcolor=black}
\psset{arrowsize=1 2,arrowlength=1,arrowinset=0.25,tbarsize=0.7 5,bracketlength=0.15,rbracketlength=0.15}
\begin{pspicture}(0,0)(30.31,27)
\psline[dotsize=1.3 2.5]{-*}(8,7)(16,15)
\psline(16,15)(24,7)
\psline(16,15)(16,24)
\rput{0}(16,8.5){\psellipticarc[linewidth=0.25,linestyle=dashed,dash=0.5 0.5](0,0)(8.5,8.5){-118.07}{-61.93}}
\rput{0}(18,12.69){\psellipticarc[linewidth=0.2,linestyle=dashed,dash=0.5 0.5](0,0)(12.3,12.3){-8.57}{65.17}}
\rput{0}(13.58,13.15){\psellipticarc[linewidth=0.2,linestyle=dashed,dash=0.5 0.5](0,0)(11.78,11.78){112.89}{190.51}}
\rput(5,6){$j_i$}
\rput(27,6){$j_j$}
\psline(16,21)(21,21)
\psline{->}(21,21)(18,21)
\rput(21,23){$m$}
\rput(13,19){$j_k$}
\rput(14,26){$c$}
\end{pspicture}
\end{array}  \nonumber \\
        &=& \sum_c \,d_c\, \frac{l_0^3}{4} \,  
                \, \sum_{\beta} V{}_{j_k}{}^\beta (j_i,j_j,m,c)
           \;
            \begin{array}{c}
	\ifx\JPicScale\undefined\def\JPicScale{1}\fi
\psset{unit=\JPicScale mm}
\psset{linewidth=0.3,dotsep=1,hatchwidth=0.3,hatchsep=1.5,shadowsize=1,dimen=middle}
\psset{dotsize=0.7 2.5,dotscale=1 1,fillcolor=black}
\psset{arrowsize=1 2,arrowlength=1,arrowinset=0.25,tbarsize=0.7 5,bracketlength=0.15,rbracketlength=0.15}
\begin{pspicture}(0,0)(30.31,27)
\psline[dotsize=1.3 2.5]{-*}(8,7)(16,15)
\psline(16,15)(24,7)
\psline(16,15)(16,24)
\rput{0}(16,8.5){\psellipticarc[linewidth=0.25,linestyle=dashed,dash=0.5 0.5](0,0)(8.5,8.5){-118.07}{-61.93}}
\rput{0}(18,12.69){\psellipticarc[linewidth=0.2,linestyle=dashed,dash=0.5 0.5](0,0)(12.3,12.3){-8.57}{65.17}}
\rput{0}(13.58,13.15){\psellipticarc[linewidth=0.2,linestyle=dashed,dash=0.5 0.5](0,0)(11.78,11.78){112.89}{190.51}}
\rput(5,6){$j_i$}
\rput(27,6){$j_j$}
\psline(16,21)(21,21)
\psline{->}(21,21)(18,21)
\rput(21,23){$m$}
\rput(13,19){$\beta$}
\rput(14,26){$c$}
\end{pspicture}
\end{array} 
     \nonumber \\
      &=&  \sum_{c,\beta} \,d_c\, \frac{l_0^3}{4} \,  
                \, V{}_{j_k}{}^\beta (j_i,j_j,m,c)
           \;
            \begin{array}{c}
	\ifx\JPicScale\undefined\def\JPicScale{0.6}\fi
\psset{unit=\JPicScale mm}
\psset{linewidth=0.3,dotsep=1,hatchwidth=0.3,hatchsep=1.5,shadowsize=1,dimen=middle}
\psset{dotsize=0.7 2.5,dotscale=1 1,fillcolor=black}
\psset{arrowsize=1 2,arrowlength=1,arrowinset=0.25,tbarsize=0.7 5,bracketlength=0.15,rbracketlength=0.15}
\begin{pspicture}(20,0)(90,74)
\psline[linewidth=0.2,linestyle=dashed,dash=0.5 0.5](10,17)(36,43)
\psline[linewidth=0.2,linestyle=dashed,dash=0.5 0.5](64,43)(90,17)
\psline[linewidth=0.2,linestyle=dashed,dash=0.5 0.5](57,74)(57,56)
\psline[linewidth=0.2,linestyle=dashed,dash=0.5 0.5](43,74)(43,56)
\psline[linewidth=0.2,linestyle=dashed,dash=0.5 0.5](46,33)(30,17)
\psline[linewidth=0.2,linestyle=dashed,dash=0.5 0.5](54,33)(70,17)
\psline[dotsize=1.3 2.5]{-*}(20,17)(50,47)
\psline(50,47)(80,17)
\psline(50,47)(50,74)
\rput{0}(50,40.5){\psellipticarc[linewidth=0.25,linestyle=dashed,dash=0.5 0.5](0,0)(8.5,8.5){-118.07}{-61.93}}
\rput{0}(52,44.7){\psellipticarc[linewidth=0.2,linestyle=dashed,dash=0.5 0.5](0,0)(12.3,12.3){-8.57}{65.17}}
\rput{0}(47.58,45.15){\psellipticarc[linewidth=0.2,linestyle=dashed,dash=0.5 0.5](0,0)(11.78,11.78){112.89}{190.51}}
\rput(30,31){$\scr{j_i}$}
\rput(70,31){$\scr{j_j}$}
\rput(46,69){$\scr{j_k}$}
\rput(55,65){$\scr{m}$}
\psline(50,53)(55,53)
\psline(50,63)(55,63)
\psline{->}(55,53)(52,53)
\psline{<-}(53,63)(50,63)
\rput(55,55){$\scr{m}$}
\rput(46,51){$\scr{\beta}$}
\rput(47,59){$\scr{c}$}
\end{pspicture}
\end{array}
\end{eqnarray}

\subsection{5-valent not gauge invariant case}
\label{5-valent not gauge}

The volume on a 5-valent, not gauge invariant is constructed from the basic triple grasping operator $\hat{W}_{[IJK]}$ acting on the all the four triples of real edges:
\begin{equation}
  \label{grapheqn_for_W5}
   {\hat{W}}_{[j_ij_jj_kc]} 
    \;\; \begin{array}{c}
\ifx\JPicScale\undefined\def\JPicScale{1}\fi
\psset{unit=\JPicScale mm}
\psset{linewidth=0.3,dotsep=1,hatchwidth=0.3,hatchsep=1.5,shadowsize=1,dimen=middle}
\psset{dotsize=0.7 2.5,dotscale=1 1,fillcolor=black}
\psset{arrowsize=1 2,arrowlength=1,arrowinset=0.25,tbarsize=0.7 5,bracketlength=0.15,rbracketlength=0.15}
\begin{pspicture}(0,0)(35,29.04)
\psline(8,6)(11,15)
\psline(20,15)(25,6)
\psline(11,15)(7,25)
\psline(20,15)(30,12)
\psline[linewidth=0.6,linestyle=dashed,dash=1 1](11,15)(20,15)
\rput{90}(17.2,14.54){\psellipticarc[linewidth=0.2,linestyle=dashed,dash=0.5 0.5](0,0)(14.5,-14){-27.38}{80.22}}
\rput{0}(17,15){\psellipticarc[linewidth=0.2,linestyle=dashed,dash=0.5 0.5](0,0)(14,-14){67.38}{111.04}}
\rput{50.55}(17.1,15.02){\psellipticarc[linewidth=0.2,linestyle=dashed,dash=0.5 0.5](0,0)(14.06,-13.96){-151.8}{-104.8}}
\rput(-3,27){}
\rput(6,4){$\scr{j_i}$}
\rput(27,4){$\scr{j_j}$}
\rput(35,10){$\scr{j_k}$}
\rput(7,17){$\scr{j_l}$}
\rput(16,18){$\scr{i_n}$}
\psline(9,21)(14,23)
\rput(16,25){$\scr{m}$}
\rput(5,27){$\scr{c}$}
\end{pspicture}
\end{array}
=
 \sum_{\alpha \beta} W^{(5)}_{[j_ij_jj_kc]}(j_i,j_j,j_k,m,c){}_{i_n, \,j_l}{}^{\alpha,\,\beta}
    \;\; \begin{array}{c}
\ifx\JPicScale\undefined\def\JPicScale{1}\fi
\psset{unit=\JPicScale mm}
\psset{linewidth=0.3,dotsep=1,hatchwidth=0.3,hatchsep=1.5,shadowsize=1,dimen=middle}
\psset{dotsize=0.7 2.5,dotscale=1 1,fillcolor=black}
\psset{arrowsize=1 2,arrowlength=1,arrowinset=0.25,tbarsize=0.7 5,bracketlength=0.15,rbracketlength=0.15}
\begin{pspicture}(0,0)(35,29.04)
\psline(8,6)(11,15)
\psline(20,15)(25,6)
\psline(11,15)(7,25)
\psline(20,15)(30,12)
\psline[linewidth=0.6,linestyle=dashed,dash=1 1](11,15)(20,15)
\rput{90}(17.2,14.54){\psellipticarc[linewidth=0.2,linestyle=dashed,dash=0.5 0.5](0,0)(14.5,-14){-27.38}{80.22}}
\rput{0}(17,15){\psellipticarc[linewidth=0.2,linestyle=dashed,dash=0.5 0.5](0,0)(14,-14){67.38}{111.04}}
\rput{50.55}(17.1,15.02){\psellipticarc[linewidth=0.2,linestyle=dashed,dash=0.5 0.5](0,0)(14.06,-13.96){-151.8}{-104.8}}
\rput(-3,27){}
\rput(6,4){$\scr{j_i}$}
\rput(27,4){$\scr{j_j}$}
\rput(35,10){$\scr{j_k}$}
\rput(7,17){$\scr{\alpha}$}
\rput(16,18){$\scr{\beta}$}
\psline(9,21)(14,23)
\rput(16,25){$\scr{m}$}
\rput(5,27){$\scr{c}$}
\end{pspicture}
\end{array}
\end{equation}

The triple grasping operator is the sum of four terms
\be
\hat{W}_{[j_ij_jj_k c]}=\hat{W}_{[j_ij_jj_k]}+\hat{W}_{[j_ij_jc]}+\hat{W}_{[j_ij_kc]}+\hat{W}_{[j_ij_jc]}
\ee
The first reads 
\be
\begin{split}
\hspace{3em}&\hspace{-3em}\begin{array}{c}
\ifx\JPicScale\undefined\def\JPicScale{.9}\fi
\psset{unit=\JPicScale mm}
\psset{linewidth=0.3,dotsep=1,hatchwidth=0.3,hatchsep=1.5,shadowsize=1,dimen=middle}
\psset{dotsize=0.7 2.5,dotscale=1 1,fillcolor=black}
\psset{arrowsize=1 2,arrowlength=1,arrowinset=0.25,tbarsize=0.7 5,bracketlength=0.15,rbracketlength=0.15}
\begin{pspicture}(0,0)(38,30)
\psline(11,9)(14,18)
\psline(14,18)(10,28)
\psline(23,18)(33,15)
\psline[linewidth=0.6,linestyle=dashed,dash=1 1](14,18)(23,18)
\rput(0,30){}
\rput(9,7){$\scr{j_i}$}
\rput(30,7){$\scr{j_j}$}
\rput(38,13){$\scr{j_k}$}
\rput(10,20){$\scr{j_l}$}
\rput(19,21){$\scr{i_n}$}
\psline(12,24)(17,26)
\rput(19,28){$\scr{m}$}
\rput(8,30){$\scr{c}$}
\psline{(-}(12,13)(20,1)
\psline{-)}(20,1)(32,15)
\psline[border=0.4](23,18)(28,9)
\psline{-)}(20,1)(26,13)
\rput(15,4){$\scr{1}$}
\rput(25,4){$\scr{1}$}
\rput(21,8){$\scr{1}$}
\end{pspicture}	
\end{array}
=\\&
=N^{j_i} N^{j_j} N^{j_k}
\;\sum_{\alpha\beta} d_{\alpha} \; d_{\beta}
\begin{array}{c}
\ifx\JPicScale\undefined\def\JPicScale{.9}\fi
\psset{unit=\JPicScale mm}
\psset{linewidth=0.3,dotsep=1,hatchwidth=0.3,hatchsep=1.5,shadowsize=1,dimen=middle}
\psset{dotsize=0.7 2.5,dotscale=1 1,fillcolor=black}
\psset{arrowsize=1 2,arrowlength=1,arrowinset=0.25,tbarsize=0.7 5,bracketlength=0.15,rbracketlength=0.15}
\begin{pspicture}(0,0)(75,43)
\psline(12,26)(14,31)
\psline(23,31)(32,28)
\psline[linewidth=0.6,linestyle=dashed,dash=1 1](14,31)(23,31)
\rput(14,11){$\scr{j_i}$}
\rput(31,13){$\scr{j_j}$}
\rput(37,21){$\scr{j_k}$}
\rput(11,35){$\scr{j_l}$}
\rput(19,34){$\scr{i_n}$}
\rput(6,18){$\scr{m}$}
\rput(2,24){$\scr{c}$}
\psline(12,26)(20,14)
\psline(20,14)(32,28)
\psline[border=0.4](23,31)(29,21)
\psline(20,14)(26,26)
\rput(15,17){$\scr{1}$}
\rput(25,17){$\scr{1}$}
\rput(21,21){$\scr{1}$}
\rput(8,2){$\scr{\alpha}$}
\rput(28,2){$\scr{\beta}$}
\psline(48,11)(51,20)
\psline(60,20)(65,11)
\psline(51,20)(47,30)
\psline(60,20)(70,17)
\psline[linewidth=0.6,linestyle=dashed,dash=1 1](51,20)(60,20)
\rput(46,9){$\scr{j_i}$}
\rput(67,9){$\scr{j_j}$}
\rput(75,15){$\scr{j_k}$}
\rput(47,22){$\scr{\alpha}$}
\rput(56,23){$\scr{\beta}$}
\psline(49,26)(54,28)
\rput(56,30){$\scr{m}$}
\rput(45,32){$\scr{c}$}
\rput(14,28){$\scr{j_i}$}
\rput(22,28){$\scr{j_j}$}
\rput(30,32){$\scr{j_k}$}
\psline(32,28)(39,4)
\psline(3,1)(14,37)
\psline[border=0.3](2,4)(39,4)
\psline(0,9)(3,1)
\psline[border=0.3](14,31)(10,43)
\psline(0,9)(10,43)
\psline(12,37)(14,37)
\psline(29,21)(39,4)
\psline(12,4)(12,26)
\end{pspicture}\end{array}\\
&=N^{j_i} N^{j_j} N^{j_k}
\;\sum_{\beta} \; d_{\beta}
\begin{array}{c}
	\ifx\JPicScale\undefined\def\JPicScale{.9}\fi
\psset{unit=\JPicScale mm}
\psset{linewidth=0.3,dotsep=1,hatchwidth=0.3,hatchsep=1.5,shadowsize=1,dimen=middle}
\psset{dotsize=0.7 2.5,dotscale=1 1,fillcolor=black}
\psset{arrowsize=1 2,arrowlength=1,arrowinset=0.25,tbarsize=0.7 5,bracketlength=0.15,rbracketlength=0.15}
\begin{pspicture}(0,0)(75,37)
\psline(12,26)(14,31)
\psline(23,31)(32,28)
\psline[linewidth=0.6,linestyle=dashed,dash=1 1](14,31)(23,31)
\rput(14,11){$\scr{j_i}$}
\rput(31,13){$\scr{j_j}$}
\rput(37,21){$\scr{j_k}$}
\rput(5,22){$\scr{j_l}$}
\rput(19,34){$\scr{i_n}$}
\psline(12,26)(20,14)
\psline(20,14)(32,28)
\psline[border=0.4](23,31)(29,21)
\psline(20,14)(26,26)
\rput(15,17){$\scr{1}$}
\rput(25,17){$\scr{1}$}
\rput(21,21){$\scr{1}$}
\rput(28,2){$\scr{\beta}$}
\psline(48,11)(51,20)
\psline(60,20)(65,11)
\psline(51,20)(47,30)
\psline(60,20)(70,17)
\psline[linewidth=0.6,linestyle=dashed,dash=1 1](51,20)(60,20)
\rput(46,9){$\scr{j_i}$}
\rput(67,9){$\scr{j_j}$}
\rput(75,15){$\scr{j_k}$}
\rput(47,22){$\scr{j_l}$}
\rput(56,23){$\scr{\beta}$}
\psline(49,26)(54,28)
\rput(56,30){$\scr{m}$}
\rput(45,32){$\scr{c}$}
\rput(14,28){$\scr{j_i}$}
\rput(22,28){$\scr{j_j}$}
\rput(30,32){$\scr{j_k}$}
\psline(32,28)(39,4)
\psline[border=0.3](2,4)(39,4)
\psline[border=0.3](14,31)(12,37)
\psline(2,4)(12,37)
\psline(29,21)(39,4)
\psline(12,4)(12,26)
\end{pspicture}
\end{array}
\\
&=N^{j_i} N^{j_j} N^{j_k}
\;\sum_{\beta} \; d_{\beta}
\left\{\begin{array}{ccc}                      \beta  & 1 & i_n \\                      j_i & j_l & j_i                  \end{array}\right\} 	
\begin{array}{c}
	\ifx\JPicScale\undefined\def\JPicScale{.9}\fi
\psset{unit=\JPicScale mm}
\psset{linewidth=0.3,dotsep=1,hatchwidth=0.3,hatchsep=1.5,shadowsize=1,dimen=middle}
\psset{dotsize=0.7 2.5,dotscale=1 1,fillcolor=black}
\psset{arrowsize=1 2,arrowlength=1,arrowinset=0.25,tbarsize=0.7 5,bracketlength=0.15,rbracketlength=0.15}
\begin{pspicture}(0,0)(64,31)
\psline(12,28)(21,25)
\psline[linewidth=0.6,linestyle=dashed,dash=1 1](3,28)(12,28)
\rput(20,10){$\scr{j_j}$}
\rput(26,18){$\scr{j_k}$}
\rput(8,31){$\scr{i_n}$}
\psline(3,28)(9,11)
\psline(9,11)(21,25)
\psline[border=0.4](12,28)(18,18)
\psline(9,11)(15,23)
\rput(5,16){$\scr{1}$}
\rput(14,14){$\scr{1}$}
\rput(10,18){$\scr{1}$}
\rput(17,-1){$\scr{\beta}$}
\psline(37,8)(40,17)
\psline(49,17)(54,8)
\psline(40,17)(36,27)
\psline(49,17)(59,14)
\psline[linewidth=0.6,linestyle=dashed,dash=1 1](40,17)(49,17)
\rput(35,6){$\scr{j_i}$}
\rput(56,6){$\scr{j_j}$}
\rput(64,12){$\scr{j_k}$}
\rput(36,19){$\scr{j_l}$}
\rput(45,20){$\scr{\beta}$}
\psline(38,23)(43,25)
\rput(45,27){$\scr{m}$}
\rput(34,29){$\scr{c}$}
\rput(11,25){$\scr{j_j}$}
\rput(19,29){$\scr{j_k}$}
\psline(21,25)(28,1)
\psline[border=0.3](1,1)(28,1)
\psline(18,18)(28,1)
\psline(1,1)(3,28)
\end{pspicture}
\end{array}
\\
&=
N^{j_i} N^{j_j} N^{j_k}
\;\sum_{\beta} \; d_{\beta}
\left\{\begin{array}{ccc}                      \beta  & 1 & i_n \\                      j_i & j_l & j_i                  \end{array}\right\}
\left\{
\begin{array}{ccc}                    
  1  & 1 & 1 \\     
   j_j & j_k & \beta \\ 
      j_j & j_k & i_n
\end{array}
\right\} 	
\begin{array}{c}
	\ifx\JPicScale\undefined\def\JPicScale{.9}\fi
\psset{unit=\JPicScale mm}
\psset{linewidth=0.3,dotsep=1,hatchwidth=0.3,hatchsep=1.5,shadowsize=1,dimen=middle}
\psset{dotsize=0.7 2.5,dotscale=1 1,fillcolor=black}
\psset{arrowsize=1 2,arrowlength=1,arrowinset=0.25,tbarsize=0.7 5,bracketlength=0.15,rbracketlength=0.15}
\begin{pspicture}(0,0)(38,24)
\psline(11,3)(14,12)
\psline(23,12)(28,3)
\psline(14,12)(10,22)
\psline(23,12)(33,9)
\psline[linewidth=0.6,linestyle=dashed,dash=1 1](14,12)(23,12)
\rput(0,24){}
\rput(9,1){$\scr{j_i}$}
\rput(30,1){$\scr{j_j}$}
\rput(38,7){$\scr{j_k}$}
\rput(10,14){$\scr{j_l}$}
\rput(19,15){$\scr{\beta}$}
\psline(12,18)(17,20)
\rput(19,22){$\scr{m}$}
\rput(8,24){$\scr{c}$}
\end{pspicture}
\end{array}
\label{volume5valente1}
\end{split}
\ee

the second is

\be
\begin{split}
&\begin{array}{c}
	\ifx\JPicScale\undefined\def\JPicScale{.9}\fi
\psset{unit=\JPicScale mm}
\psset{linewidth=0.3,dotsep=1,hatchwidth=0.3,hatchsep=1.5,shadowsize=1,dimen=middle}
\psset{dotsize=0.7 2.5,dotscale=1 1,fillcolor=black}
\psset{arrowsize=1 2,arrowlength=1,arrowinset=0.25,tbarsize=0.7 5,bracketlength=0.15,rbracketlength=0.15}
\begin{pspicture}(0,0)(38,30)
\psline(14,18)(10,28)
\psline(23,18)(33,15)
\psline[linewidth=0.6,linestyle=dashed,dash=1 1](14,18)(23,18)
\rput(0,30){}
\rput(9,7){$\scr{j_i}$}
\rput(30,7){$\scr{j_j}$}
\rput(38,13){$\scr{j_k}$}
\rput(10,20){$\scr{j_l}$}
\rput(19,21){$\scr{i_n}$}
\psline(12,24)(17,26)
\rput(19,28){$\scr{m}$}
\rput(8,30){$\scr{c}$}
\psline[border=0.4](23,18)(28,9)
\rput(15,3){$\scr{1}$}
\rput(25,4){$\scr{1}$}
\rput(21,8){$\scr{1}$}
\pscustom[]{\psbezier{-}(20,1)(2,17)(10,25)(10,25)
\psline{-)}(10,25)(11,26)
}
\psline[border=0.4](11,9)(14,18)
\psline{-)}(20,1)(13,14)
\psline{-)}(20,1)(26,13)
\end{pspicture}
\end{array}
=
N^{c} N^{j_i} N^{j_j}
\;\sum_{\alpha\beta} d_{\alpha} \; d_{\beta}
\begin{array}{c}
\ifx\JPicScale\undefined\def\JPicScale{.9}\fi
\psset{unit=\JPicScale mm}
\psset{linewidth=0.3,dotsep=1,hatchwidth=0.3,hatchsep=1.5,shadowsize=1,dimen=middle}
\psset{dotsize=0.7 2.5,dotscale=1 1,fillcolor=black}
\psset{arrowsize=1 2,arrowlength=1,arrowinset=0.25,tbarsize=0.7 5,bracketlength=0.15,rbracketlength=0.15}
\begin{pspicture}(0,0)(75,43)
\psline(12,26)(14,31)
\psline(23,31)(32,28)
\psline[linewidth=0.6,linestyle=dashed,dash=1 1](14,31)(23,31)
\rput(14,11){$\scr{j_i}$}
\rput(31,13){$\scr{j_j}$}
\rput(11,35){$\scr{j_l}$}
\rput(19,34){$\scr{i_n}$}
\rput(6,18){$\scr{m}$}
\rput(2,24){$\scr{c}$}
\psline[border=0.4](23,31)(29,21)
\rput(8,2){$\scr{\alpha}$}
\rput(28,2){$\scr{\beta}$}
\psline(48,11)(51,20)
\psline(60,20)(65,11)
\psline(51,20)(47,30)
\psline(60,20)(70,17)
\psline[linewidth=0.6,linestyle=dashed,dash=1 1](51,20)(60,20)
\rput(46,9){$\scr{j_i}$}
\rput(67,9){$\scr{j_j}$}
\rput(75,15){$\scr{j_k}$}
\rput(47,22){$\scr{\alpha}$}
\rput(56,23){$\scr{\beta}$}
\psline(49,26)(54,28)
\rput(56,30){$\scr{m}$}
\rput(45,32){$\scr{c}$}
\rput(15,28){$\scr{j_i}$}
\rput(22,28){$\scr{j_j}$}
\rput(30,32){$\scr{j_k}$}
\psline(32,28)(39,4)
\psline(3,1)(14,37)
\psline[border=0.3](2,4)(39,4)
\psline(0,9)(3,1)
\psline[border=0.3](14,31)(10,43)
\psline(0,9)(10,43)
\psline(12,37)(14,37)
\psline(29,21)(39,4)
\psline(12,4)(12,26)
\rput(15,17){$\scr{1}$}
\rput(25,18){$\scr{1}$}
\rput(18,22){$\scr{1}$}
\pscustom[border=0.3]{\psbezier(20,15)(2,31)(10,39)(10,39)
\psline(10,39)(11,40)
\stroke[linewidth=0.9,linecolor=white]}
\psline(20,15)(13,28)
\psline(20,15)(26,27)
\rput(13,39){$\scr{c}$}
\end{pspicture}\end{array}
\\
&=(-1)^{2c+1} N^{c} N^{j_i} N^{j_j}
\;\sum_{\alpha\beta} d_{\alpha} \; d_{\beta} \left\{
\begin{array}{ccc}
1  & \alpha & j_l \\
m & c & c
\end{array}
\right\}
\begin{array}{c}
\ifx\JPicScale\undefined\def\JPicScale{.9}\fi
\psset{unit=\JPicScale mm}
\psset{linewidth=0.3,dotsep=1,hatchwidth=0.3,hatchsep=1.5,shadowsize=1,dimen=middle}
\psset{dotsize=0.7 2.5,dotscale=1 1,fillcolor=black}
\psset{arrowsize=1 2,arrowlength=1,arrowinset=0.25,tbarsize=0.7 5,bracketlength=0.15,rbracketlength=0.15}
\begin{pspicture}(0,0)(75,34)
\psline(12,26)(14,31)
\psline(23,31)(32,28)
\psline[linewidth=0.6,linestyle=dashed,dash=1 1](14,31)(23,31)
\rput(14,11){$\scr{j_i}$}
\rput(31,13){$\scr{j_j}$}
\rput(6,31){$\scr{j_l}$}
\rput(19,34){$\scr{i_n}$}
\psline[border=0.4](23,31)(29,21)
\rput(5,14){$\scr{\alpha}$}
\rput(28,2){$\scr{\beta}$}
\psline(48,11)(51,20)
\psline(60,20)(65,11)
\psline(51,20)(47,30)
\psline(60,20)(70,17)
\psline[linewidth=0.6,linestyle=dashed,dash=1 1](51,20)(60,20)
\rput(46,9){$\scr{j_i}$}
\rput(67,9){$\scr{j_j}$}
\rput(75,15){$\scr{j_k}$}
\rput(47,22){$\scr{\alpha}$}
\rput(56,23){$\scr{\beta}$}
\psline(49,26)(54,28)
\rput(56,30){$\scr{m}$}
\rput(45,32){$\scr{c}$}
\rput(15,28){$\scr{j_i}$}
\rput(22,28){$\scr{j_j}$}
\rput(30,32){$\scr{j_k}$}
\psline(32,28)(39,4)
\psline[border=0.3](12,4)(39,4)
\psline[border=0.3](14,31)(2,27)
\psline(29,21)(39,4)
\psline(12,4)(12,26)
\rput(16,15){$\scr{1}$}
\rput(25,18){$\scr{1}$}
\rput(18,22){$\scr{1}$}
\psline(20,15)(13,28)
\psline(20,15)(26,27)
\psline[border=0.3](20,15)(2,27)
\psline(2,27)(12,4)
\end{pspicture}
\end{array}\\
&=(-1)^{2c+1} N^{c} N^{j_i} N^{j_j}
\sum_{\alpha\beta} d_{\alpha} \; d_{\beta} \left\{
\begin{array}{ccc}
1  & \alpha & j_l \\
m & c & c
\end{array}
\right\}
\left\{\begin{array}{ccc}
1  & \beta & i_n \\
j_k & j_j & j_j
\end{array}
\right\}\hspace{-1em}
\begin{array}{c}
\ifx\JPicScale\undefined\def\JPicScale{.9}\fi
\psset{unit=\JPicScale mm}
\psset{linewidth=0.3,dotsep=1,hatchwidth=0.3,hatchsep=1.5,shadowsize=1,dimen=middle}
\psset{dotsize=0.7 2.5,dotscale=1 1,fillcolor=black}
\psset{arrowsize=1 2,arrowlength=1,arrowinset=0.25,tbarsize=0.7 5,bracketlength=0.15,rbracketlength=0.15}
\begin{pspicture}(0,0)(65,34)
\psline(12,26)(14,31)
\psline[linewidth=0.6,linestyle=dashed,dash=1 1](14,31)(23,31)
\rput(14,11){$\scr{j_i}$}
\rput(6,31){$\scr{j_l}$}
\rput(19,34){$\scr{i_n}$}
\rput(5,14){$\scr{\alpha}$}
\rput(19,2){$\scr{\beta}$}
\psline(38,10)(41,19)
\psline(50,19)(55,10)
\psline(41,19)(37,29)
\psline(50,19)(60,16)
\psline[linewidth=0.6,linestyle=dashed,dash=1 1](41,19)(50,19)
\rput(36,8){$\scr{j_i}$}
\rput(57,8){$\scr{j_j}$}
\rput(65,14){$\scr{j_k}$}
\rput(37,21){$\scr{\alpha}$}
\rput(46,22){$\scr{\beta}$}
\psline(39,25)(44,27)
\rput(46,29){$\scr{m}$}
\rput(35,31){$\scr{c}$}
\rput(15,28){$\scr{j_i}$}
\psline[border=0.3](12,4)(23,4)
\psline[border=0.3](14,31)(2,27)
\psline(12,4)(12,26)
\rput(16,15){$\scr{1}$}
\rput(20,24){$\scr{1}$}
\rput(18,21){$\scr{1}$}
\psline(20,15)(13,28)
\psline(20,15)(23,31)
\psline[border=0.3](20,15)(2,27)
\psline(2,27)(12,4)
\psline(23,31)(23,4)
\end{pspicture}
\end{array}\\
&=
(-1)^{2c+1} N^{c} N^{j_i} N^{j_j}
\;\sum_{\alpha\beta} d_{\alpha} \; d_{\beta} \left\{
\begin{array}{ccc}
1  & \alpha & j_l \\
m & c & c
\end{array}
\right\}
\left\{\begin{array}{ccc}
1  & \beta & i_n \\
j_k & j_j & j_j
\end{array}\right\}
\left\{\begin{array}{ccc}
\alpha  & 1 & j_l \\
\beta & 1 & i_n \\
j_i  & 1 & j_i
\end{array}\right\}
\hspace{-2em}
\begin{array}{c}
	\ifx\JPicScale\undefined\def\JPicScale{.9}\fi
\psset{unit=\JPicScale mm}
\psset{linewidth=0.3,dotsep=1,hatchwidth=0.3,hatchsep=1.5,shadowsize=1,dimen=middle}
\psset{dotsize=0.7 2.5,dotscale=1 1,fillcolor=black}
\psset{arrowsize=1 2,arrowlength=1,arrowinset=0.25,tbarsize=0.7 5,bracketlength=0.15,rbracketlength=0.15}
\begin{pspicture}(0,0)(38,24)
\psline(11,3)(14,12)
\psline(23,12)(28,3)
\psline(14,12)(10,22)
\psline(23,12)(33,9)
\psline[linewidth=0.6,linestyle=dashed,dash=1 1](14,12)(23,12)
\rput(0,24){}
\rput(9,1){$\scr{j_i}$}
\rput(30,1){$\scr{j_j}$}
\rput(38,7){$\scr{j_k}$}
\rput(10,14){$\scr{\alpha}$}
\rput(19,15){$\scr{\beta}$}
\psline(12,18)(17,20)
\rput(19,22){$\scr{m}$}
\rput(8,24){$\scr{c}$}
\end{pspicture}
\end{array}
\end{split}
\ee
The third term $\hat{W}_[j_i,j_k,c]$ is just the previous one with the exchange $j_j, j_k$ in the coefficients, and 
the last $\hat{W}_[j_j,j_k,c]$ is
\be
\begin{split}
&\begin{array}{c}
\ifx\JPicScale\undefined\def\JPicScale{.9}\fi
\psset{unit=\JPicScale mm}
\psset{linewidth=0.3,dotsep=1,hatchwidth=0.3,hatchsep=1.5,shadowsize=1,dimen=middle}
\psset{dotsize=0.7 2.5,dotscale=1 1,fillcolor=black}
\psset{arrowsize=1 2,arrowlength=1,arrowinset=0.25,tbarsize=0.7 5,bracketlength=0.15,rbracketlength=0.15}
\begin{pspicture}(0,0)(39,30)
\psline(15,18)(11,28)
\psline(24,18)(34,15)
\psline[linewidth=0.6,linestyle=dashed,dash=1 1](15,18)(24,18)
\rput(1,30){}
\rput(10,7){$\scr{j_i}$}
\rput(31,7){$\scr{j_j}$}
\rput(39,13){$\scr{j_k}$}
\rput(11,20){$\scr{j_l}$}
\rput(20,21){$\scr{i_n}$}
\psline(13,24)(18,26)
\rput(20,28){$\scr{m}$}
\rput(9,30){$\scr{c}$}
\psline{-)}(21,1)(33,15)
\psline[border=0.4](24,18)(29,9)
\psline{-)}(21,1)(27,13)
\rput(16,3){$\scr{1}$}
\rput(26,4){$\scr{1}$}
\rput(22,8){$\scr{1}$}
\pscustom[]{\psbezier{-}(21,1)(3,17)(11,25)(11,25)
\psline{-)}(11,25)(12,26)
}
\psline[border=0.4](12,9)(15,18)
\end{pspicture}
\end{array}=
\\
&=(-1)^{2c+1}N^{j_j} N^{j_k} N^{c}
\;\sum_{\alpha\beta} \; d_{\alpha}d_{\beta}
\left\{\begin{array}{ccc}                      \alpha  & 1 & j_l \\                      c & m & c                  \end{array}\right\}
\left\{\begin{array}{ccc}                      i_n  & \beta & 1 \\                      \alpha & j_l & j_j                  \end{array}\right\}
\left\{
\begin{array}{ccc}                    
  \beta  & j_j & j_k \\     
   1 & 1 & 1 \\ 
      i_n & j_j & j_k
\end{array}
\right\} \hspace{-2em}
\begin{array}{c}
	\ifx\JPicScale\undefined\def\JPicScale{.9}\fi
\psset{unit=\JPicScale mm}
\psset{linewidth=0.3,dotsep=1,hatchwidth=0.3,hatchsep=1.5,shadowsize=1,dimen=middle}
\psset{dotsize=0.7 2.5,dotscale=1 1,fillcolor=black}
\psset{arrowsize=1 2,arrowlength=1,arrowinset=0.25,tbarsize=0.7 5,bracketlength=0.15,rbracketlength=0.15}
\begin{pspicture}(0,0)(38,24)
\psline(11,3)(14,12)
\psline(23,12)(28,3)
\psline(14,12)(10,22)
\psline(23,12)(33,9)
\psline[linewidth=0.6,linestyle=dashed,dash=1 1](14,12)(23,12)
\rput(0,24){}
\rput(9,1){$\scr{j_i}$}
\rput(30,1){$\scr{j_j}$}
\rput(38,7){$\scr{j_k}$}
\rput(10,14){$\scr{j_l}$}
\rput(19,15){$\scr{\beta}$}
\psline(12,18)(17,20)
\rput(19,22){$\scr{m}$}
\rput(8,24){$\scr{c}$}
\end{pspicture}
\end{array}
\end{split}
\ee
The final volume matrix elements are obtained summing the previous operator and diagonalizing them to extract the square root (this last operation is the one that prevents us from having a close analytic expression).

%\bibliographystyle{hunsrt}%{abbrv}{abuser}
%\bibliography{BiblioCarlo}

%\bibliographystyle{plain}

%\end{document}

\end{document}